\documentclass[journal,a4paper,oneside,onecolumn]{IEEEtran} 
\usepackage{amsmath,amsfonts,amssymb}
\usepackage{algorithm}
\usepackage{algpseudocode}
\makeatletter
\let\OldStatex\Statex
\renewcommand{\Statex}[1][3]{%
 \setlength\@tempdima{\algorithmicindent}%
 \OldStatex\hskip\dimexpr#1\@tempdima\relax}
\makeatother

\usepackage{pict2e}

\newtheorem{thm}{Theorem}

\newtheorem{lem}{Lemma}
\newtheorem{rem}{Remark}

\newcommand{\lrB}[1]{\left[{#1}\right]}
\newcommand{\lrb}[1]{\left\{{#1}\right\}}
\newcommand{\lrsb}[1]{\left({#1}\right)}
\newcommand{\lrbar}[1]{\left|{#1}\right|}

\newcommand{\e}{\varepsilon}
\newcommand{\qed}{$\blacksquare$}
\newcommand{\limn}{\lim_{n\to\infty}}

\newcommand{\Error}{\mathrm{Error}}
\newcommand{\Prob}{\mathrm{Prob}}
\newcommand{\fMAP}{f_{\mathrm{MAP}}}

\newcommand{\I}{\mathcal{I}}
\newcommand{\U}{\mathcal{U}}
\newcommand{\V}{\mathcal{V}}
\newcommand{\cP}{\mathcal{P}}
\newcommand{\X}{\mathcal{X}}
\newcommand{\Y}{\mathcal{Y}}

\newcommand{\ff}{\boldsymbol{f}}

\newcommand{\uu}{\boldsymbol{u}}
\newcommand{\UU}{\boldsymbol{U}}
\newcommand{\vv}{\boldsymbol{v}}
\newcommand{\xx}{\boldsymbol{x}}
\newcommand{\XX}{\boldsymbol{X}}
\newcommand{\yy}{\boldsymbol{y}}
\newcommand{\YY}{\boldsymbol{Y}}
\newcommand{\ttheta}{\boldsymbol{\theta}}

\newcommand{\sfA}{\mathsf{A}}
\newcommand{\sfC}{\mathsf{C}}
\newcommand{\sfE}{\mathsf{E}}
\newcommand{\sfQ}{\mathsf{Q}}
\newcommand{\sfSigma}{\mathsf{\Sigma}}

\newcommand{\hu}{\widehat{u}}
\newcommand{\hv}{\widehat{v}}
\newcommand{\hV}{\widehat{V}}
\newcommand{\hx}{\widehat{x}}
\newcommand{\hcV}{\widehat{\V}}
\newcommand{\hcP}{\widehat{\cP}}
\newcommand{\huu}{\widehat{\uu}}
\newcommand{\hxx}{\widehat{\xx}}

\newcommand{\tU}{\widetilde{U}}

\newcommand{\usfQ}{\underline{\sfQ}}
\newcommand{\osfQ}{\overline{\sfQ}}
\newcommand{\ueps}{\underline{\e}}
\newcommand{\oeps}{\overline{\e}}
\newcommand{\udelta}{\underline{\delta}}
\newcommand{\odelta}{\overline{\delta}}
\newcommand{\otheta}{\overline{\theta}}

\allowdisplaybreaks

\begin{document}

\title{
 Binary Polar Codes Based on Bit Error Probability
}
\author{
 Jun~Muramatsu~\IEEEmembership{Senior Member,~IEEE}
 \thanks{J.~Muramatsu is with
  NTT Communication Science Laboratories, NTT Corporation,
  2-4, Hikaridai, Seika-cho, Soraku-gun, Kyoto 619-0237, Japan
  (E-mail: jun.muramatsu@ieee.org).
}}

\maketitle

\begin{abstract}
 This paper introduces techniques to construct
 binary polar source/channel codes
 based on the bit error probability of successive-cancellation decoding.
 The polarization lemma is reconstructed
 based on the bit error probability
 and then techniques to compute the bit error probability are introduced.
 These techniques can be applied to the construction of polar codes
 and the computation of lower and upper bounds
 of the block decoding error probability.
\end{abstract}

\section{Introduction}
Polar source/channel codes were introduced by Ar\i{}kan~\cite{A09,A10,A11}.
When these codes are applied to source coding with decoder side information
for joint stationary memoryless sources,
the coding rate achieves a fundamental limit called the conditional entropy.
When applied to the channel coding for a symmetric channel,
the coding rate achieves a fundamental limit called the channel capacity.
Ar\i{}kan introduced successive-cancellation decoding,
which can be implemented with computational
complexity $O(N\log_2 N)$ with respect to block length $N$.

One theoretical element of polar codes is the notion of polarization.
The polarization theorem is usually introduced
using mutual information, conditional entropy,
and the Battacharyya parameters~\cite{A09,A10,S12}.

In this paper,
by assembling scattered ideas
in the previous results~\cite{A15,AT09,GR20,GHU12,GYB18,MT14,S12,TV13},
we directly analyze the bit error probability
of successive-cancellation decoding instead of the Battacharyya parameter
to simplify the analysis of the block decoding error probability
and introduce techniques to compute the bit error probability.
With these techniques, we have the lower and upper bounds
of the block decoding error probability
tighter than the one in terms of the Battacharyya parameter
or the conditional entropy.

Main contribution of this paper is the introduction of
novel loss functions for a greedy degrading-/upgrading-merge algorithm
and the evaluation of the degrading/upgrading loss
after a single polar transform.
With this technique, we can reduce
the computational complexity of the construction of polar codes.
When the joint probability distribution consists of rational numbers,
the computation can also be precisely performed using the arithmetic of
rational numbers if sufficient memory is available.
In contrast,
we had to focused on the precision of real numbers
to compute irrational (square root or logarithmic) functions
in conventional techniques.

The rest of this paper is organized as follows.
Section~\ref{sec:definition} introduces the definitions and notations.
Section~\ref{sec:polar-code} reviews polar codes based on bit error probability.
Section~\ref{sec:construction} introduces
their construction based on bit error probability,
where a novel loss function for the greedy degrading-/upgrading-merge algorithm
is introduced.
Section~\ref{sec:experiment} introduces experimental results
on the computation of bit error probability.
Theorems and lemmas are proved in Section~\ref{sec:proof}.
A remark is mentioned in Section~\ref{sec:conclusion}.
Some lemmas are shown in Appendix.

\section{Definitions and Notations}
\label{sec:definition}

Throughout this paper, we use the following definitions and notations.

For given $n$, let $2^n$ denote the block length.
We use the bit-indexing approach introduced in \cite{A09}.
The indexes of a $2^n$-dimensional vector are represented by
$n$-bit sequences as $X^{2^n}\equiv(X_{0^n},\ldots,X_{1^n})$,
where $0^n$/$1^n$ is the $n$-bit all zero/one sequence.
For given $\I\subset\{0,1\}^n$, we define the sub-sequences of $X^{2^n}$ as
\begin{equation*}
 X_{\I}
 \equiv\{X_{b^n}\}_{b^n\in\I}.
\end{equation*}

For a given $b^n\equiv(b_1,\ldots,b_n)$,
we use notations
\begin{align*}
 [0^n:b^n]
 &\equiv
 \{0^n,\ldots,b^n\}
 \\
 [0^n:b^n)
 &\equiv
 [0^n:b^n]\setminus\{b^n\}
 \\
 [b^n:1^n]
 &\equiv
 [0^n:1^n]\setminus[0^n:b^n)
 \\
 (b^n:1^n]
 &\equiv
 [0^n:1^n]\setminus[0^n:b^n]
\end{align*}
to represent an interval of integers.
Let $bb^k,b^kb\in\{0,1\}^{k+1}$ be the concatenations of 
$b\in\{0,1\}$ and $b^k\in\{0,1\}^k$.
For given $b\in\{0,1\}$ and $b^k\in\{0,1\}^k$,
define $b[0:b^k],b[0:b^k)\subset\{0,1\}^{k+1}$ as
\begin{align*}
 b[0:b^k]
 &\equiv
 \{bb'^k: b'^k\in[0^k:b^k]\}
 \\
 b[0:b^k)
 &\equiv
 \{bb'^k: b'^k\in[0^k:b^k)\}.
\end{align*}

Bipolar-binary conversions $\pm_b$ and $\mp_b$ of $b\in\{0,1\}$ are defined as
\begin{align*}
 \pm_b
 &\equiv
 \begin{cases}
  +
  &\text{if $b=1$}
  \\
  -
  &\text{if $b=0$}
 \end{cases}
 \\
 \mp_b
 &\equiv
 \begin{cases}
  -
  &\text{if $b=1$}
  \\
  +
  &\text{if $b=0$}.
 \end{cases}
\end{align*}

\section{Binary Polar Codes}
\label{sec:polar-code}
In this section, we revisit the binary polar source/channel codes introduced
in previous works~\cite{A09,A10,A11,S12}.

Assume that $\X=\{0,1\}$ is the binary finite field.
For given positive integer $n$, polar transform $G$ is defined as
\begin{equation*}
 G\equiv
 \begin{pmatrix}
  1 & 0
  \\
  1 & 1
 \end{pmatrix}^{\otimes n}
 \Pi_{\mathrm{BR}},
\end{equation*}
where $\otimes n$ denotes the $n$-th Kronecker power
and $\Pi_{\mathrm{BR}}$ is the previously defined
bit-reversal permutation matrix~\cite{A09}.
Then vector $\uu\in\X^{2^n}$ is defined as $\uu\equiv \xx G$
for given vector $\xx\in\X^{2^n}$.
Let $\{\I_0,\I_1\}$ be a partition of $\{0,1\}^n$,
satisfying $\I_0\cap\I_1=\emptyset$ and $\I_0\cup\I_1=\{0,1\}^n$.
We will define $\{\I_0,\I_1\}$ in Section~\ref{sec:polarization}.

Let $\XX\equiv(X_{0^n},\ldots,X_{1^n})$ and $\YY\equiv(Y_{0^n},\ldots,Y_{1^n})$
be random variables and let $\UU\equiv(U_{0^n},\ldots,U_{1^n})$
be a random variable defined as $\UU\equiv \XX G$.
Then joint distribution $P_{U_{\I_0}U_{\I_1}\YY}$ of $(U_{\I_0},U_{\I_1},\YY)$
is defined using joint distribution $P_{\XX\YY}$ of $(\XX,\YY)$ as
\begin{equation*}
 P_{U_{\I_0}U_{\I_1}\YY}(u_{\I_0},u_{\I_1},\yy)
 \equiv P_{\XX\YY}((u_{\I_1},u_{\I_0})G^{-1},\yy),
\end{equation*}
where the elements in $(u_{\I_1},u_{\I_0})$ are sorted in
index order before operating $G^{-1}$.

Let $P_{U_{b^n}|U_{[0^n:b^n)}\YY}$ be the conditional probability distribution,
defined as
\begin{equation}
 P_{U_{b^n}|U_{[0^n:b^n)}\YY}(u_{b^n}|u_{[0^n:b^n)},\yy)
 \equiv
 \frac{\sum_{u_{(b^n:1^n]}}P_{U_{\I_0}U_{\I_1}\YY}(u_{\I_0},u_{\I_1},\yy)}
 {\sum_{u_{[b^n:1^n]}}P_{U_{\I_0}U_{\I_1}\YY}(u_{\I_0},u_{\I_1},\yy)}.
 \label{eq:muCgCY}
\end{equation}
For vector $u_{\I_1}$ and side information $\yy\in\Y^{2^n}$,
output $\huu\equiv \ff(u_{\I_1},\yy)$ of
successive-cancellation (SC) decoder $\ff$ is defined recursively
\begin{equation*}
 \hu_{b^n}
 \equiv
 \begin{cases}
  f_{b^n}(\hu_{[0^n:b^n)},\yy)
  &\text{if}\ b^n\in\I_0
  \\
  u_{b^n}
  &\text{if}\ b^n\in\I_1
 \end{cases}
\end{equation*}
using functions $\{f_{b^n}\}_{b^n\in\I_0}$ defined as
\begin{equation*}
 f_{b^n}(u_{[0^n:b^n)},\yy)
 \equiv
 \arg\max_{u_{b^n}\in\U}P_{U_{b^n}|U_{[0^n:b^n)}\YY}(u_{b^n}|u_{[0^n:b^n)},\yy),
\end{equation*}
which is the maximum a posteriori decision rule
after observation $(u_{[0^n:b^n)},\yy)$.

For a polar source code (with decoder side information),
$\xx\in\X^{2^n}$ is a source output, $u_{\I_1}$ is a codeword,
and $\yy\in\Y^{2^n}$ is a side information output.
The decoder reproduces source output $\hxx\equiv\ff(u_{\I_1},\yy)G^{-1}$
from codeword $u_{\I_1}$ and $\yy$.
The (block) decoding error probability is given as
$\Prob(\ff(U_{\I_1},\YY)G^{-1}\neq \XX)$.

For a systematic polar channel code \cite{A11},
let $\I'_0$ and $\I'_1$ be defined as
\begin{align}
 \I'_0
 &\equiv\lrb{
  b_0b_1\cdots b_{n-1}: b_{n-1}\cdots b_1b_0\in\I_0
 }
 \label{eq:I'0}
 \\
 \I'_1
 &\equiv\lrb{
  b_0b_1\cdots b_{n-1}: b_{n-1}\cdots b_1b_0\in\I_1
 }
 \label{eq:I'1}
\end{align}
for given $(\I_0,\I_1)$.
We assume that encoder and decoder share a vector $u_{\I_1}$.
The encoder computes $(x_{\I'_1},u_{\I_0})$
from a message $x_{\I'_0}$ and a shared vector $u_{\I_1}$
so that $(x_{\I'_1},x_{\I'_0})=(u_{\I_1},u_{\I_0})G^{-1}$,
where the elements in $(x_{\I'_1},x_{\I'_0})$ and $(u_{\I_1},u_{\I_0})$
are sorted in index order before operating $G^{-1}$.
Then the encoder generates channel input $\xx\equiv(x_{\I'_0},x_{\I'_1})$,
where the elements in $(x_{\I'_1},x_{\I'_0})$ are sorted in index order.
The decoder reproduces a pair of vectors
$(u_{\I_1},\hu_{\I_0})\equiv \ff(u_{\I_1},\yy)$
from channel output $\yy\in\Y^{2^n}$ and shared vector $u_{\I_1}$.
Then $\hxx\equiv\huu G^{-1}$ is reproduced
and $\hx_{\I_0}$ is a reproduction of the message.
The (block) decoding error probability is also given as
$\Prob(\ff(U_{\I_1},\YY)G^{-1}\neq \XX)$.

For a non-systematic polar channel code, $u_{\I_0}$ is a message
and vector $u_{\I_1}$ is shared by the encoder and decoder.
The encoder generates channel input $\xx\in\X^{2^n}$
as $\xx\equiv(u_{\I_1},u_{\I_0})G^{-1}$,
where the elements in $(u_{\I_1},u_{\I_0})$ are sorted in
index order before operating $G^{-1}$.
The decoder reproduces a pair of vectors as
$(u_{\I_1},\hu_{\I_0})\equiv \ff(u_{\I_1},\yy)$
from channel output $\yy\in\Y^{2^n}$ and shared vector $u_{\I_1}$.
Then $\hu_{\I_0}$ is a reproduction of the message.
The (block) decoding error probability is given as
$\Prob\lrsb{\ff(U_{\I_1},\YY)\neq (U_{\I_0},U_{\I_1})}$.

We have the following lemma,
which can be shown similarly
to a previous proof~\cite{A09}, \cite[Eq.\ (13)]{KSU10}, \cite[Eq.~(1)]{TV13}.

\begin{lem}[{\cite[Eq.\ (13)]{GHU12}, \cite[Lemmas 1, 2, and 3]{SCD}}]
\label{lem:sc}
\begin{align*}
 &\max_{b^n\in\I_0}
 \Prob(f_{b^n}(U_{[0^n:b^n)},\YY)\neq U_{b^n})
 \notag
 \\*
 &\leq
 \Prob(\ff(U_{\I_1},\YY)G^{-1}\neq \XX)
 \notag
 \\
 &=
 \Prob\lrsb{\ff(U_{\I_1},\YY)\neq (U_{\I_0},U_{\I_1})}
 \notag
 \\
 &\leq
 \sum_{b^n\in\I_0}
 \Prob(f_{b^n}(U_{[0^n:b^n)},\YY)\neq U_{b^n}).
\end{align*}
\end{lem}

In the following, we call $\Prob(f_{b^n}(U_{[0^n:b^n)},\YY)\neq U_{b^n})$
the {\em bit error probability}.

In~\cite{KSU10,MT14},
the decoding error probability is bounded\footnote{
 In \cite[Proposition 2]{A09}, \cite[Proposition 2.7]{S12},
 the factor $1/2$ of the upper bound in (\ref{eq:Zbound}) is missing.
 We have the upper bound in (\ref{eq:Zbound}) from \cite[Lemma 22]{MT14}.
}
in terms of Battacharyya parameter $Z$ as
\begin{equation}
 \frac12\lrB{1-\sqrt{1-Z(U_{b^n}|U_{[0^n:b^n)},\YY)^2}}
 \leq
 \Prob(f_{b^n}(U_{[0^n:b^n)},\YY)\neq U_{b^n})
 \leq
 \frac{Z(U_{b^n}|U_{[0^n:b^n)},\YY)}2,
 \label{eq:Zbound}
\end{equation}
which is implicitly used in the proof.
Furthermore,
we have
\begin{equation}
 h^{-1}(H(U_{b^n}|U_{[0^n:b^n)},\YY))
 \leq
 \Prob(f_{b^n}(U_{[0^n:b^n)},\YY)\neq U_{b^n})
 \leq
 \frac{H(U_{b^n}|U_{[0^n:b^n)},\YY)}2,
 \label{eq:Hbound}
\end{equation}
where $H$ is the conditional entropy
and $h^{-1}$ is the inverse of the binary entropy function $h$ defined as
\begin{align}
 h(\xi)
 &\equiv
 -\xi\log_2(\xi)-[1-\xi]\log_2(1-\xi)
 \label{eq:h}
\end{align}
for $\xi\in[0,1/2]$,
the lower bound comes from the Fano inequality
$H(U_{b^n}|U_{[0^n:b^n)},\YY)
\leq h(\Prob(f_{b^n}(U_{[0^n:b^n)},\YY)\neq U_{b^n}))$,
and the upper bound comes from the fact that
$2\Prob(f_{b^n}(U_{[0^n:b^n)},\YY)\neq U_{b^n})
\leq H(U_{b^n}|U_{[0^n:b^n)},\YY)$ shown in~\cite{CC66}.
Then Lemma~\ref{lem:sc} provides a tighter
bounds than that in terms of
the Battacharyya parameter and the conditional entropy.

From the above lemma,
we can make a decoding error probability close to zero
by specifying index set $\I_0$
so that $\Prob(f_{b^n}(U_{[0^n:b^n)},\YY)\neq U_{b^n})$ is close to zero
for all $b^n\in\I_0$.
One of our goals is computing $\Prob(f_{b^n}(U_{[0^n:b^n)},\YY)\neq U_{b^n})$
for every $b^n\in\{0,1\}^n$
to improve the bounds of the decoding error probability.

\subsection{Symmetric Parameterization of Joint Sources}

This section introduces the symmetric parameterization of joint sources
to simplify the analysis of the bit error probability.
This approach is introduced in~\cite[Chapter 5]{A15} to analyze
the likelihood ratio of the polar channel code.
Constructions of the successive-cancellation decoder
and the successive-cancellation list decoder are introduced based on this approach \cite{SCL}.

Let $(U,V)$ be a pair of random variables on $\U\times\V$,
where $\U\equiv\{0,1\}$.
Let $P_{V}$, and $P_{U|V}$ be the distribution of $V$
and the conditional distribution of $U$ for given $V$.
We define a list $\cP(U|V)\equiv\{(\mu_v,\theta_v)\}_{v\in\V}$ as
\begin{align}
 \mu_v
 &\equiv P_V(v)
 \label{eq:mu}
 \\
 \theta_v
 &\equiv P_{U|V}(0|v)-P_{U|V}(1|v),
 \label{eq:theta}
\end{align}
where $\theta_v\in[-1,1]$.
We have relation
\begin{equation}
 P_{U|V}(u|v)
 =\frac{1\mp_u\theta_v}2
 \label{eq:PUV}
\end{equation}
from the definition of $\theta_v$ and relation $P_{U|V}(0|v)+P_{U|V}(1|v)=1$,
where $\mp_u$ is the bipolar-binary conversion of $u$.
Then list $\cP(U|V)$ represents the joint probability distribution of $(U,V)$.
Note that expression $\{(\mu_v,\theta_v)\}_{v\in\V}$
does not represent a set;
it represents a list.
That is, we treat $(\mu_v,\theta_v)$ and $(\mu_{v'},\theta_{v'})$ differently
when $v\neq v'$.
Furthermore, when bijection $\pi:\V\to\V$ exists,
we identify list
$\cP(U|\pi(V))\equiv\{(\mu_{\pi(v)},\theta_{\pi(v)})\}_{v\in\V}$
with $\cP(U|V)\equiv\{(\mu_v,\theta_v)\}_{v\in\V}$.
For this reason, we denote $\cP\equiv\cP(U|V)$
to represent an identical list when the corresponding random variables
are not essential. We denote\footnote{In~\cite{MT14,GYB18},
 such equivalence is denoted by $\cP'\stackrel{P}{\sim}\cP$.
 We use notation $\cP'=\cP$ because $\cP$ and $\cP'$
 are identical as multisets.}
$\cP'=\cP$
when $\cP'$ is identical with $\cP$ in the sense mentioned above.
Note such an equivalence was previously 
introduced \cite[Definition 3]{GYB18}, \cite[Definition 4]{MT14}.

When we guess the outcome of $U$ by observing that for $V$'s outcome,
the decision error probability is defined as
\begin{equation*}
 \Error(f)\equiv\sum_{u\in\U}\sum_{v\in\V}P_{U|V}(u|v)P_V(v)\chi(f(v)\neq u)
\end{equation*}
for given decision rule $f$, where $\chi$ is the support function defined as
\begin{equation*}
 \chi(\text{statement})
 \equiv
 \begin{cases}
  1
  &\text{if the statement is true}
  \\
  0
  &\text{if the statement is false}.
 \end{cases}
\end{equation*}
The maximum a posteriori (MAP) decision rule
\begin{equation*}
 \fMAP(u|v)\equiv\arg\max_{u\in\U} P_{U|V}(u|v)
\end{equation*}
minimizes the decision error probability.
The MAP decision error probability $\Error(U|V)$ is given as
\begin{align}
 \Error(U|V)
 &\equiv
 \Error(\fMAP)
 \notag
 \\
 &=
 \sum_{v\in\V}P_V(v)\min\{P_{U|V}(0|v),P_{U|V}(1|v)\}
 \notag
 \\
 &=
 \sum_{v\in\V}\frac{\mu_v[1-|\theta_v|]}2
 \notag
 \\
 &=
 \frac12\lrB{1-\sum_{v\in\V}\mu_v|\theta_v|}
 \notag
 \\
 &=
 \frac12\lrB{1-M(\cP(U|V))},
 \label{eq:Error-fMAP}
\end{align}
where 
\begin{equation}
 M(\cP)
 \equiv
 \sum_{v\in\V}
 \mu_v|\theta_v|,
 \label{eq:MP}
\end{equation}
and the second equality comes from (\ref{eq:PUV}).
Note that $M(\cP)$ is invariant under any permutation $\pi:\V\to\V$.
Symmetric parameterization is a binary version of the Fourier transform
of the a posteriori probability~\cite[Definitions 24 and 25]{MT14}.

In the following, we focus on value $M(\cP)$.

\begin{rem}
Value $M(\cP(U|V))$ equals the variational distance
between joint distribution $P_{UV}$ and distribution
$P_{\tU}\times P_V$, where $\tU$ is the uniform distribution on $\U$.
To show this fact, we assume 
$P_{U|V}(0|v)\geq 1/2 \geq P_{U|V}(1|v)$ without loss of generality.
Then we have
\begin{align}
 M(\cP(U|V))
 &=
 \frac 12\sum_{v\in\V}P_V(v)\lrB{P_{U|V}(0|v)-P_{U|V}(1|v)}
 \notag
 \\
 &=
 \frac 12
 \sum_{v\in\V}P_V(v)\lrB{P_{U|V}(0|v)-P_{\tU}(0)+P_{\tU}(1)-P_{U|V}(1|v)}
 \notag
 \\
 &=
 \frac 12
 \sum_{v\in\V}\sum_{u\in\U}P_V(v)\lrbar{P_{U|V}(u|v)-P_{\tU}(u)}
 \notag
 \\
 &=
 \frac 12
 \sum_{v\in\V}\sum_{u\in\U}\lrbar{P_{UV}(u,v)-P_{\tU}\times P_V(u,v)},
\end{align}
where the first equality comes from (\ref{eq:mu}), (\ref{eq:theta}),
(\ref{eq:MP}), and the assumption,
the second and third equalities come from the assumption
and the fact that $P_{\tU}(0)=P_{\tU}(1)=1/2$,
and the right hand side of the last inequality is the definition of the variational distance.
Note that $M(\cP(U|V))$ is also relevant to the polar lossy source codes
and the polar channel codes for asymmetric channels~\cite{HY13,KU10}.
\end{rem}

\subsection{Plus and Minus Operations}

In this section, we define plus and minus operations
based on symmetric parameterization.

In the basic polar transform,
a pair of binary random variables $(U_0,U_1)$ is transformed into
$B_0\equiv U_0\oplus U_1$ and $B_1\equiv U_1$, where $\oplus$ denotes
the addition on the binary finite field.
Assume that random variables $U_0,U_1,V_0,V_1$
satisfy that $U_0,U_1\in\{0,1\}$, $V_0,V_1\in\V$,
and $(U_0,V_0)$ and $(U_1,V_1)$ are independent,
but $U_b$ and $V_b$ are allowed to be correlated for each $b\in\{0,1\}$.

Regarding probability distribution $P_{V^2}$
and conditional probability distribution $P_{B_0|V^2}$,
we have
\begin{align}
 P_{V^2}(v_0,v_1)
 &=\mu_{v_0}\mu_{v_1}
 \\
 P_{B_0|V^2}(b_0|v_0,v_1)
 &=
 \frac{1\mp_{b_0}\theta_{v_0}\theta_{v_1}}2,
\end{align}
where $\mp_{b_0}$ is bipolar-binary conversion of $b_0$.
To describe the joint probability distribution of $(B_0,V^2)$,
we introduce minus operation
$\cP^-\equiv\{(\mu_{v_0,v_1}^-,\theta_{v_0,v_1}^-)\}_{(v_0,v_1)\in\V\times\V}$
as
\begin{align}
 \mu_{v_0,v_1}^-
 &\equiv
 \mu_{v_0}\mu_{v_1}
 \label{eq:P-mu}
 \\
 \theta_{v_0,v_1}^-
 &\equiv
 \theta_{v_0}\theta_{v_1}.
 \label{eq:P-theta}
\end{align}
Then we have relation
\begin{equation}
 \cP^-=\cP(U_0\oplus U_1|V_0,V_1).
 \label{eq:PS-}
\end{equation}

Regarding probability distribution $P_{B_0V^2}$
and conditional probability distribution $P_{B_1|B_0V^2}$,
we have
\begin{align*}
 P_{B_0V^2}(b_0,v_0,v_1)
 &=
 \frac{\mu_{v_0}\mu_{v_1}[1\mp_{b_0}\theta_{v_0}\theta_{v_1}]}2
 \\
 P_{B_1|B_0V^2}(b_1|b_0,v_0,v_1)
 &=
 \frac{1\mp_{b_1}[\theta_{v_1}\mp_{b_0}\theta_{v_0}]/[1\mp_{b_0}\theta_{v_0}\theta_{v_1}]}
 2,
\end{align*}
where $\mp_{b_0}$ and $\mp_{b_1}$ are
the bipolar-binary conversions of $b_0$ and $b_1$.
To describe the joint probability distribution of $(B_1,(B_0,V^2))$,
we introduce plus operation
$\cP^+
\equiv
\{(\mu_{u,v_0,v_1}^+,\theta_{u,v_0,v_1}^+)\}_{(u,v_0,v_1)\in\U\times\V\times\V}$
as
\begin{align}
 \mu_{u,v_0,v_1}^+
 &\equiv
 \frac{\mu_{v_0}\mu_{v_1}[1\mp_u\theta_{v_0}\theta_{v_1}]}2
 \label{eq:P+mu}
 \\
 \theta_{u,v_0,v_1}^+
 &\equiv
 \frac{\theta_{v_1}\mp_u\theta_{v_0}}{1\mp_u\theta_{v_0}\theta_{v_1}},
 \label{eq:P+theta}
\end{align}
where we define $(\mu_{u,v_0,v_1}^+,\theta_{u,v_0,v_1}^+)\equiv(0,0)$
when $1\mp_u\theta_{v_0}\theta_{v_1}=0$.
Then we have relation
\begin{equation}
 \cP^+=\cP(U_1|U_0\oplus U_1,V_0,V_1).
 \label{eq:PS+}
\end{equation}

Recursively define $U^{(n)}_{b^n}$ as
\begin{align}
 U^{(0)}
 &\equiv
 X
 \label{eq:U(0)}
 \\
 U^{(k+1)}_{b^k0}
 &\equiv
 U^{(k)}_{0b^k}\oplus U^{(k)}_{1b^k}
 \label{eq:minus-k}
 \\
 U^{(k+1)}_{b^k1}
 &\equiv
 U^{(k)}_{1b^k},
 \label{eq:plus-k}
\end{align}
where  $U^{(k)}_{0b^k}$ and $U^{(k)}_{1b^k}$
are two independent copies of $U^{(k)}_{b^k}$.
Then we have $U^{(n)}_{[0^n:1^n]}=X^{2^n}G$,
which is the polar transform of $X^{2^n}$.
Recursively define $Y_{[0^k:b^k]}$ as
\begin{align*}
 Y^{(0)}
 &\equiv
 Y
 \\
 Y^{(k+1)}_{[0^{k+1}:1^{k+1}]}
 &\equiv
 \lrsb{Y^{(k)}_{0[0^k:1^k]},Y^{(k)}_{1[0^k:1^k]}},
\end{align*}
where $Y^{(0)}\equiv Y$ is correlated with $U^{(0)}\equiv X$,
and $Y^{(k)}_{0[0^k:1^k]}$ and $Y^{(k)}_{1[0^k:1^k]}$
are two independent copies of $Y^{(k)}_{[0^k:1^k]}$.
Then we have
\begin{align}
 \cP\lrsb{\left.
   U^{(k+1)}_{b^k0}
  \right|U^{(k+1)}_{[0^{k+1}:b^k0)},Y^{(k+1)}_{[0^{k+1}:1^{k+1}]}
 }
 &=
 \cP\lrsb{\left.
   U^{(k+1)}_{b^k0}
  \right|
  U^{(k)}_{0[0^k:b^k)}, Y^{(k)}_{0[0^k:1^{k}]}, U^{(k)}_{1[0^k:b^k)},
  Y^{(k)}_{1[0^k:1^{k}]
 }}
 \notag
 \\
 &=
 \cP\lrsb{\left.
   U^{(k)}_{0b^k}\oplus U^{(k)}_{1b^k}
  \right|
  U^{(k)}_{0[0^k:b^k)}, Y^{(k)}_{0[0^k:1^{k}]}, U^{(k)}_{1[0^k:b^k)},
  Y^{(k)}_{1[0^k:1^{k}]
 }}
 \notag
 \\
 &=
 \cP\lrsb{\left.
   U^{(k)}_{b^k}
  \right|
  U^{(k)}_{[0^k:b^k)}, Y^{(k)}_{[0^k:1^{k}]}
 }^-,
 \label{eq:U(k)-}
\end{align}
where the first equality comes from the fact that we can construct a bijection
\begin{equation*}
 \pi\lrsb{
  U^{(k)}_{0[0^k:b^k)}, Y^{(k)}_{0[0^k:1^{k}]},
  U^{(k)}_{1[0^k:b^k)}, Y^{(k)}_{1[0^k:1^{k}]}
 }
 \equiv
 \lrsb{U^{(k+1)}_{[0^{k+1}:b^k0)},Y^{(k+1)}_{[0^{k+1}:1^{k+1}]}}
\end{equation*}
using relations (\ref{eq:minus-k}) and (\ref{eq:plus-k}),
the second equality comes from (\ref{eq:minus-k}),
and the last one comes from (\ref{eq:PS-}).
Similarly, we have
\begin{align}
 \cP\lrsb{\left.
   U^{(k+1)}_{b^k1}
  \right|U^{(k+1)}_{[0^{k+1}:b^k1)},Y^{(k+1)}_{[0^{k+1}:1^{k+1}]}
 }
 &=
 \cP\lrsb{\left.
   U^{(k+1)}_{b^k1}
  \right|
  U^{(k+1)}_{b^k0},
  U^{(k)}_{0[0^k:b^k)}, Y^{(k)}_{0[0^k:1^{k}]},
  U^{(k)}_{1[0^k:b^k)}, Y^{(k)}_{1[0^k:1^{k}]
 }}
 \notag
 \\
 &=
 \cP\lrsb{\left.
   U^{(k)}_{1b^k}
  \right|
  U^{(k)}_{0b^k}\oplus U^{(k)}_{1b^k},
  U^{(k)}_{0[0^k:b^k)}, Y^{(k)}_{0[0^k:1^{k}]},
  U^{(k)}_{1[0^k:b^k)}, Y^{(k)}_{1[0^k:1^{k}]
 }}
 \notag
 \\
 &=
 \cP\lrsb{\left.
   U^{(k)}_{b^k}
  \right|
  U^{(k)}_{[0^k:b^k)}, Y^{(k)}_{[0^k:1^{k}]}
 }^+,
 \label{eq:U(k)+}
\end{align}
where the second equality comes from (\ref{eq:minus-k}) and (\ref{eq:plus-k}),
and the last one comes from (\ref{eq:PS+}).
From (\ref{eq:U(k)-}) and (\ref{eq:U(k)+}), we have
\begin{equation}
 \cP\lrsb{\left.
   U^{(k+1)}_{b^kb}
  \right|U^{(k+1)}_{[0^{k+1}:b^kb)},Y^{(k+1)}_{[0^{k+1}:1^{k+1}]}
 }
 =
 \cP\lrsb{\left.
   U^{(k)}_{b^k}
  \right|
  U^{(k)}_{[0^k:b^k)}, Y^{(k)}_{[0^k:1^{k}]}
 }^{\pm_b},
 \label{eq:Ppm}
\end{equation}
where $\pm_b$ is the bipolar-binary conversion of $b$.

\subsection{Polarization}
\label{sec:polarization}

In this section, we revisit the polarization lemma.
We prepare some lemmas,
where proofs are given in Sections~\ref{sec:proof-MP+-}--\ref{sec:proof-HV}.
\begin{lem}
\label{lem:MP+-}
Let $\cP\equiv\{(\mu_v,\theta_v)\}_{v\in\V}$. Then we have
\begin{align}
 M(\cP^-)
 &=
 \sum_{v_0\in\V}\sum_{v_1\in\V}
 \mu_{v_0}\mu_{v_1}|\theta_{v_0}\theta_{v_1}|
 \label{eq:MP-}
 \\
 M(\cP^+)
 &=
 \sum_{v_0\in\V}\sum_{v_1\in\V}
 \mu_{v_0}\mu_{v_1}\max\{|\theta_{v_0}|,|\theta_{v_1}|\}.
 \label{eq:MP+}
\end{align}
\end{lem}
\begin{lem}
\label{lem:M}
Let $\cP\equiv\{(\mu_v,\theta_v)\}_{v\in\V}$. Then we have
\begin{gather}
 M(\cP^-)
 \leq
 M(\cP)
 \leq
 M(\cP^+)
 \label{eq:-V+}
 \\
 M(\cP^-)
 =
 M(\cP)^2
 \label{eq:V-}
 \\
 M(\cP^-)+M(\cP^+)
 \leq
 2M(\cP).
 \label{eq:V-+}
\end{gather}
\end{lem}

\begin{lem}
\label{lem:HV}
For $\cP(U|V)\equiv\{(\mu_v,\theta_v)\}_{v\in\V}$,
define $h(\xi)$ by (\ref{eq:h}) and $H(U|V)$ as
\begin{align*}
 H(U|V)
 &\equiv
 \sum_{v\in\V}
 \mu_v h\lrsb{\frac{1-|\theta_v|}2}.
\end{align*}
Then
\begin{equation*}
 0\leq 1-M(\cP(U|V))\leq H(U|V)\leq h\lrsb{\frac{1-M(\cP(U|V))}2}\leq 1.
\end{equation*}
From this inequality, we have 
\begin{align}
 H(U|V)\to 1
 &\Leftrightarrow
 M(\cP(U|V))\to 0
 \label{eq:H1M0}
 \\
 H(U|V)\to 0
 &\Leftrightarrow
 M(\cP(U|V))\to 1.
 \label{eq:H0M1}
\end{align}
\end{lem}

Recursively define joint distribution $\cP^{\pm_{b_1}\cdots\pm_{b_n}}$ as
\begin{equation*}
 \cP^{\pm_{b_1}\cdots\pm_{b_n}}
 \equiv
 \lrB{\cP^{\pm_{b_1}\cdots \pm_{b_{n-1}}}}^{\pm_{b_n}}
\end{equation*}
for $(b_1,\ldots,b_n)\in\{0,1\}^n$.
Then from (\ref{eq:Error-fMAP}) and (\ref{eq:Ppm}), we have 
\begin{align}
 \Prob(f_{b^n}(U_{[0^n:b^n)},\YY)\neq U_{b^n})
 &=
 \frac12\lrB{
  1-M(\cP(U_{b^n}|U_{[0^n:b^n)},Y_{[0^n:1^n]}))
 }
 \notag
 \\
 &=
 \frac12\lrB{
  1-M(\cP^{\pm_{b_1}\cdots \pm_{b_n}})
 }
 \label{eq:bit-error-probability}
\end{align}
by assuming that joint source $\{(X_{b^n},Y_{b^n})\}_{b^n\in\{0,1\}^n}$
is stationary memoryless subject to joint distribution $P_{XY}$,
where $M$ is defined by (\ref{eq:MP}),
$\cP\equiv\{(\mu_y,\theta_y)\}_{y\in\Y}$ is defined as
\begin{align*}
 \mu_y
 &\equiv
 P_Y(y)
 \\
 \theta_y
 &\equiv
 P_{X|Y}(0|y)-P_{X|Y}(1|y),
\end{align*}
and $\pm_{b_i}$ is the bipolar-binary conversion of $b_i$
for each $i\in\{1,\ldots,n\}$.
Then we have
\begin{equation*}
 \Prob\lrsb{\limn H(\cP^{S_1\cdots S_n}) = 1}
 =
 1-\Prob\lrsb{\limn H(\cP^{S_1\cdots S_n}) = 0}
 =
 H(X|Y)
\end{equation*}
from \cite[Theorem 2.4]{S12}\footnote{
 When $H_n\equiv H(\cP^{S_1\cdots S_n})$
 and $H^{\pm}$ is defined by \cite[Eq. (2.9)]{S12},
 we have the relation \cite[Eq. (2.11)]{S12}
 from (\ref{eq:PS-}), (\ref{eq:PS+}),
 and the definition of $H(\cP)$.},
where $(S_1,\ldots,S_n)$ is the sequence of the stationary memoryless
random variables subject to uniform distribution on $\{-,+\}$.
Furthermore, from (\ref{eq:H1M0}) and (\ref{eq:H0M1}), we have 
\begin{equation}
 \Prob\lrsb{\limn M(\cP^{S_1\cdots S_n})=0}
 =
 1-\Prob\lrsb{\limn M(\cP^{S_1\cdots S_n})=1}=H(X|Y).
 \label{eq:limM}
\end{equation}
Then from the above equalities, (\ref{eq:V-}), (\ref{eq:V-+}), (\ref{eq:limM}),
and~\cite[Theorem 1]{AT09},\cite[Lemma 2.10]{S12},
we have the following polarization lemma,
where the Battacharyya parameter in~\cite[Theorem 4.10]{S12}
is replaced by $M$, which is related directly to the bit error probability.
Note that it is unnecessary to introduce
the Battacharyya parameter for the proof of this lemma.
Note that similar results
have been obtained in \cite{GR20,MT14}.
\begin{lem}
\label{lem:psc}
Define $\I_0$ and $\I_1$ as
\begin{align*}
 \I_0
 &\equiv\lrb{
  b^n\in\{0,1\}^n:
  M(\cP^{\pm_{b_1}\cdots\pm_{b_n}})\geq 1-2^{-2^{n\beta}}
 }
 \\
 \I_1
 &\equiv\lrb{
  b^n\in\{0,1\}^n:
  M(\cP^{\pm_{b_1}\cdots\pm_{b_n}})< 1-2^{-2^{n\beta}}
 }.
\end{align*}
Then we have 
\begin{align}
 \lim_{n\to\infty}\frac{|\I_0|}{2^n}
 &=
 1-H(X|Y)
 \label{eq:limI0}
 \\
 \lim_{n\to\infty}\frac{|\I_1|}{2^n}
 &=
 H(X|Y)
 \label{eq:limI1}
\end{align}
for any $\beta\in[0,1/2)$.
Note that $\I_0$ corresponds to the low entropy set
($H(\cP^{\pm_{b_1}\cdots\pm_{b_n}})\to 0$)
and $\I_1$ corresponds to the high entropy set
($H(\cP^{\pm_{b_1}\cdots\pm_{b_n}})\to 1$).
\end{lem}

For the polar source code,
(\ref{eq:limI1}) implies that encoding rate $|\I_1|/2^n$ approaches $H(X|Y)$,
suggesting that the fundamental limit is achievable with the polar source code.
For the polar channel code, (\ref{eq:limI0}) implies that
encoding rate $|\I_0|/2^n$ approaches $1-H(X|Y)=I(X;Y)$
by assuming that $P_X$ is the uniform distribution on $\{0,1\}$.
This implies that
symmetric capacity can be achieved using the polar channel code.
From Lemma~\ref{lem:sc} and (\ref{eq:bit-error-probability}),
we can evaluate the lower and upper bounds
of the decoding error probability:
\begin{align}
 &
 \max_{b^n\in\I_0}
 \frac12\lrB{
  1-M(\cP^{\pm_{b_1}\cdots \pm_{b_n}})
 }
 \notag
 \\*
 &\leq
 \Prob(\ff(U_{\I_1},\YY)G^{-1}\neq \XX)
 \notag
 \\
 &=
 \Prob\lrsb{\ff(U_{\I_1},\YY)\neq (U_{\I_0},U_{\I_1})}
 \notag
 \\
 &\leq
 \sum_{b^n\in\I_0}
 \frac12\lrB{
  1-M(\cP^{\pm_{b_1}\cdots \pm_{b_n}})
 },
 \label{eq:error}
\end{align}
which goes to zero as $n$ goes to infinity
by letting $\I_0$ and $\I_1$ be defined as those in
Lemma~\ref{lem:psc}.

\section{Construction of Polar Codes Based on Bit Error Probability}
\label{sec:construction}

Based on (\ref{eq:error}), we expect that the decoding error probability
is minimized by establishing set $\I_0$ so that it maximizes
value $\sum_{b^n\in\I_0}M(\cP^{\pm^n})$,
where $\pm^n\equiv(\pm_{b_1},\ldots,\pm_{b_n})$.
To this end, it is sufficient to compute value
$M(\cP^{\pm^n})$ for all $b^n\in\{0,1\}^n$.
However, the computation seems hard
because the cardinality of
$\cP^{\pm^n}$ grows rapidly with respect to block length $2^n$.
In fact, when $\cP$ is given,
we have $|\cP^-|=|\cP|^2$ and $|\cP^+|=2|\cP|^2$
from the definition of $\cP^{\pm}$.

In the following, we introduce two techniques
to reduce the computational complexity.
The first one reduces the cardinality of set $\cP^{\pm^n}$,
where $M(\cP^{\pm^n})$ is invariant under this reduction.
In the second, we introduce a novel quantization loss function
to obtain an approximation of $M(\cP^{\pm^n})$.

\subsection{Canonical Representation of Joint Probability Distribution}

In this section, we introduce the reduction of list $\cP$,
where $M(\cP^{\pm^n})$ is invariant under this reduction.

First, we focus on the fact that $M$ is an even function.
We introduce operator $\sfA$
applied to list $\cP\equiv\{(\mu_v,\theta_v)\}_{v\in\V}$.
The proof is given in~\ref{sec:proof-MPA}.
\begin{lem}
\label{lem:MPA}
Define $\cP^{\sfA}$ as
\begin{equation*}
 \cP^{\sfA}
 \equiv
 \lrb{
  (\mu_v,|\theta_v|): v\in\V
 }.
\end{equation*}
Then we have
\begin{equation*}
 M(\cP)=M(\cP^{\sfA}).
\end{equation*}
\end{lem}

Next we introduce the following lemma.
The proof is given in~\ref{sec:proof-ApmA}.
\begin{lem}
\label{lem:ApmA}
Let $\cP^{\sfA\pm\sfA}$ and $\cP^{\pm\sfA}$ be defined as
\begin{align*}
 \cP^{\sfA\pm\sfA}
 &\equiv
 [[\cP^{\sfA}]^\pm]^{\sfA}
 \\
 \cP^{\pm\sfA}
 &\equiv
 [\cP^\pm]^{\sfA},
\end{align*}
where $\pm$ denotes a `$+$' or `$-$' operator
and the same operator appears on both sides.  
Then we have
\begin{equation*}
 \cP^{\sfA\pm\sfA}=\cP^{\pm\sfA}.
\end{equation*}
\end{lem}

Next we introduce another operator, $\sfSigma$,
which is applied to $\cP$ to reduce set $\V$.
The proof is given in~\ref{sec:proof-MPS}.
\begin{lem}
\label{lem:MPS}
Define $\cP^{\sfSigma}$ as
\begin{equation}
 \cP^{\sfSigma}
 \equiv
 \lrb{
  \lrsb{\sum_{v:\theta_v=\theta}\mu_v,\theta}: \theta\in\Theta
 },
 \label{eq:S}
\end{equation}
where
\begin{equation}
 \Theta\equiv\{\theta_v: v\in\V\ \text{s.t.}\ \mu_v>0\}.
 \label{eq:Theta}
\end{equation}
Then we have
\begin{equation*}
 M(\cP)=M(\cP^{\sfSigma}).
\end{equation*}
\end{lem}

Next we show the following lemma.
The proof is given in~\ref{sec:proof-SpmS}.
\begin{lem}
\label{lem:SpmS}
Let $\cP^{\sfSigma\pm\sfSigma}$ and $\cP^{\pm\sfSigma}$ be defined as
\begin{align*}
 \cP^{\sfSigma\pm\sfSigma}
 &\equiv
 [[\cP^{\sfSigma}]^\pm]^{\sfSigma}
 \\
 \cP^{\pm\sfSigma}
 &\equiv
 [\cP^\pm]^{\sfSigma},
\end{align*}
where $\pm$ denotes a `$+$' or `$-$' operator
and the same operator appears on both sides.  
Then we have
\begin{equation*}
 \cP^{\sfSigma\pm\sfSigma}=\cP^{\pm\sfSigma}.
\end{equation*}
\end{lem}

Next we introduce the following lemma
regarding operations $\sfA$ and $\sfSigma$.
The proof is given in~\ref{sec:proof-SAS}.
\begin{lem}
\label{lem:SAS}
\begin{equation*}
 \cP^{\sfSigma\sfA\sfSigma}=\cP^{\sfA\sfSigma}.
\end{equation*}
\end{lem}

Here we introduce the following theorem.
The proof is given in~\ref{sec:proof-canonical}.
\begin{thm}
\label{thm:canonical}
Define operator $\sfC$ as
\begin{equation*}
 \sfC\equiv \sfA\sfSigma.
\end{equation*}
Then we have
\begin{gather}
 M(\cP^{\sfC})
 =
 M(\cP)
 \label{eq:MPC}
 \\
 \cP^{\sfC\pm\sfC}
 =
 \cP^{\pm\sfC}
 \label{eq:CpmC}
 \\
 M(\cP^{\sfC\pm})=M(\cP^{\sfC\pm\sfC})=M(\cP^{\pm}),
 \label{eq:MPCpm}
\end{gather}
where $\pm$ denotes a `$+$' or `$-$' operator
and the same operator appears on both sides.  
\end{thm}

By repeatedly applying the above theorem, we have
\begin{align}
 \cP^{\sfC\pm_{b_1}\sfC\pm_{b_2}\sfC\cdots\sfC\pm_{b_n}\sfC}
 &=
 \cP^{\pm_{b_1}\sfC\pm_{b_2}\sfC\cdots\sfC\pm_{b_n}\sfC}
 \notag
 \\
 &=
 \cP^{\pm_{b_1}\pm_{b_2}\sfC\cdots\sfC\pm_{b_n}\sfC}
 \notag
 \\
 &
 \quad\vdots
 \notag
 \\
 &=
 \cP^{\pm_{b_1}\pm_{b_2}\cdots \pm_{b_n}\sfC}
\end{align}
and
\begin{align}
 M(\cP^{\sfC\pm_{b_1}\sfC\pm_{b_2}\sfC\cdots\sfC\pm_{b_n}\sfC})
 &=
 M(\cP^{\pm_{b_1}\pm_{b_2}\cdots\pm_{b_n}\sfC})
 \notag
 \\
 &=
 M(\cP^{\pm_{b_1}\pm_{b_2}\cdots\pm_{b_n}})
 \label{eq:MPCPMC}
\end{align}
for all $(b_1,b_2,\ldots,b_n)\in\{0,1\}^n$,
where we can reduce parameter set $\V$ to $\Theta$
by applying operation $\sfSigma$.
This implies that
computing $M(\cP^{\sfC\pm_{b_1}\sfC\pm_{b_2}\sfC\cdots\sfC\pm_{b_n}\sfC})$
might be easier than directly computing
$M(\cP^{\pm_{b_1}\pm_{b_2}\cdots\pm_{b_n}})$.

Based on (\ref{eq:CpmC}), call $\sfC$ the {\em canonicalizing operator}
and $\cP^{\sfC}$ the {\em canonical\footnote{ `Canonical' reflects our
  conjecture that $\cP^{\sfC}$ is the smallest representation of $\cP$
  satisfying (\ref{eq:CpmC}).}
 representation}
of $\cP$.
Then the left hand side of (\ref{eq:MPCPMC})
represents the computation of $M(\cP^{\pm_{b_1}\pm_{b_2}\cdots \pm_{b_n}})$
based on the canonical representation.

\begin{rem}
Let
\begin{align*}
 H(U|V)
 &\equiv
 \sum_{v\in\V}\mu_v\lrB{
  \frac{1+\theta_v}2\log_2\frac2{1+\theta_v}
  +
  \frac{1-\theta_v}2\log_2\frac2{1-\theta_v}
 }
 \notag
 \\
 &=
 \sum_{\theta}\sum_{v:\theta_v=\theta}\mu_v
 \lrB{
  \frac{1+|\theta|}2\log_2\frac2{1+|\theta|}
  +
  \frac{1-|\theta|}2\log_2\frac2{1-|\theta|}
 }
 \\
 Z(U|V)
 &\equiv
 2\sum_{v\in\V}\mu_v\sqrt{\frac{1+\theta_v}2\cdot\frac{1-\theta_v}2}
 \notag
 \\
 &=
 \sum_{\theta}\sum_{v:\theta_v=\theta}\mu_v\sqrt{1-|\theta|^2}
 \\
 H_{\mathrm{Q}}(U|V)
 &\equiv
 2\sum_{v\in\V}\mu_v\cdot\frac{1+\theta_v}2\cdot\frac{1-\theta_v}2
 \notag
 \\
 &=
 \frac 12\lrB{
  1-
  \sum_{\theta}\sum_{v:\theta_v=\theta}\mu_v|\theta|^2
 },
\end{align*}
where $H(U|V)$ is the conditional entropy,
$Z(U|V)$ is the source Battacharyya parameters,
and $H_{\mathrm{Q}}(U|V)$ is the quadratic entropy.
We have similar results regarding these quantities
because it is sufficient to assume a function of $|\theta|$
(i.e.\ even function) in general.
\end{rem}

\subsection{Computation of $M(\cP^{\pm})$}

In this section, we introduce techniques for the computation of $M(\cP^{\pm})$.
When $\cP\equiv\{(\mu_v,\theta_v)\}_{v\in\V}$ is given,
we have $|\cP^-|=|\cP|^2$ and $|\cP^+|=2|\cP|^2$
from the definition of $\cP^{\pm}$.
It seems that the computational complexity
for the computation of $M(\cP^{\pm})$ is $O(|\cP|^2)$.
However, we can reduce it as follows.

For given $\cP$, we can compute $M(\cP)$ with complexity $O(|\cP|)$ and then
\begin{equation*}
 M(\cP^-)=M(\cP)^2
\end{equation*}
by applying (\ref{eq:V-}).
In the following, we introduce the algorithm for the computation of $M(\cP^+)$.

First, we apply operator $\sfC$ introduced in the last section
to obtain canonical list $\cP^{\sfC}\equiv\{(\mu_v,\theta_v)\}_{v=1}^{C}$,
satisfying
\begin{gather}
 \mu_v>0\ \text{for all}\ v\in\{1,\ldots,C\}
 \label{eq:canonical0}
 \\
 \sum_{v=1}^{C}\mu_v = 1
 \label{eq:canonical1}
 \\
 0\leq \theta_1<\theta_2<\cdots<\theta_{C}\leq 1.
 \label{eq:canonical<}
\end{gather}
We introduce Algorithm \ref{alg:canonical} to compute $\cP^{\sfC}$
for given $\cP\equiv\{(\mu_v,\theta_v)\}_{v=1}^{|\V|}$.
The computational complexity of Algorithm \ref{alg:canonical}
is $O(|\cP|\log_2|\cP|)$,
which corresponds to that of the sorting elements of list $\cP$.

\begin{algorithm}
 \caption{Computation of $\cP^{\sfC}$}
 \label{alg:canonical}
 \begin{algorithmic}[1]
  \State Sort list $\cP^{\sfA}\equiv\{(\mu_v,|\theta_v|)\}_{v=1}^{|\V|}$
  in ascending order of $|\theta_v|$.
  \State   $v\leftarrow 1$, $C\leftarrow 0$.
  \While{$v\leq |\V|$}
  \If{$\mu_v>0$}
  \State $C\leftarrow C+1$.
  \State $(\mu_{C},\theta_{C})\leftarrow (\mu_v,\theta_v)$.
  \State $v\leftarrow v+1$.
  \While{$\theta_v=\theta_{C}$}
  \State $\mu_{C}\leftarrow \mu_{C}+\mu_v$.
  \State $v\leftarrow v+1$.
  \EndWhile
  \Else
  \State $v\leftarrow v+1$.
  \EndIf
  \EndWhile
  \State
  Output $\cP^{\sfC}\equiv\{(\mu_v,\theta_v)\}_{v=1}^{C}$.
 \end{algorithmic}
\end{algorithm}

Next we introduce the following lemma.
The proof is given in Section~\ref{sec:proof-laststep+}.
\begin{lem}
\label{lem:laststep+}
Assume that $\cP^{\sfC}\equiv\{(\mu_v,\theta_v)\}_{v=1}^{C}$ satisfies
(\ref{eq:canonical1}) and (\ref{eq:canonical<}).
Then we have
\begin{align}
 M(\cP^+)
 &=
 M(\cP^{\sfC+})
 \notag
 \\
 &=
 \sum_{v=1}^{C}\mu_v\theta_v\lrB{\mu_v+2\sum_{v'=1}^{v-1}\mu_{v'}}.
\end{align}
\end{lem}

Based on this lemma, we introduce Algorithm \ref{alg:Mcanonical},
which computes $M(\cP^{\sfC+})$ for given $\cP^{\sfC}$.
In Algorithm \ref{alg:Mcanonical}, we have
\begin{equation*}
 \tau = 2\sum_{v'=1}^{i-1}\mu_{v'}
\end{equation*}
just before line 4.
At line 4, we have
\begin{equation*}
 \tau = \mu_v+2\sum_{v'=1}^{i-1}\mu_{v'}.
\end{equation*}
At line 5, we have
\begin{equation*}
 M = \sum_{v=1}^i \mu_v\theta_v\lrB{\mu_v+2\sum_{v'=1}^{v-1}\mu_{v'}}
\end{equation*}
and finally we have output
\begin{equation*}
 M = \sum_{v=1}^C \mu_v\theta_v\lrB{\mu_v+2\sum_{v'=1}^{v-1}\mu_{v'}}.
\end{equation*}
From (\ref{eq:MPCpm}), we have
\begin{equation*}
 M(\cP^+)=M(\cP^{\sfC+}).
\end{equation*}
The computational complexity of Algorithm \ref{alg:Mcanonical} is $O(|\cP|)$.

\begin{algorithm}
 \caption{Computation of $M(\cP^{\sfC+})$}
 \label{alg:Mcanonical}
 \begin{algorithmic}[1]
  \State Start from list $\cP^{\sfC}\equiv\{(\mu_v,\theta_v)\}_{v=1}^{C}$.
  \State $M\leftarrow 0$, $\tau\leftarrow 0$.
  \For{$i\in\{1,\ldots,C\}$}
  \State $\tau\leftarrow \tau+\mu_i$.
  \State $M\leftarrow M+\mu_i\theta_i\tau$.
  \State $\tau\leftarrow \tau+\mu_i$.
  \EndFor
  \State
  Output $M$.
 \end{algorithmic}
\end{algorithm}

To obtain value $M(\cP^{\pm_{b_1}\cdots\pm_{b_n}})$
for a given $(b_1,\ldots,b_n)\in\{0,1\}^n$,
we compute
\begin{equation*}
 \cP^{\sfC\pm_{b_1}\sfC\pm_{b_2}\cdots\sfC\pm_{b_{n-1}}\sfC}
\end{equation*}
using (\ref{eq:P-mu})--(\ref{eq:P+theta}),
and Algorithm \ref{alg:canonical},
and then compute
\begin{equation}
 M(\cP^{\pm_{b_1}\cdots\pm_{b_n}})
 =
 M([\cP^{\sfC\pm_{b_1}\sfC\pm_{b_2}\cdots\sfC\pm_{b_{n-1}}\sfC}]^{\pm_{b_n}})
 \label{eq:exact}
\end{equation}
using (\ref{eq:V-}) when $\pm_{b_n}=-$
and using Algorithm \ref{alg:Mcanonical} when $\pm_{b_n}=+$.

\subsection{Approximation of $M(\cP^{\pm^n})$}

In the last section, we introduced techniques
for the exact computation of $M(\cP^{\pm^n})$
based on the canonical expression of $\cP$
to reduce the computational complexity.
However, they are not sufficient for a large $n$.
In this section, we introduce techniques to obtain
an approximation of $M(\cP^{\pm^n})$ by the quantization of $\cP$.
The quantization is based on the previously introduced
greedy degrading-/upgrading-merge algorithm \cite{PHTT11}\cite{TV13},
where we introduce a novel loss function
instead of the difference of mutual information.
We can compute this loss function
using the four fundamental rules of arithmetic,
which is much easier than computing the difference of the mutual
information, including logarithmic functions.
In addition, when initial value $\cP\equiv\{(\mu_y,\theta_y)\}_{y\in\Y}$
consists of rational numbers,
the computation can be exactly performed using the arithmetic of
rational numbers, assuming sufficient memory.
In contrast, we had to focus on the precision of the mutual information
in conventional techniques.

First we introduce the notions of degradation and upgradation
in terms of expression $\cP(U|V)\equiv\{(\mu_v,\theta_v)\}_{v\in\V}$.
Let $P_{UV}$ be the joint distribution of $(U,V)$.
When $U-V-\hV$ forms a Markov chain,
there is a conditional probability distribution $P_{\hV|V}$ such that
the joint distribution of $(U,V,\hV)$ is given as
\begin{equation*}
 P_{UV\hV}(u,v,\hv)=P_{\hV|V}(\hv|v)P_{UV}(u,v)
\end{equation*}
for each $(u,v,\hv)\in\U\times\V\times\hcV$,
where $\U$, $\V$, and $\hcV$ are the alphabets of respective
random variables $U$, $V$, and $\hV$.
The channel between $U$ and $\hV$ is (physically) {\em degraded}
with respect to the channel between $U$ and $V$.
Conversely, 
the channel between $U$ and $V$ is (physically) {\em upgraded}
with respect to the channel between $U$ and $\hV$.
Note that physical degradation/upgradation implies 
the stochastic degradation/upgradation \cite{PHTT11}\cite{TV13};
conditional probability distribution $P_{\hV|V}$ exists, such that
\begin{equation*}
 P_{\hV|U}(\hv|u) = \sum_{v\in\V}P_{\hV|V}(\hv|v)P_{V|U}(v|u)
\end{equation*}
for all $(u,\hv)\in\U\times\hcV$.

Here we consider a particular case of degradation/upgradation
where $\hcV\subset\V$ and $P_{\hV|V}(\hv|v)\in\{0,1\}$
for all $(v,\hv)\in\V\times\hcV$.
In this case, $\hcV$ is a {\em quantization} of $\V$,
and $\V$ is a {\em extension} of $\hcV$.
We define surjection $q:\V\to\hcV$ called {\em quantization map} as
\begin{equation*}
 q(v)\equiv\hv\quad\text{if}\ P_{\hV|V}(\hv|v)=1,
\end{equation*}
where for all $v\in\V$ there is a unique $\hv\in\hcV$ such that
$P_{\hV|V}(\hv|v)=1$ because $P_{\hV|V}(\cdot|v)$ is a probability distribution.
The conditional probability distribution $P_{\hV|V}$
can be represented in terms of $q$,
and $\{q^{-1}(\hv)\}_{\hv\in\hcV}$ forms a partition of $\V$.
Probability distributions $P_{U\hV}$, $P_{\hV}$ and $P_{U|\hV}$ are given as
\begin{align}
 P_{U\hV}(u,\hv)
 &=
 \sum_{v\in\V}P_{\hV|V}(\hv|v)P_{UV}(u,v)
 \notag
 \\
 &=
 \sum_{v\in q^{-1}(\hv)}P_{UV}(u,v)
 \notag
 \\
 &=
 \sum_{v\in q^{-1}(\hv)}\mu_v\cdot\frac{1\mp_u\theta_v}2
 \notag
 \\
 &=
 \frac 12\lrB{
  \sum_{v\in q^{-1}(\hv)}\mu_v\mp_u\sum_{v\in q^{-1}(\hv)}\mu_v\theta_v
 }
 \\
 P_{\hV}(\hv)
 &=
 \sum_{u\in\U}\sum_{v\in\V}
 P_{\hV|V}(\hv|v)P_{UV}(u,v)
 \notag
 \\
 &=
 \sum_{u\in\U}\sum_{v\in q^{-1}(\hv)}
 P_{UV}(u,v)
 \notag
 \\
 &=
 \sum_{v\in q^{-1}(\hv)}
 P_{V}(v)
 \notag
 \\
 &=
 \sum_{v\in q^{-1}(\hv)}\mu_v
 \\
 P_{U|\hV}(u|\hv)
 &=
 \frac{P_{U\hV}(u,\hv)}{P_{\hV}(\hv)}
 \notag
 \\
 &=
 \frac 12\lrB{1\mp_u
  \frac{\sum_{v\in q^{-1}(\hv)}\mu_v\theta_v}
  {\sum_{v\in q^{-1}(\hv)}\mu_v}
 },
\end{align}
where $\mp_u$ is the bipolar-binary conversion of $u$.
Then the joint distribution is represented as
\begin{equation}
 \hcP
 \equiv
 \lrb{\lrsb{
   \sum_{v\in q^{-1}(\hv)}\mu_v,
   \frac{\sum_{v\in q^{-1}(\hv)}\mu_v\theta_v}
   {\sum_{v\in q^{-1}(\hv)}\mu_v}
 }}_{\hv\in\hcV}.
 \label{eq:PV'}
\end{equation}

Let $\usfQ$ denote a quantize operator that is applied to $\cP$,
defined as
\begin{equation*}
 \cP^{\usfQ},
 \equiv
 \hcP
\end{equation*}
where $\usfQ$ depends on quantization map $q$.
Then we call $\cP^{\usfQ}$ a {\em degrading merger} of $\cP$.
Let $\sfE$ denote an extend operator applied to $\hcP$,
defined as
\begin{equation*}
 \hcP^{\sfE}\equiv \cP,
\end{equation*}
where $\hcP$ is a quantization of $\cP$ that satisfies (\ref{eq:PV'}).
Note that $\sfE$ depends on quantization map $q$.
Then {\em upgrading merger} $\hcP^{\osfQ}$ of $\hcP$ is defined as
\begin{align}
 \hcP^{\osfQ}
 &\equiv
 \hcP^{\sfE\sfSigma},
\end{align}
where $\sfSigma$ is defined by (\ref{eq:S}).
Note that the cardinality of the list $\hcP^{\osfQ}$
is reduced to that of $\Theta$ defined by (\ref{eq:Theta})
when $|\Theta|<|\hcP|$,
but it may be increased when $|\Theta|>|\hcP|$.

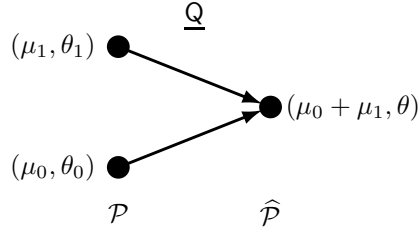
\begin{figure}
\begin{center}
 \unitlength = 1mm
 \begin{picture}(80,34)(0,0)
  \put(30,26){\circle*{3}}
  \put(30,10){\circle*{3}}
  \put(50,18){\circle*{3}}
  \linethickness{1.1pt}
  \put(30,26){\vector(20,-8){19}}
  \put(30,10){\vector(20,8){19}}
  \put(40,30){\makebox(0,0){$\usfQ$}}
  \put(27,26){\makebox(0,0)[r]{$(\mu_1,\theta_1)$}}
  \put(27,10){\makebox(0,0)[r]{$(\mu_0,\theta_0)$}}
  \put(52,18){\makebox(0,0)[l]{$(\mu_0+\mu_1,\theta)$}}
  \put(30,4){\makebox(0,0){$\cP$}}
  \put(50,4){\makebox(0,0){$\hcP$}}
 \end{picture}
\end{center}
\caption{Degrading-merge operation: We define
 $\mu\equiv\mu_0+\mu_1$ and
 $\theta\equiv\frac{\mu_0\theta_0+\mu_1\theta_1}{\mu_0+\mu_1}$,
 where $(\mu,\theta)$ is a member of $\hcP$.}
\label{fig:degradation}
\end{figure}

\begin{figure}
\begin{center}
 \unitlength = 1mm
 \begin{picture}(80,74)(0,0)
  \put(20,58){\circle*{3}}
  \put(20,34){\circle*{3}}
  \put(20,10){\circle*{3}}
  \put(40,58){\circle*{3}}
  \put(40,42){\circle*{3}}
  \put(40,26){\circle*{3}}
  \put(40,10){\circle*{3}}
  \put(60,50){\circle*{3}}
  \put(60,18){\circle*{3}}
  \linethickness{1.1pt}
  \put(20,58){\vector(1,0){19}}
  \put(20,34){\vector(20,8){19}}
  \put(20,34){\vector(20,-8){19}}
  \put(20,10){\vector(1,0){19}}
  \put(40,58){\vector(20,-8){19}}
  \put(40,42){\vector(20,8){19}}
  \put(40,26){\vector(20,-8){19}}
  \put(40,10){\vector(20,8){19}}
  \put(30,62){\makebox(0,0){$\sfE$}}
  \put(50,62){\makebox(0,0){$\sfSigma$}}
  \put(40,70){\makebox(0,0){$\osfQ$}}
  \put(20,66){\vector(1,0){40}}
  \put(20,4){\makebox(0,0){$\hcP$}}
  \put(40,4){\makebox(0,0){$\cP$}}
  \put(60,4){\makebox(0,0){$\Theta$}}
  \put(18,58){\makebox(0,0)[r]{$(\mu_1,\theta_1)$}}
  \put(18,34){\makebox(0,0)[r]{$(\mu,\theta)$}}
  \put(18,10){\makebox(0,0)[r]{$(\mu_0,\theta_0)$}}
  \put(38,55){\makebox(0,0)[r]{$(\mu_1,\theta_1)$}}
  \put(38,44){\makebox(0,0)[r]{$(\mu'_1,\theta_1)$}}
  \put(38,24){\makebox(0,0)[r]{$(\mu'_0,\theta_0)$}}
  \put(38,13){\makebox(0,0)[r]{$(\mu_0,\theta_0)$}}
  \put(63,50){\makebox(0,0)[l]{$(\mu_1+\mu'_1,\theta_1)$}}
  \put(63,18){\makebox(0,0)[l]{$(\mu_0+\mu'_0,\theta_0)$}}
 \end{picture}
\end{center}
\caption{Upgrading-merge operation:
 For given $(\mu,\theta)$ and $(\theta_0,\theta_1)$,
 a pair $(\mu'_0,\mu'_1)$ is defined so that
 $\mu=\mu'_0+\mu'_1$ and
 $\theta=\frac{\mu'_0\theta_0+\mu'_1\theta_1}{\mu'_0+\mu'_1}$.}
\label{fig:upgradation}
\end{figure}
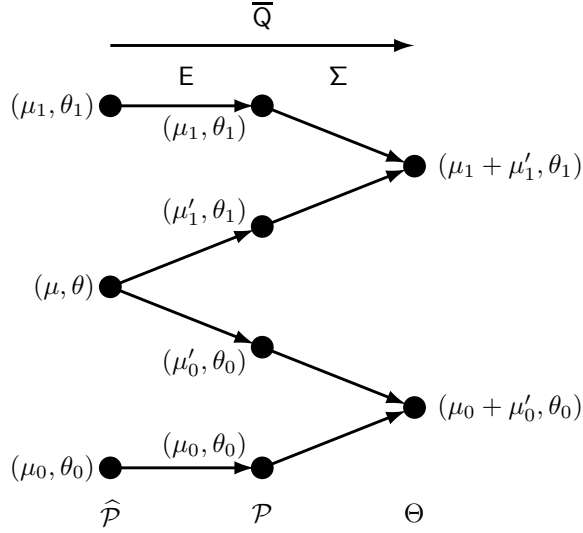

Particular cases of degrading- and upgrading-merge operations
are shown in Figs.~\ref{fig:degradation} and~\ref{fig:upgradation},
and details will be presented in
Sections \ref{sec:degrade} and \ref{sec:upgrade}.

We introduce the following lemma, which asserts that
$M$ is invariant of any quantize operations.
The proof is given in Section~\ref{sec:proof-quantization}.
\begin{lem}
\label{lem:quantization}
Assume that $\cP\equiv\{(\mu_v,\theta_v)\}_{v\in\V}$
satisfies that $\mu_v>0$ and $\theta_v\geq 0$ for all $v\in\V$.
Define $\hcP$ by (\ref{eq:PV'}).
Then 
\begin{align}
 M(\hcP)
 &=
 M(\cP)
 \label{eq:invariance}
 \\
 M(\cP^{\usfQ})
 &=
 M(\cP)
 \label{eq:uQinvariance}
 \\
 M(\hcP^{\osfQ})
 &=
 M(\hcP).
 \label{eq:oQinvariance}
\end{align}
\end{lem}

The above lemma implies that
$M$ is inappropriate for the quantization loss function
because $M$ is unchanged by any degrading/upgrading operations.
Here we introduce a lemma that motivates our quantization loss function.
The proof is given in Section~\ref{sec:proof-MP-k+}.
\begin{lem}
\label{lem:MP-k+}
For given $\cP\equiv\{(\mu_v,\theta_v)\}_{v\in\V}$, we have
\begin{gather}
 M(\cP^{-^k})
 =
 M(\cP)^{2^k}
 \label{eq:MP-k}
 \\
 M(\cP)M(\cP^+)
 \leq
 M(\cP^{-+})
 \leq
 M(\cP^{+-})
 \label{eq:MP-+}
 \\
 M(\cP)^{2^k-1}M(\cP^+)
 \leq
 M(\cP^{-^k+})
 \leq
 M(\cP^+)^{2^k}.
 \label{eq:MP-k+}
\end{gather}
\end{lem}

Assume that $\cP$ is canonical.
From (\ref{eq:MP-k}), we can compute value $M(\cP^{-^k})=M(\cP)^{2^k}$.
This implies that the quantization is unnecessary when $\cP$ is given.
Thus, it is enough to consider the computation of $M(\cP^{-^k+})$.

Here assume that $M(\cP^{-^k+})>M(\cP^{\usfQ-^k+})$,
which corresponds to the case of degradation.
Then quantization loss $M(\cP^{-^k+})-M(\cP^{\usfQ-^k+})$
can be minimized by maximizing $M(\cP^{\usfQ-^k+})$.
Furthermore, we have
\begin{equation*}
 M(\cP)^{2^k-1}M(\cP^{\usfQ+})
 \leq
 M(\cP^{\usfQ-^k+})
\end{equation*}
from (\ref{eq:uQinvariance}) and (\ref{eq:MP-k+}).
This inequality suggests that
the lower bound of $M(\cP^{\usfQ-^k+})$
can be maximized by maximizing $M(\cP^{\usfQ+})$.

Similarly, assume that $M(\cP^{-^k+})<M(\cP^{\osfQ-^k+})$,
which is the case of upgradation.
Then quantization loss $M(\cP^{\osfQ-^k+})-M(\cP^{-^k+})$
can be minimized by minimizing $M(\cP^{\osfQ-^k+})$.
Furthermore, we have
\begin{equation*}
 M(\cP^{\osfQ-^k+})
 \leq
 M(\cP^{\osfQ+})^{2^k}
\end{equation*}
from (\ref{eq:oQinvariance}) and (\ref{eq:MP-k+}).
This inequality suggests that
the upper bound of $M(\cP^{\osfQ-^k+})$
can be minimized by minimizing $M(\cP^{\osfQ+})$.

In the following sections,
we introduce the greedy degrading-/upgrading-merge algorithm
to reduce quantization loss $|M(\cP^{\sfQ+})-M(\cP^+)|$,
where $\sfQ\in\{\usfQ,\osfQ\}$.
Although the computational complexity of the algorithms introduced below
is $O(|\cP|^2)$,
it can be reduced to $O(|\cP|\log_2|\cP|)$
using a doubly linked list and a heap, as in \cite[Sec.~IV-A]{TV13}.

To state the theorems in the next sections,
we define
\begin{align}
 \Error(\cP)
 &\equiv
 \frac 12[1-M(\cP)]
 \label{eq:EP}
 \\
 \otheta(\cP)
 &\equiv
 \max_{v\in\V}\theta_v,
 \label{eq:max_theta}
\end{align}
where (\ref{eq:EP}) comes from (\ref{eq:Error-fMAP}).
For given lists $\cP$ and $\cP'$,
$\cP\setminus\cP'$ denotes the deletion of elements in $\cP'$ from list $\cP$,
and $\cP\cup\cP'$ denotes the addition of elements in $\cP'$ to $\cP$.
We assume that $\cP\equiv\{(\mu_v,\theta_v)\}_{v=1}^{C}$ is canonical,
that is, it satisfies (\ref{eq:canonical0})--(\ref{eq:canonical<}).

\begin{rem}
Computing $\otheta(\cP^{\pm^n})$
for given $\cP\equiv\{(\mu_v,\theta_v)\}_{v\in\V}$,
$(b_1,\ldots,b_n)\in\{0,1\}^n$,
and $\pm^n\equiv(\pm_{b_1},\ldots\pm_{b_n})$ is easy.
Define $t^+$ and $t^-$ as
\begin{align*}
 t^-(\theta)
 &\equiv
 \theta^2
 \\
 t^+(\theta)
 &\equiv
 \frac{2\theta}{1+\theta^2},
\end{align*}
where the first equality comes from 
\begin{equation*}
 \max_{v_0\in\V,v_1\in\V}\theta_{v_0}\theta_{v_1}
 =
 \lrB{\max_{v\in\V}\theta_v}^2
\end{equation*}
and the second one comes from
\begin{align}
 \max_{\substack{
   v_0\in\V,v_1\in\V:
 }}
 \max\lrb{
  \lrbar{\frac{\theta_{v_1}+\theta_{v_0}}
   {1+\theta_{v_0}\theta_{v_1}}},
  \lrbar{\frac{\theta_{v_1}-\theta_{v_0}}
   {1-\theta_{v_0}\theta_{v_1}}}
 }
 &=
 \max_{\substack{
   v_0\in\V,v_1\in\V:
 }}
 \frac{\theta_{v_1}+\theta_{v_0}}
 {1+\theta_{v_0}\theta_{v_1}}
 \notag
 \\
 &=
 \frac{2\max_{v\in\V}\theta_v}
 {1+\lrB{\max_{v\in\V}\theta_v}^2}.
\end{align}
Then we can compute $\otheta(\cP^{\pm^n})$ with recursion
\begin{align*}
 \otheta(\cP^{\pm^{i+1}})
 &=
 t^{\pm_{b_{i+1}}}(\otheta(\cP^{\pm^i}))
\end{align*}
starting from $\otheta(\cP)$.
Note that 
$\otheta(\cP^{\pm^n})\in[0,1)$ if $\otheta(\cP)\in[0,1)$.
\end{rem}

\subsection{Greedy Degrading-Merge Algorithm} 
\label{sec:degrade}

This section introduces a greedy degrading-merge algorithm
based on the one introduced in~\cite{PHTT11}.

Assume that $\cP\equiv\{(\mu_v,\theta_v)\}_{v=1}^{C}$ is canonical satisfying
(\ref{eq:canonical0})--(\ref{eq:canonical<}).
The algorithm iterates the merging of two elements into one
(Fig.~\ref{fig:degradation})
until the cardinality of $\cP$ is reduced to desired number $Q$.

For $i\in\{C,C-1,\ldots,Q\}$,
let $\cP_i\equiv\{(v,\theta_v)\}_{v=1}^i$ be the list
where the cardinality of the list is $i$,
$\cP_{C}\equiv\cP$, and $\cP_{Q}\equiv\cP^{\usfQ}$.
Assume that
two elements, $(\mu_j,\theta_j)$ and $(\mu_{j+1},\theta_{j+1})$,
are merged into the element
\begin{equation*}
 \lrsb{
  \mu_j+\mu_{j+1},
  \frac{\mu_j\theta_j+\mu_{j+1}\theta_{j+1}}
  {\mu_j+\mu_{j+1}}
 }
\end{equation*}
for given $j\in\{1,\ldots i-1\}$,
that is, $\cP_{i-1}$ is defined as
\begin{equation*}
 \cP_{i-1}
 \equiv
 \lrB{\cP_i\setminus\{(\mu_j,\theta_j),(\mu_{j+1},\theta_{j+1})\}}
 \cup
 \lrb{
  \lrsb{
   \mu_j+\mu_{j+1},
   \frac{\mu_j\theta_j+\mu_{j+1}\theta_{j+1}}
   {\mu_j+\mu_{j+1}}
  }
 }.
\end{equation*}
From Lemma~\ref{lem:laststep+}, we have
\begin{align*}
 M(\cP_i^+)
 &=
 \sum_{v=1}^i\mu_v\theta_v\lrB{\mu_v+2\sum_{v'=1}^{v-1}\mu_{v'}}
\end{align*}
and
\begin{align}
 M(\cP_{i-1}^+)
 &=
 \sum_{v=1}^{j-1}\mu_v\theta_v\lrB{\mu_v+2\sum_{v'=1}^{v-1}\mu_{v'}}
 +
 [\mu_j+\mu_{j+1}]
 \cdot
 \frac{\mu_j\theta_j+\mu_{j+1}\theta_{j+1}}
 {\mu_j+\mu_{j+1}}
 \cdot
 \lrB{
  [\mu_j+\mu_{j+1}]
  +2\sum_{v'=1}^{j-1}\mu_{v'}
 }
 \notag
 \\*
 &\quad
 +
 \sum_{v=j+2}^i\mu_v\theta_v
 \lrB{\mu_v
  +2\lrB{
   \sum_{v'=1}^{j-1}\mu_{v'}
   +[\mu_j+\mu_{j+1}]
   +\sum_{v'=j+2}^{v-1}\mu_{v'}
 }}
 \notag
 \\
 &=
 \sum_{v=1}^{j-1}\mu_v\theta_v\lrB{\mu_v+2\sum_{v'=1}^{v-1}\mu_{v'}}
 +
 \lrB{\mu_j\theta_j+\mu_{j+1}\theta_{j+1}}
 \lrB{
  \mu_j+\mu_{j+1}
  +2\sum_{v'=1}^{j-1}\mu_{v'}
 }
 \notag
 \\*
 &\quad
 +
 \sum_{v=j+2}^i\mu_v\theta_v
 \lrB{\mu_v
  +2\sum_{v'=1}^{v-1}\mu_{v'}
 }.
\end{align}
Then we have
\begin{align}
 &
 M(\cP_i^+)-M(\cP_{i-1}^+)
 \notag
 \\*
 &=
 \mu_j\theta_j\lrB{\mu_j+2\sum_{v'=1}^{j-1}\mu_{v'}}
 +
 \mu_{j+1}\theta_{j+1}\lrB{\mu_{j+1}+2\sum_{v'=1}^j\mu_{v'}}
 -
 \lrB{\mu_j\theta_j+\mu_{j+1}\theta_{j+1}}
 \lrB{\mu_j+\mu_{j+1}+2\sum_{v'=1}^{j-1}\mu_{v'}}
 \notag
 \\
 &=
 \mu_j\mu_{j+1}\lrB{\theta_{j+1}-\theta_j}.
 \label{eq:Mdiff-d}
\end{align}
Using the condition (\ref{eq:canonical<}),
we have $M(\cP_i^+)-M(\cP_{i-1}^+)>0$ for every $j\in\{1,\ldots,i-1\}$,
which implies that the degrading-merge operation
reduces the $M$-value after the plus operation, that is,
\begin{equation}
 M(\cP^+)>M(\cP^{\usfQ+}).
 \label{eq:d}
\end{equation}
Then (\ref{eq:Mdiff-d}) introduces a greedy degrading-merge algorithm
as Algorithm \ref{alg:degrading-merge}.
\begin{algorithm}
 \caption{Greedy Degrading-Merge Algorithm}
 \label{alg:degrading-merge}
 \begin{algorithmic}[1]
  \State Start from a canonical list
  $\cP\equiv\{(\mu_v,\theta_v)\}_{v=1}^{C}$.
  \State $i\leftarrow C$.
  \While{$i> Q$}
  \State Find
  \begin{equation*}
   j\equiv \arg\min_{v\in\{1,\ldots,i-1\}}
   \mu_v\mu_{v+1}[\theta_{v+1}-\theta_v].
  \end{equation*}
  \State Merge two pairs $(\mu_{j},\theta_{j})$ and $(\mu_{j+1},\theta_{j+1})$
  into a single pair $(\mu_{j},\theta_{j})$:
  \begin{align*}
   \mu_{j}
   &\leftarrow
   \mu_{j}+\mu_{j+1}
   \\
   \theta_{j}
   &\leftarrow
   \frac{\mu_{j}\theta_{j}+\mu_{j+1}\theta_{j+1}}
   {\mu_{j}+\mu_{j+1}}.
  \end{align*}
  \State Obtain a new list $\{(\mu_v,\theta_v)\}_{v=1}^{i-1}$:
  \begin{align*}
   \mu_v
   &\leftarrow
   \mu_{v+1}
   \\
   \theta_v
   &\leftarrow
   \theta_{v+1}
  \end{align*}
  \Statex[1] for $v\in\{j+1,\ldots i-1\}$, where we do nothing when $j=i-1$.
  \State $i\leftarrow i-1$.
  \EndWhile
  \State Output $\cP^{\usfQ}\equiv\{(\mu_v,\theta_v)\}_{v=1}^{Q}$.
 \end{algorithmic}
\end{algorithm}

Regarding the quantization loss, we have the following theorem,
where (\ref{eq:abs-M-d}) and (\ref{eq:abs-error-d}) show the absolute precision
and (\ref{eq:rel-M-d}) and (\ref{eq:rel-error-d}) show the relative precision,
which is relevant when the value is close to $0$.
The proof is given in Section~\ref{sec:proof-degrade}.
\begin{thm}
\label{thm:degrade}
Let $M$ be defined by (\ref{eq:MP}),
let $\otheta(\cP)$ be defined by (\ref{eq:max_theta}),
and let $\Error(\cP)$ be defined by (\ref{eq:EP}).
Let $Q\equiv|\cP^{\usfQ+}|$.
Then we have
\begin{gather}
 M(\cP^+)-\frac{\otheta(\cP)}{Q^2}\lrB{1-\frac{Q^2}{2C^2}}
 \leq
 M(\cP^{\usfQ+})
 <
 M(\cP^+)
 \label{eq:abs-M-d}
 \\
 M(\cP^+)\lrB{1-\frac{\otheta(\cP)}{M(\cP)Q^2}\lrB{1-\frac{Q^2}{2C^2}}}
 \leq
 M(\cP^{\usfQ+})
 <
 M(\cP^+)
 \label{eq:rel-M-d}
 \\
 \Error(\cP^+)
 <
 \Error(\cP^{\usfQ+})
 \leq
 \Error(\cP^+)
 +\frac{\otheta(\cP)}{2Q^2}
 \lrB{1-\frac{Q^2}{2C^2}}
 \label{eq:abs-error-d}
 \\
 \Error(\cP^+)
 <
 \Error(\cP^{\usfQ+})
 \leq
 \Error(\cP^+)
 \lrB{
  1
  +\frac{\otheta(\cP)}{[1-\otheta(\cP)]Q^2}\lrB{1-\frac{Q^2}{2C^2}}
 }.
 \label{eq:rel-error-d}
\end{gather}
\end{thm}

The following theorem provides tighter bounds
when $M(\cP)$ and $\Error(\cP)$ are close to either $0$ or $1$.
\begin{thm}
\label{thm:degrade1}
We have
\begin{gather}
 M(\cP^+)-\frac{M(\cP)[1-M(\cP)]}{Q-1}\lrB{1-\frac{Q-1}{C-1}}
 \leq
 M(\cP^{\usfQ+})
 <
 M(\cP^+)
 \label{eq:abs-M-d-1}
 \\
 M(\cP^+)\lrB{1-\frac{1-M(\cP)}{Q-1}\lrB{1-\frac{Q-1}{C-1}}}
 \leq
 M(\cP^{\usfQ+})
 <
 M(\cP^+)
 \label{eq:rel-M-d-1}
 \\
 \Error(\cP^+)
 <
 \Error(\cP^{\usfQ+})
 \leq
 \Error(\cP^+)
 +\frac{M(\cP)[1-M(\cP)]}{2[Q-1]}\lrB{1-\frac{Q-1}{C-1}}
 \label{eq:abs-error-d-1}
 \\
 \Error(\cP^+)
 <
 \Error(\cP^{\usfQ+})
 \leq
 \Error(\cP^+)
 \lrB{
  1
  +\frac{M(\cP)[1-M(\cP)]}{[1-\otheta(\cP)][Q-1]}
  \lrB{1-\frac{Q-1}{C-1}}
 }.
 \label{eq:rel-error-d-1}
\end{gather}
\end{thm}

\begin{rem}
When $\otheta(\cP)=1$, the right hand side inequality of
(\ref{eq:rel-error-d}) and (\ref{eq:rel-error-d-1})
are trivial because the right hand side is $\infty$.
In this case, we can modify the algorithm as follows:
apply it to $\{(\mu_v,\theta_v)\}_{v=1}^{C-1}$,
where $(\mu_C,\theta_C)$ is excluded
and return $(\mu_C,\theta_C)$ to the output.
Then we have $\cP^{\usfQ}$ satisfying
\begin{align*}
 \Error(\cP^{\usfQ^+})
 \leq
 \Error(\cP^+)
 \lrB{
  1
  +\frac{\theta_{C-1}}{[1-\theta_{C-1}][Q-1]^2}
  \lrB{1-\frac{[Q-1]^2}{2[C-1]^2}}
 }
 \\
 \Error(\cP^{\usfQ+})
 \leq
 \Error(\cP^+)
 \lrB{
  1
  +\frac{M(\cP)[1-M(\cP)]}{[1-\theta_{C-1}][Q-2]}
  \lrB{1-\frac{Q-2}{C-2}}
 }.
\end{align*}
\end{rem}

\subsection{Greedy Upgrading-Merge Algorithm}
\label{sec:upgrade}

This section introduces a greedy upgrading-merge algorithm
based on the one introduced in~\cite{PHTT11}.

Assume that $\cP\equiv\{(\mu_v,\theta_v)\}_{v=1}^C$ is canonical
satisfying (\ref{eq:canonical0})--(\ref{eq:canonical<}).
The algorithm iterates the merging of three elements into two elements
(Fig.~\ref{fig:upgradation})
until the cardinality of $\cP$ is reduced to desired number $Q$.

For $i\in\{C,C-1,\ldots,Q\}$,
let $\cP_i\equiv\{(v,\theta_v)\}_{v=1}^i$ be the list
obtained by the algorithm,
where the cardinality of the list is $i$,
$\cP_C\equiv\cP$, and $\cP_Q\equiv\cP^{\usfQ}$.
Here assume that,
for given $j\in\{2,\ldots,i-1\}$,
an element $(\mu_j,\theta_j)$ is extended to two elements
$(\mu'_{j-1},\theta_{j-1})$ and $(\mu'_{j+1},\theta_{j+1})$,
and then four elements
$(\mu_{j-1},\theta_{j-1})$,
$(\mu'_{j-1},\theta_{j-1})$,
$(\mu'_{j+1},\theta_{j+1})$,
and $(\mu_{j+1},\theta_{j+1})$
are merged into two elements
$(\mu_{j-1}+\mu'_{j-1},\theta_{j-1})$ and
$(\mu_{j+1}+\mu'_{j+1},\theta_{j+1})$,
that is, $\cP_{i-1}$ is defined as
\begin{equation*}
 \cP_{i-1}
 \equiv
 \lrB{\cP_i\setminus\{
   (\mu_{j-1},\theta_{j-1}), (\mu_j,\theta_j),
   (\mu_{j+1},\theta_{j+1})
   \}
 }
 \cup
 \lrb{
  (\mu_{j-1}+\mu'_{j-1},\theta_{j-1}),
  (\mu_{j+1}+\mu'_{j+1},\theta_{j+1})
 }.
\end{equation*}
Then $\mu'_{j-1}$ and $\mu'_{j+1}$ should satisfy
\begin{align}
 \mu_j
 &=
 \mu'_{j-1}+\mu'_{j+1}
 \label{eq:mu_j}
 \\
 \theta_j
 &=
 \frac{\mu'_{j-1}\theta_{j-1}+\mu'_{j+1}\theta_{j+1}}
 {\mu'_{j-1}+\mu'_{j+1}}.
 \label{eq:theta_j}
\end{align}
From the above two equalities, we have
\begin{align}
 \mu'_{j-1}
 &=
 \frac{\mu_j[\theta_j-\theta_{j-1}]}{\theta_{j+1}-\theta_{j-1}}
 \label{eq:mu'-1}
 \\
 \mu'_{j+1}
 &=
 \frac{\mu_j[\theta_{j+1}-\theta_j]}{\theta_{j+1}-\theta_{j-1}}.
 \label{eq:mu'+1}
\end{align}
From Lemma~\ref{lem:laststep+}, we have
\begin{align}
 M(\cP^+_{i-1})
 &=
 \sum_{v=1}^{j-2}\mu_v\theta_v\lrB{\mu_v+2\sum_{v'=1}^{v-1}\mu_{v'}}
 +
 \lrB{\mu_{j-1}+\mu'_{j-1}}
 \theta_{j-1}
 \lrB{
  [\mu_{j-1}+\mu'_{j-1}]
  +
  2\sum_{v'=1}^{j-2}\mu_{v'}
 }
 \notag
 \\*
 &\quad
 +
 \lrB{\mu_{j+1}+\mu'_{j+1}}
 \theta_{j+1}
 \lrB{
  [\mu_{j+1}+\mu'_{j+1}]
  +
  2\lrB{
   \sum_{v'=1}^{j-2}\mu_{v'}
   +
   [\mu_{j-1}+\mu'_{j-1}]
  }
 }
 +
 \sum_{v=j+2}^C\mu_v\theta_v
 \lrB{
  \mu_v
  +2\sum_{v'=1}^{v-1}\mu_{v'}
 }
 \notag
 \\
 &=
 \sum_{v=1}^{j-1}\mu_v\theta_v\lrB{\mu_v+2\sum_{v'=1}^{v-1}\mu_{v'}}
 +
 \mu'_{j-1}
 \theta_{j-1}
 \lrB{
  \mu_j-\mu'_{j+1}
  +
  2\sum_{v'=1}^{j-1}\mu_{v'}
 }
 \notag
 \\*
 &\quad
 +
 \mu'_{j+1}
 \theta_{j+1}
 \lrB{
  \mu_j+\mu'_{j-1}
  +
  2\sum_{v'=1}^{j-1}\mu_{v'}
 }
 +
 \sum_{v=j+1}^C\mu_v\theta_v
 \lrB{
  \mu_v
  +2\sum_{v'=1}^{v-1}\mu_{v'}
 },
\end{align}
where we use (\ref{eq:mu_j}) in the last equality.
Then we have
\begin{align}
 M(\cP^+_{i-1})-M(\cP^+_i)
 &=
 \mu'_{j-1}
 \theta_{j-1}
 \lrB{
  \mu_j-\mu'_{j+1}
  +
  2\sum_{v'=1}^{j-1}\mu_{v'}
 }
 +
 \mu'_{j+1}\theta_{j+1}
 \lrB{
  \mu_j
  +
  \mu'_{j-1}
  +
  2\sum_{v'=1}^{j-1}\mu_{v'}
 }
 -
 \mu_j\theta_j\lrB{\mu_j+2\sum_{v'=1}^{j-1}\mu_{v'}}
 \notag
 \\
 &=
 \mu'_{j-1}\mu'_{j+1}[\theta_{j+1}-\theta_{j-1}],
 \label{eq:Mdiff-u}
\end{align}
where we use (\ref{eq:mu_j}) and (\ref{eq:theta_j}) in the last equality.
By using the condition (\ref{eq:canonical<}),
we have $M(\cP_{i-1}^+)-M(\cP_i^+)>0$ for every $j\in\{2,\ldots,i-1\}$,
which implies that the upgrading-merge operation
increases the $M$-value after the plus operation, that is,
\begin{equation}
 M(\cP^+)<M(\cP^{\osfQ+}).
 \label{eq:u}
\end{equation}
Then (\ref{eq:Mdiff-u}) introduces a greedy upgrading-merge algorithm as
Algorithm \ref{alg:upgrading-merge}.
\begin{algorithm}
 \caption{Greedy Upgrading-Merge Algorithm}
 \label{alg:upgrading-merge}
 \begin{algorithmic}[1]
  \State Start from a canonical list $\cP\equiv\{(\mu_v,\theta_v)\}_{v=1}^C$.
  \State $i\leftarrow C$.
  \While{$i> Q$}
  \State Find
  \begin{equation}
   j\equiv \arg\min_{v\in\{2,\ldots,i-1\}}
   \mu'_{v-1}\mu'_{v+1}[\theta_{v+1}-\theta_{v-1}],
   \label{eq:argmin_u}
  \end{equation}
  \Statex[1] where $\mu'_{v-1}$ and $\mu'_{v+1}$
  are respectively defined by (\ref{eq:mu'-1}) and (\ref{eq:mu'+1}).
  \State Split the pair $(\mu_{j},\theta_{j})$ into
  two pairs $(\mu'_{j-1},\theta_{j-1})$ and $(\mu'_{j+1},\theta_{j+1})$
  and merge into the respective pairs
  \Statex[1] $(\mu_{j-1},\theta_{j-1})$ and $(\mu_{j+1},\theta_{j+1})$:
  \begin{align*}
   \mu_{j-1}
   &\leftarrow
   \mu_{j-1}+\mu'_{j-1}
   \\
   \mu_{j+1}
   &\leftarrow
   \mu_{j+1}+\mu'_{j-1}.
  \end{align*}
  \State Obtain new list $\{(\mu_v,\theta_v)\}_{v=1}^{i-1}$:
  \begin{align*}
   \mu_v
   &\leftarrow
   \mu_{v+1}
   \\
   \theta_v
   &\leftarrow
   \theta_{v+1}
  \end{align*}
  \Statex[1] for $v\in\{j+1,\ldots i-1\}$.
  \State $i\leftarrow i-1$.
  \EndWhile
  \State Output $\cP^{\osfQ}\equiv\{(\mu_v,\theta_v)\}_{v=1}^Q$.
 \end{algorithmic}
\end{algorithm}

Regarding the quantization loss, we have the following theorem,
where (\ref{eq:abs-M-u}) and (\ref{eq:abs-error-u}) show the absolute precision
and (\ref{eq:rel-M-u}) and (\ref{eq:rel-error-u}) show the relative precision,
which is relevant when the value is close to $0$.
The proof is given in Section~\ref{sec:proof-upgrade}.
\begin{thm}
\label{thm:upgrade}
Let $M$ be defined by (\ref{eq:MP}),
let $\otheta(\cP)$ be defined by (\ref{eq:max_theta}),
and let $\Error(\cP)$ be defined by (\ref{eq:EP}).
Let $Q\equiv|\cP^{\osfQ+}|$.
Then we have
\begin{gather}
 M(\cP^+)
 <
 M(\cP^{\osfQ+})
 \leq
 M(\cP^+)
 +\frac{\otheta(\cP)}{[Q-1]^2}
 \lrB{1-\frac{[Q-1]^2}{2[C-1]^2}}
 \label{eq:abs-M-u}
 \\
 M(\cP^+)
 <
 M(\cP^{\osfQ+})
 \leq
 M(\cP^+)\lrB{
  1
  +\frac{\otheta(\cP)}{M(\cP)[Q-1]^2}
  \lrB{1-\frac{[Q-1]^2}{2[C-1]^2}}
 }
 \label{eq:rel-M-u}
 \\
 \Error(\cP^+)
 -\frac{\otheta(\cP)}{2[Q-1]^2}
 \lrB{1-\frac{[Q-1]^2}{2[C-1]^2}}
 \leq
 \Error(\cP^{\osfQ+})
 <
 \Error(\cP^+)
 \label{eq:abs-error-u}
 \\
 \Error(\cP^+)
 \lrB{
  1
  -\frac{\otheta(\cP)}{[1-\otheta(\cP)][Q-1]^2}
  \lrB{1-\frac{[Q-1]^2}{2[C-1]^2}}
 }
 \leq
 \Error(\cP^{\osfQ+})
 <
 \Error(\cP^+).
 \label{eq:rel-error-u}
\end{gather}
\end{thm}

The following theorem provides tighter bounds when $M(\cP)$ is close
to either $0$ or $1$.
\begin{thm}
\label{thm:upgrade1}
We have
\begin{gather}
 M(\cP^+)
 <
 M(\cP^{\osfQ+})
 \leq
 M(\cP^+)
 +
 \frac{\min\lrb{M(\cP),1-M(\cP)}}{Q-2}
 \lrB{1-\frac{Q-2}{C-2}}
 \label{eq:abs-M-u-1}
 \\
 M(\cP^+)
 <
 M(\cP^{\osfQ+})
 \leq
 M(\cP^+)\lrB{
  1
  +\frac{1}{Q-2}
  \lrB{1-\frac{Q-2}{C-2}}
 }
 \label{eq:rel-M-u-1}
 \\
 \Error(\cP^+)
 -\frac{\min\lrb{M(\cP),1-M(\cP)}}{2[Q-2]}
 \lrB{1-\frac{Q-2}{C-2}}
 \leq
 \Error(\cP^{\osfQ+})
 <
 \Error(\cP^+)
 \label{eq:abs-error-u-1}
 \\
 \Error(\cP^+)
 \lrB{
  1
  -\frac{1-M(\cP)}{[1-\otheta(\cP)][Q-2]}
  \lrB{1-\frac{Q-2}{C-2}}
 }
 \leq
 \Error(\cP^{\osfQ+})
 <
 \Error(\cP^+).
 \label{eq:rel-error-u-1}
\end{gather}
\end{thm}

\begin{rem}
When $\otheta(\cP)=1$, the left hand side inequalities of
(\ref{eq:rel-error-u}) and (\ref{eq:rel-error-u-1})
are trivial because the left hand side is -$\infty$.
In this case, we can modify the algorithm:
apply it to $\{(\mu_v,\theta_v)\}_{v=1}^{C-1}$,
where $(\mu_C,\theta_C)$ is excluded,
and return $(\mu_C,\theta_C)$ to the output.
Then we have $\cP^{\osfQ}$ satisfying
\begin{align*}
 \Error(\cP^{\osfQ+})
 \geq
 \Error(\cP^+)
 \lrB{1
  -\frac{\theta_{C-1}}{[1-\theta_{C-1}][Q-2]^2}
  \lrB{1-\frac{[Q-2]^2}{2[C-2]^2}}
 }
 \\
 \Error(\cP^{\osfQ+})
 \geq
 \Error(\cP^+)
 \lrB{1
  -\frac{1-M(\cP)}{[1-\theta_{C-1}][Q-3]^2}
   \lrB{1-\frac{Q-3}{C-3}}
 }.
\end{align*}
\end{rem}

\section{Experimental Results}
\label{sec:experiment}

In this section, we introduce our experimental results.
In them, we calculate the bit error probability of
the polar source code without decoder side information,
where we define $P_X(1)\equiv 0.11$.
The initial joint probability distribution is $\cP\equiv\{(1,\theta)\}$,
where $\theta\equiv P_X(0)-P_X(1)= 0.78$.

Our results can be applied to the polar channel code
for the binary symmetric channel,
where the crossover probability is $P_{Y|X}(1|0)=P_{Y|X}(0|1)=0.11$.
Although we can reduce the linear channel code to linear
source code without decoder side information
by considering the syndrome decoding of additive noise,
we introduce an alternative approach.
Assume that $\e = P_{Y|X}(1|0)=P_{Y|X}(0|1)$
and the channel input distribution is uniform,
where $P_X(0)=P_X(1)=1/2$.
Then we have $P_Y(0)=P_Y(1)=1/2$ and $P_{X|Y}(x|y)=P_{Y|X}(x|y)$ for all $x,y\in\{0,1\}$.
The initial joint probability distribution is given as
$\cP\equiv\{(1/2,\theta_0),(1/2,\theta_1)\}$,
where $\theta_y$ is defined as
\begin{align}
 \theta_y
 &\equiv
 P_{X|Y}(0|y)-P_{X|Y}(1|y)
 \notag
 \\
 &=
 \mp_y [1-2\e]
\end{align}
using bipolar-binary conversion $\mp_y$ of $y\in\{0,1\}$.
Since $|\theta_0|=|\theta_1|=|1-2\e|$,
we have canonical representation $\cP^{\sfC}$ of joint probability as
\begin{equation*}
 \cP^{\sfC}=\{(1,|1-2\e|)\}.
\end{equation*}
By applying (\ref{eq:MPC}), 
we can compute the bit error probability
of the polar channel code
with initial distribution $\cP^{\sfC}=\{(1,|1-2\e|)\}$,
which is the same as the initial joint probability distribution
of the polar source code.
By letting $\e\equiv 0.11$, we have $\cP^{\sfC}=\{(1,0.78)\}$.

In the experimental results,
we consider the polar codes with $n\in\{8,10\}$,
where the block length is $2^n\in\{256,1024\}$.
Tables~\ref{table:n=8} and \ref{table:n=10}
show the bit error probabilities,
where the set $[0^n:1^n]$ of indexes
is sorted in descending order of the bit error probability.
A negative value, $-\gamma$ ($\gamma>0$),
means that the bit error probability is $0.5-\gamma$.

When $n=8$, we exactly compute the bit error probabilities as (\ref{eq:exact}),
where we use only Algorithms \ref{alg:canonical} and \ref{alg:Mcanonical}.
Since the initial joint distribution consists of rational numbers,
we can employ algorithms with infinite precision
using the four fundamental rules of arithmetic.
After the exact computation of the bit error probability,
the fifth digit is rounded up to obtain the upper bound
in Table~\ref{table:n=8}.

When $n=10$, we compute the approximation of the bit error probability
with the greedy degrading-/upgrading-merge algorithms
(Algorithms \ref{alg:degrading-merge} and \ref{alg:upgrading-merge})
before the $\pm$ operations.
Although $Q\equiv 128$ is sufficient to determine the order of the bit error probabilities, we set $Q\equiv 512$ to improve the approximation.
In Algorithm \ref{alg:upgrading-merge},
we reduce the objective function of minimization (\ref{eq:argmin_u}) to
\begin{align}
 \mu'_{v-1}\mu'_{v+1}[\theta_{v+1}-\theta_{v-1}]
 &=
 \frac{\mu_v[\theta_v-\theta_{v-1}]}{\theta_{v+1}-\theta_{v-1}}
 \cdot
 \frac{\mu_v[\theta_{v+1}-\theta_v]}{\theta_{v+1}-\theta_{v-1}}
 \cdot
 [\theta_{v+1}-\theta_{v-1}]
 \notag
 \\
 &=
 \frac{\mu_v^2[\theta_v-\theta_{v-1}][\theta_{v+1}-\theta_v]}
 {\theta_{v+1}-\theta_{v-1}}
 \notag
 \\
 &=
 \frac{\mu_v^2[\theta_v-\theta_{v-1}][\theta_{v+1}-\theta_v]}
 {[\theta_{v+1}-\theta_v]+[\theta_v-\theta_{v-1}]}
 \notag
 \\
 &=
 \frac{\mu_v^2}
 {\frac1{\theta_v-\theta_{v-1}}+\frac1{\theta_{v+1}-\theta_v}},
\end{align}
where the first equality comes from (\ref{eq:mu'-1}) and (\ref{eq:mu'+1}).
Using this relation, we can reduce the number of multiplications,
where the computation of the inverse of a rational number is easy.
The fifth digit of the bit error probability
is suitably truncated or rounded up in Table~\ref{table:n=10}
to make the margin of the upper and lower bounds.

\begin{rem}
\label{rem:same}
Both in Tables~\ref{table:n=8} and~\ref{table:n=10},
the bit error probability corresponding to $b^n=0^{k}10^{n-k-1}$
($\pm^n=-^{k}+-^{n-k-1}$)
is constant for all $k\in\{0,1,\dots,n-1\}$.
In the following, we show that this fact is generally true.
When $\cP\equiv\{(1,\theta)\}$, we have
\begin{equation*}
 \cP^{-^k}
 =
 \{(1,\theta^{2^k})\}
\end{equation*}
from (\ref{eq:P-mu}) and (\ref{eq:P-theta}).
Then we have
\begin{equation*}
 \cP^{-^k+}
 =
 \lrb{
  \lrsb{
   \frac{1+\theta^{2^{k+1}}}2,
   \frac{2\theta^{2^k}}{1+\theta^{2^{k+1}}}
  },
  \lrsb{
   \frac{1-\theta^{2^{k+1}}}2, 0
  }
 }
\end{equation*}
from (\ref{eq:P+mu}), (\ref{eq:P+theta}) and
\begin{equation*}
 M(\cP^{-^k+})= \theta^{2^k}
\end{equation*}
from (\ref{eq:MP}).
Then we have
\begin{align}
 M(\cP^{-^k+-^{n-k-1}})
 &=
 [\theta^{2^k}]^{2^{n-k-1}}
 \notag
 \\
 &=
 \theta^{2^{n-1}},
\end{align}
which does not depend on $k\in\{0,1,\ldots,n-1\}$,
where the first equality comes from (\ref{eq:MP-k}).
\end{rem}

Here, we compare the proposed method
with the conventional method in \cite{PHTT11,TV13}.
To this end, let us describe the implementation of the conventional method.
The symmetric parameterization of the binary probability distribution
is used in the implementation of the conventional method,
where the plus and minus operations are employed with infinite precision
using the four fundamental rules of arithmetic
but Algorithm \ref{alg:Mcanonical} is not used.
Because Algorithm \ref{alg:Mcanonical} is not used
in the conventional method,
we have to use degrading-/upgrading-merge algorithms
for the both case of $n=8$ and $n=10$.
In the degrading-/upgrading-merge algorithms,
the loss function in the conventional method
is the differnce $|H(\cP_i)-H(\cP_{i-1})|$
of the conditional entropies, where
\begin{equation*}
 H(\cP)
 \equiv
 \sum_{v\in\V}\mu_v
 \lrB{
  \frac{1+|\theta_v|}2\log_2\frac2{1+|\theta_v|}
  +
  \frac{1-|\theta_v|}2\log_2\frac2{1-|\theta_v|}
 }
\end{equation*}
for $\cP\equiv\{(\mu_v,\theta_v)\}_{v=1}^C$.
It should be noted that the condional entropy corresponds
to the capacity $C(\cP)\equiv 1-H(\cP)$,
which is the mutual informathion between the uniform input and the output,
of the binary symmetric channel.
In the computation of the conditional entropy,
parameters and results are rounded/trancated to floting point numbers
with $256$ bits mantissa when $n=8$ and $1024$ bits mattissa when $n=10$
so that the results are stable in the sense
that they are the same with double-bits mantissa
($512$ bits when $n=8$, $2048$ bits when $n=10$).

Table \ref{table:vs} shows
the maximum relative error and the number of unclarified relations.
Let $\ueps_{n,Q}(i)$ and $\oeps_{n,Q}(i)$ 
be the lower and the upper bounds, respectively, of
the $i$-th highest bit error probability
for given number of polar transforms, $n$,
and number of quantization points $Q$.
Then the maximum relative errors $\odelta_{n,Q}$ and $\udelta_{n,Q}$
are given as
\begin{align*}
 \odelta_{n,Q}
 &\equiv 
 \max_{i\in\{1,\ldots,2^n\}}
 \frac{\oeps_{n,Q}(i)-\e_n(i)}{\e_n(i)}
 \\
 \udelta_{n,Q}
 &\equiv 
 \max_{i\in\{1,\ldots,2^n\}}
 \frac{\e_{n}(i)-\ueps_{n,Q}(i)}{\e_n(i)},
\end{align*}
where $\e_8(i)$ is defined by Table \ref{table:n=8}
obtained by using the proposed method without quantization ($Q=\infty$)
and $\e_{10}(i)$ is defined as
the harmonic mean of $\ueps_{10,512}(i)$ and $\oeps_{10,512}(i)$
both of which appear in Table \ref{table:n=10}.
The number of unclarified relations $N_{n,Q}$ is given as
\[
 N_{n,Q}
 \equiv
 \lrbar{
  \lrb{
   i\in\{1,\ldots,2^n-1\}:
   \oeps_{n,Q}(i) > \ueps_{n,Q}(i+1)
   \ \text{and}\ b^n(i),b^n(i+1)\notin\{0^k10^{n-k-1}:k\in\{0,\ldots,n-1\}
  }
 },
\]
where $b^n(i)$ is the $i$-th binary sequence
representing plus/minus transforms
and the condition $b^n(i),b^n(i+1)\in\{0^k10^{n-k-1}:k\in\{0,\ldots,n-1\}\}$
corresponds to the case $\e_{n,Q}(i)=\e_{n,Q}(i+1)$
mentioned in Remark \ref{rem:same}.
It is shown that the proposed method outperforms the conventional one.

Table \ref{table:vs-cpu} shows the CPU time,
which is obtained by using $\mathtt{clock}()$ in the C standard library,
of the computation,
where Algorithm \ref{alg:Mcanonical} in the proposed method
and the computation of $M(\cP)$ in the conventional method
are employed with the floting point numbers
with $256$ bits mantissa when $n=8$ and $1024$ bits mantissa when $n=10$
so that the results are stable.
In the both proposed and conventional degrading-/upgrading-merge algorithms,
a doubly linked list and a heap are used as in \cite[Sec.~IV-A]{TV13}.
It is shown that the CPU time of the proposed method
is much smaller than that of the conventional method,
where the dominant reason for the difference
seems to be the usage of Algorithm \ref{alg:Mcanonical}.

\begin{rem}
 To compare loss functions of the proposed and the conventional methods
 without computing the Battacharyya parameters,
 the algorithm introduced in \cite[Algorithm D]{TV13}
 is not used in the experiments.
 In fact, when $Q$ is small, 
 we can obtain better upper bound of the bit error probability
 for the both proposed and conventional methods
 by computing the upper bound\footnote{
  It should be noted that 
  we can improve \cite[Line 11 of Algorithm D]{TV13}
  by using the upper bound in (\ref{eq:Zbound}).
 }
 in terms of the Battacharyya parameter.
 Furthermore, we may obtain better upper bound of $M(\cP^+)$
 corresponding to the lower bound of the bit error probability
 in the proposed method similarly to \cite[Algorithm D]{TV13}
 by using the relations
 \begin{align*}
  M(\cP^-)&= M(\cP)^2
  \\
  M(\cP^+)&\leq 2M(\cP) - M(\cP)^2,
 \end{align*}
 which come from (\ref{eq:V-}) and (\ref{eq:V-+}).
\end{rem}

\section{Proofs}
\label{sec:proof}

\subsection{Proof of Lemma~\ref{lem:MP+-}}
\label{sec:proof-MP+-}
The first equality is shown directly
from (\ref{eq:MP}), (\ref{eq:P-mu}), and (\ref{eq:P-theta}).
The second equality is shown as
\begin{align}
 M(\cP^+)
 &=
 \sum_{v_0\in\V}
 \sum_{v_1\in\V}
 \frac{\mu_{v_0}\mu_{v_1}[1+\theta_{v_0}\theta_{v_1}]}2
 \lrbar{\frac{\theta_{v_1}+\theta_{v_0}}{1+\theta_{v_0}\theta_{v_1}}}
 +
 \sum_{v_0\in\V}
 \sum_{v_1\in\V}
 \frac{\mu_{v_0}\mu_{v_1}[1-\theta_{v_0}\theta_{v_1}]}2
 \lrbar{\frac{\theta_{v_1}-\theta_{v_0}}{1-\theta_{v_0}\theta_{v_1}}}
 \notag
 \\
 &=
 \sum_{v_0\in\V}
 \sum_{v_1\in\V}
 \frac{\mu_{v_0}\mu_{v_1}\lrbar{\theta_{v_1}+\theta_{v_0}}}2
 +
 \sum_{v_0\in\V}
 \sum_{v_1\in\V}
 \frac{\mu_{v_0}\mu_{v_1}\lrbar{\theta_{v_1}-\theta_{v_0}}}2
 \notag
 \\
 &=
 \sum_{v_0\in\V}\sum_{v_1\in\V}
 \mu_{v_0}\mu_{v_1}\max\{|\theta_{v_0}|,|\theta_{v_1}|\},
\end{align}
where the first equality comes from
(\ref{eq:MP}), (\ref{eq:P+mu}), and (\ref{eq:P+theta}),
the second equality comes from the fact that
$\theta_{v_0},\theta_{v_1}\in[-1,1]$
and $|\theta_{v_1}\pm\theta_{v_0}|=0$ when $1\pm\theta_{v_1}\theta_{v_0}=0$,
and the last equality comes from Lemma~\ref{lem:max} in Appendix.
\hfill\qed

\subsection{Proof of Lemma~\ref{lem:M}}
\label{sec:proof-M}
Although the lemma was previously given \cite[Proposition 5.1]{A15},
we show it directly with Lemma~\ref{lem:MP+-}
and the fact that $\theta_v\in[-1,1]$.
Inequalities (\ref{eq:-V+}) are shown as
\begin{align}
 M(\cP^-)
 &=
 \sum_{v_0\in\V}\sum_{v_1\in\V}
 \mu_{v_0}\mu_{v_1}|\theta_{v_0}\theta_{v_1}|
 \notag
 \\
 &\leq
 \sum_{v_0\in\V}\sum_{v_1\in\V}
 \mu_{v_0}\mu_{v_1}|\theta_{v_0}|
 \notag
 \\
 &=
 \sum_{v_0\in\V}
 \mu_{v_0}|\theta_{v_0}|
 \notag
 \\
 &=
 M(\cP)
\end{align}
and
\begin{align}
 M(\cP^+)
 &=
 \sum_{v_0\in\V}\sum_{v_1\in\V}
 \mu_{v_0}\mu_{v_1}\max\{|\theta_{v_0}|,|\theta_{v_1}|\}
 \notag
 \\
 &\geq
 \sum_{v_0\in\V}\sum_{v_1\in\V}
 \mu_{v_0}\mu_{v_1}|\theta_{v_0}|
 \notag
 \\
 &=
 \sum_{v_0\in\V}
 \mu_{v_0}|\theta_{v_0}|
 \notag
 \\
 &=
 M(\cP).
\end{align}
Equality (\ref{eq:V-}) is shown as
\begin{align}
 M(\cP^-)
 &=
 \sum_{v_0\in\V}\sum_{v_1\in\V}
 \mu_{v_0}\mu_{v_1}|\theta_{v_0}\theta_{v_1}|
 \notag
 \\
 &=
 \sum_{v_0\in\V}\mu_{v_0}|\theta_{v_0}|
 \sum_{v_1\in\V}\mu_{v_1}|\theta_{v_1}|
 \notag
 \\
 &=
 \lrB{\sum_{v_0\in\V}\mu_{v_0}|\theta_{v_0}|}
 \lrB{\sum_{v_1\in\V}\mu_{v_1}|\theta_{v_1}|}
 \notag
 \\
 &=
 M(\cP)^2.
\end{align}
Inequality (\ref{eq:V-+}) is shown as
\begin{align}
 M(\cP^-)
 +M(\cP^+)
 &=
 \sum_{v_0\in\V}\sum_{v_1\in\V}
 \mu_{v_0}\mu_{v_1}|\theta_{v_0}\theta_{v_1}|
 +
 \sum_{v_0\in\V}\sum_{v_1\in\V}
 \mu_{v_0}\mu_{v_1}\max\{|\theta_{v_0}|,|\theta_{v_1}|\}
 \notag
 \\
 &\leq
 \sum_{v_0\in\V}\sum_{v_1\in\V}
 \mu_{v_0}\mu_{v_1}\min\{|\theta_{v_0}|,|\theta_{v_1}|\}
 +
 \sum_{v_0\in\V}\sum_{v_1\in\V}
 \mu_{v_0}\mu_{v_1}\max\{|\theta_{v_0}|,|\theta_{v_1}|\}
 \notag
 \\
 &=
 \sum_{v_0\in\V}\sum_{v_1\in\V}
 \mu_{v_0}\mu_{v_1}\lrB{|\theta_{v_0}|+|\theta_{v_1}|}
 \notag
 \\
 &=
 \sum_{v_0\in\V}
 \mu_{v_0}|\theta_{v_0}|
 +
 \sum_{v_1\in\V}
 \mu_{v_1}|\theta_{v_1}|
 \notag
 \\
 &=
 2M(\cP),
\end{align}
where the inequality comes from relation
$|\theta_{v_0}\theta_{v_1}|
=|\theta_{v_0}||\theta_{v_1}|
\leq\min\{|\theta_{v_0}|,|\theta_{v_1}|\}$,
which comes from the fact that $\theta_{v_0},\theta_{v_1}\in[-1,1]$.
\hfill\qed

\subsection{Proof of Lemma~\ref{lem:HV}}
\label{sec:proof-HV}
The lemma is shown immediately from
(\ref{eq:Hbound}) and the
fact that $\Error(U|V)=[1-M(\cP(U|V))]/2$ is the MAP decision error probability 
of guessing $U$ from $V$.
\hfill\qed

\subsection{Proof of Lemma~\ref{lem:MPA}}
\label{sec:proof-MPA}
From the definition (\ref{eq:MP}) of $M$,
value $M(\cP)$ does not change when $\theta_v$ is replaced by $|\theta_v|$.
\hfill\qed

\subsection{Proof of Lemma~\ref{lem:ApmA}}
\label{sec:proof-ApmA}
Relation $\cP^{\sfA-\sfA}=\cP^{-\sfA}$ is shown as
\begin{align}
 \cP^{\sfA-\sfA}
 &=
 \lrb{
  (\mu_{v_0}\mu_{v_1},\lrbar{|\theta_{v_0}||\theta_{v_1}|}): (v_0,v_1)\in\V\times\V
 }
 \notag
 \\
 &=
 \lrb{
  (\mu_{v_0}\mu_{v_1},|\theta_{v_0}\theta_{v_1}|): (v_0,v_1)\in\V\times\V
 }
 \notag
 \\
 &=
 \cP^{-\sfA}.
\end{align}
Next we show relation $\cP^{\sfA+\sfA}=\cP^{+\sfA}$.
We have
\begin{align*}
 \cP^{\sfA+\sfA}
 &=
 \lrb{
  \lrsb{
   \frac{\mu_{v_0}\mu_{v_1}[1\mp_u|\theta_{v_0}||\theta_{v_1}|]}2,
   \lrbar{
    \frac{|\theta_{v_0}|\mp_u|\theta_{v_1}|}{1\mp_u|\theta_{v_0}||\theta_{v_1}|}
   }
  }
  :
  (u,v_0,v_1)\in\U\times\V\times\V
 }
 \\
 \cP^{+\sfA}
 &=
 \lrb{
  \lrsb{
   \frac{\mu_{v_0}\mu_{v_1}[1\mp_u\theta_{v_0}\theta_{v_1}]}2,
   \lrbar{
    \frac{\theta_{v_0}\mp_u\theta_{v_1}}{1\mp_u\theta_{v_0}\theta_{v_1}}
   }
  }
  :
  (u,v_0,v_1)\in\U\times\V\times\V
 },
\end{align*}
where $\mp_u$ is the bipolar-binary conversion of $u$.
When  $\theta_{v_0}\theta_{v_1}\geq 0$,
we have the fact that 
$\theta_{v_0},\theta_{v_1}\geq 0$ or $\theta_{v_0},\theta_{v_1}\leq 0$.
Then we have relation $\cP^{\sfA+\sfA}=\cP^{+\sfA}$
immediately from the fact that
\begin{gather*}
 1\mp_u|\theta_{v_0}||\theta_{v_1}|=1\mp_u\theta_{v_0}\theta_{v_1}\geq 0
 \\
 ||\theta_{v_0}|\mp_u|\theta_{v_1}||=|\theta_{v_0}\mp_u\theta_{v_1}|.
\end{gather*}
When $\theta_{v_0}\theta_{v_1}<0$,
we have the fact that
$\theta_{v_0}<0<\theta_{v_1}$ or $\theta_{v_1}<0<\theta_{v_0}$.
Then we have
\begin{gather*}
 1\mp_u|\theta_{v_0}||\theta_{v_1}|=1\pm_u\theta_{v_0}\theta_{v_1}\geq 0
 \\
 ||\theta_{v_1}|\mp_u|\theta_{v_0}||=|\theta_{v_1}\pm_u\theta_{v_0}|,
\end{gather*}
which implies that
\begin{align*}
 \lrsb{
  \frac{\mu_{v_0}\mu_{v_1}[1\mp_u|\theta_{v_0}||\theta_{v_1}|]}2,
  \lrbar{
   \frac{|\theta_{v_0}|\mp_u|\theta_{v_1}|}{1\mp_u|\theta_{v_0}||\theta_{v_1}|}
  }
 }
 &=
 \lrsb{
  \frac{\mu_{v_0}\mu_{v_1}[1\pm_u\theta_{v_0}\theta_{v_1}]}2,
  \lrbar{
   \frac{\theta_{v_0}\pm_u\theta_{v_1}}{1\pm_u\theta_{v_0}\theta_{v_1}}
  }
 }
\end{align*}
Then we have relation $\cP^{\sfA+\sfA}=\cP^{+\sfA}$ using a bijection:
\begin{equation*}
 \pi(u,v_0,v_1)
 \equiv
 \begin{cases}
  (u,v_0,v_1)
  &\text{if}\ \theta_{v_0}\theta_{v_1}\geq 0
  \\
  (u\oplus 1,v_0,v_1)
  &\text{if}\ \theta_{v_0}\theta_{v_1}<0.
 \end{cases}
\end{equation*}
\hfill\qed

\subsection{Proof of Lemma~\ref{lem:MPS}}
\label{sec:proof-MPS}
We have 
\begin{align}
 M(\cP)
 &=
 \sum_{\substack{
   v\in\V:
   \\
   \mu_v>0
 }}
 \mu_v|\theta_v|
 \notag
 \\
 &=
 \sum_{\theta\in\Theta}\sum_{v:\theta_v=\theta}\mu_v|\theta|
 \notag
 \\
 &=
 \sum_{\theta\in\Theta}
 \lrB{\sum_{v:\theta_v=\theta}\mu_v}|\theta|
 \notag
 \\
 &=
 M(\cP^{\sfSigma}).
\end{align}
\hfill\qed

\subsection{Proof of Lemma~\ref{lem:SpmS}}
\label{sec:proof-SpmS}
First, we show relation $\cP^{\sfSigma-\sfSigma}=\cP^{-\sfSigma}$.
Define $\mu_{\theta}$ and $\Theta^-$ as
\begin{align}
 \mu_{\theta}
 &\equiv\sum_{v:\theta_v=\theta}\mu_v.
 \label{eq:mu_theta}
 \\
 \Theta^-
 &\equiv
 \{\theta_{v_0}\theta_{v_1}:
  (v_0,v_1)\in\V\times\V
  \ \text{s.t.}\ \mu_{v_0}\mu_{v_1}>0
  \}
 \notag
 \\
 &=
 \lrb{\theta_0\theta_1:
  (\theta_0,\theta_1)\in\Theta\times\Theta
  \ \text{s.t.}\ \mu_{\theta_0}\mu_{\theta_1}>0
 },
 \label{eq:Theta-}
\end{align}
where $\Theta$ is defined by (\ref{eq:Theta}).
Then we have
\begin{align}
 \cP^{\sfSigma-\sfSigma}
 &=
 \lrb{
  \lrsb{
   \sum_{\substack{
     (\theta_0,\theta_1):
     \\
     \theta_0\theta_1=\theta
   }}
   \mu_{\theta_0}\mu_{\theta_1},
   \theta
  }:
  \theta\in\Theta^-
 }
 \notag
 \\
 &=
 \lrb{
  \lrsb{
   \sum_{\substack{(\theta_0,\theta_1):\\
     \theta_0\theta_1=\theta
   }}
   \sum_{\substack{
     (v_0,v_1):
     \\
     \theta_{v_0}=\theta_0
     \\
     \theta_{v_1}=\theta_1
   }}
   \mu_{v_0}\mu_{v_1},
   \theta
  }:
  \theta\in\Theta^-
 }
 \notag
 \\
 &=
 \lrb{
  \lrsb{
   \sum_{\substack{
     (v_0,v_1):
     \\
     \theta_{v_0}\theta_{v_1}=\theta
   }}
   \mu_{v_0}\mu_{v_1},
   \theta
  }: 
  \theta\in\Theta^-
 }
 \notag
 \\
 &=
 \cP^{-\sfSigma},
\end{align}
where the second equality comes from (\ref{eq:mu_theta}) and
the third equality comes from Lemma~\ref{lem:gth},
which appears in Appendix,
by letting $\vv\equiv(v_0,v_1)$, $\ttheta\equiv(\theta_0,\theta_1)$,
and $g(\ttheta)\equiv\theta_0\theta_1$,
and the last equality comes from (\ref{eq:Theta-}).

Next we show relation $\cP^{\sfSigma+\sfSigma}=\cP^{+\sfSigma}$.
Let
\begin{align}
 \Theta^+
 &\equiv
 \lrb{\frac{\theta_{v_0}\mp_u\theta_{v_1}}{1\mp_u\theta_{v_0}\theta_{v_1}}:
  u\in\{0,1\}, (v_0,v_1)\in\V\times\V
  \ \text{s.t.}
  \ \frac{\mu_{\theta_{v_0}}\mu_{\theta_{v_1}}[1\mp_u\theta_{v_0}\theta_{v_1}]}2
  >0
 }
 \notag
 \\
 &=
 \lrb{\frac{\theta_0\mp_u\theta_1}{1\mp_u\theta_0\theta_1}:
  u\in\{0,1\}, \theta_0,\theta_1\in\Theta
  \ \text{s.t.}
  \ \frac{\mu_{\theta_0}\mu_{\theta_1}[1\mp_u\theta_0\theta_1]}2
  >0
 },
 \label{eq:Theta+}
\end{align}
where $\mp_u$ is the bipolar-binary conversion of $u$.
Then we have
\begin{align}
 \cP^{\sfSigma+\sfSigma}
 &=
 \lrb{
  \lrsb{
   \sum_{\substack{
     (u,\theta_0,\theta_1):
     \\
     \frac{\theta_1\mp_u\theta_0}{1\mp_u\theta_0\theta_1}=\theta
   }}
   \frac{\mu_{\theta_0}\mu_{\theta_1}[1\mp_u\theta_0\theta_1]}2,
   \theta
  }
  :
  \theta\in\Theta^+
 }
 \notag
 \\
 &=
 \lrb{
  \lrsb{
   \sum_{\substack{
     (u,\theta_0,\theta_1):
     \\
     \frac{\theta_1\mp_u\theta_0}{1\mp_u\theta_0\theta_1}=\theta
   }}
   \sum_{\substack{
     (v_0,v_1):
     \\
     \theta_{v_0}=\theta_0
     \\
     \theta_{v_1}=\theta_1
   }}
   \frac{\mu_{\theta_{v_0}}\mu_{\theta_{v_1}}[1\mp_u\theta_{v_0}\theta_{v_1}]}2,
   \theta
  }:
  \theta\in\Theta^+
 }
 \notag
 \\
 &=
 \lrb{
  \lrsb{
   \sum_{\substack{
     (u,v_0,v_1):
     \\
     \frac{\theta_{v_1}\mp_u\theta_{v_0}}{1\mp_u\theta_{v_0}\theta_{v_1}}=\theta
   }}
   \frac{\mu_{v_0}\mu_{v_1}[1\mp_u\theta_{v_0}\theta_{v_1}]}2,
   \theta
  }:
  \theta\in\Theta^+
 }
 \notag
 \\
 &=
 \cP^{+\sfSigma},
\end{align}
where the third equality comes from Lemma~\ref{lem:gth},
which appears in Appendix,
by letting $\vv\equiv(v_0,v_1)$, $\ttheta\equiv(\theta_0,\theta_1)$,
and
\begin{equation*}
 g(\ttheta)\equiv\frac{\theta_1\mp_u\theta_0}{1\mp_u\theta_0\theta_1}
\end{equation*}
for each $u\in\{0,1\}$,
and the last equality comes from (\ref{eq:Theta+}).
\hfill\qed

\subsection{Proof of Lemma~\ref{lem:SAS}}
\label{sec:proof-SAS}
Let
\begin{align}
 \Theta^{\sfA}
 &\equiv
 \{
  |\theta_v|: v\in\V
  \ \text{s.t.}\ \mu_v>0
  \}
 \notag
 \\
 &=
 \lrb{
  |\theta|: \theta\in\Theta
  \ \text{s.t.}\ \mu_{\theta}>0
 },
 \label{eq:ThetaA}
\end{align}
where $\Theta$ is defined by (\ref{eq:Theta})
and $\mu_{\theta}$ is defined by (\ref{eq:mu_theta}).
Then we have
\begin{align}
 \cP^{\sfSigma\sfA\sfSigma}
 &=
 \lrb{
  \lrsb{
   \sum_{\theta':|\theta'|=\theta}\mu_{\theta'},
   \theta
  }: \theta\in\Theta^{\sfA}
 }
 \notag
 \\
 &=
 \lrb{
  \lrsb{
   \sum_{\theta':|\theta'|=\theta}\sum_{v:\theta_v=\theta'}\mu_v,
   \theta
  }: \theta\in\Theta^{\sfA}
 }
 \notag
 \\
 &=
 \lrb{
  \lrsb{
   \sum_{v:|\theta_v|=\theta}\mu_v,
   \theta
  }:
  \theta\in\Theta^{\sfA}
 }
 \notag
 \\
 &=
 \cP^{\sfA\sfSigma},
\end{align}
where the third equality comes from Lemma~\ref{lem:gth} in Appendix
by letting $\vv\equiv v$, $\ttheta\equiv\theta'$,
and $g(\ttheta)\equiv|\theta'|$,
and the last equality comes from (\ref{eq:ThetaA}).
\hfill\qed

\subsection{Proof of Theorem~\ref{thm:canonical}}
\label{sec:proof-canonical}
Equation (\ref{eq:MPC}) is shown as
\begin{align}
 M(\cP^{\sfC})
 &=
 M(\cP^{\sfA\sfSigma})
 \notag
 \\
 &=
 M(\cP^{\sfA})
 \notag
 \\
 &=
 M(\cP),
\end{align}
where the second equality comes from Lemma~\ref{lem:MPS} and
the third equality comes from Lemma~\ref{lem:MPA}.
Eq. (\ref{eq:CpmC}) is shown as
\begin{align}
 \cP^{\sfC\pm\sfC}
 &=
 \cP^{\sfA\sfSigma\pm\sfA\sfSigma}
 \notag
 \\
 &=
 \cP^{\sfA\sfSigma\pm\sfSigma\sfA\sfSigma}
 \notag
 \\
 &=
 \cP^{\sfA\pm\sfSigma\sfA\sfSigma}
 \notag
 \\
 &=
 \cP^{\sfA\pm\sfA\sfSigma}
 \notag
 \\
 &=
 \cP^{\pm\sfA\sfSigma}
 \notag
 \\
 &=
 \cP^{\pm\sfC},
\end{align}
where the second equality comes from Lemma~\ref{lem:SAS},
the third comes from Lemma~\ref{lem:SpmS},
the fourth comes from Lemma~\ref{lem:SAS},
and the fifth comes from Lemma~\ref{lem:ApmA}.
Eq. (\ref{eq:MPCpm}) is shown as
\begin{align}
 M(\cP^{\sfC\pm})
 &=
 M(\cP^{\sfC\pm\sfA})
 \notag
 \\
 &=
 M(\cP^{\sfC\pm\sfA\sfSigma})
 \notag
 \\
 &=
 M(\cP^{\sfC\pm\sfC})
 \notag
 \\
 &=
 M(\cP^{\pm\sfC})
 \notag
 \\
 &=
 M(\cP^{\pm\sfA\sfSigma})
 \notag
 \\
 &=
 M(\cP^{\pm\sfA})
 \notag
 \\
 &=
 M(\cP^{\pm}),
\end{align}
where the first equality comes from Lemma~\ref{lem:MPA},
the second comes from Lemma~\ref{lem:MPS}
the fourth comes from (\ref{eq:CpmC}),
the sixth comes from Lemma~\ref{lem:MPS},
and the last one comes from Lemma~\ref{lem:MPA}.
\hfill\qed

\subsection{Proof of Lemma~\ref{lem:laststep+}}
\label{sec:proof-laststep+}
We have
\begin{align}
 M(\cP^+)
 &=
 M(\cP^{\sfC+})
 \notag
 \\
 &=
 \sum_{v_0=1}^{C}\sum_{v_1=1}^{C}
 \mu_{v_0}\mu_{v_1}\max\{\theta_{v_0},\theta_{v_1}\}
 \notag
 \\
 &=
 \sum_{v_0=1}^{C}\sum_{v_1=1}^{v_0-1}\mu_{v_0}\mu_{v_1}\theta_{v_0}
 +
 \sum_{v_0=1}^{C}\mu_{v_0}\mu_{v_0}\theta_{v_0}
 +
 \sum_{v_0=1}^{C}\sum_{v_1=v_0+1}^{C}\mu_{v_0}\mu_{v_1}\theta_{v_1}
 \notag
 \\
 &=
 \sum_{v_0=1}^{C}\sum_{v_1=1}^{v_0-1}\mu_{v_0}\mu_{v_1}\theta_{v_0}
 +
 \sum_{v_0=1}^{C}\mu_{v_0}\mu_{v_0}\theta_{v_0}
 +
 \sum_{v_1=1}^{C}\sum_{v_0=1}^{v_1-1}\mu_{v_0}\mu_{v_1}\theta_{v_1}
 \notag
 \\
 &=
 2\sum_{v=1}^{C}\sum_{v'=1}^{v-1}\mu_v\mu_{v'}\theta_v
 +
 \sum_{v=1}^{C}\mu_v\mu_v\theta_v
 \notag
 \\
 &=
 \sum_{v=1}^{C}\mu_v\theta_v\lrB{\mu_v+2\sum_{v'=1}^{v-1}\mu_{v'}},
\end{align}
where the first equality comes from (\ref{eq:MPCpm}),
the second equality comes from (\ref{eq:MP+}),
and parameters are renamed in the fifth equality.
\hfill\qed

\subsection{Proof of Lemma~\ref{lem:quantization}}
\label{sec:proof-quantization}
Equation (\ref{eq:invariance}) is shown as
\begin{align}
 M(\hcP)
 &=
 \sum_{\hv\in\hcV}
 \lrB{
  \sum_{v\in q^{-1}(\hv)}\mu_v}
 \lrbar{
  \frac{\sum_{v\in q^{-1}(\hv)}\mu_v\theta_v}
  {\sum_{v\in q^{-1}(\hv)}\mu_v}
 }
 \notag
 \\
 &=
 \sum_{\hv\in\hcV}
 \sum_{v\in q^{-1}(\hv)}\mu_v|\theta_v|
 \notag
 \\
 &=
 \sum_{v\in\V}\mu_v|\theta_v|
 \notag
 \\
 &=
 M(\cP),
\end{align}
where the first equality comes from (\ref{eq:PV'}),
the second comes from the 
fact that $\mu_v>0$ and $\theta_v\geq 0$ for all $v\in\V$,
and the third comes from the fact that
$\{q^{-1}(\hv)\}_{\hv\in\hcV}$ forms a partition of $\V$.
Eq. (\ref{eq:uQinvariance}) is immediately shown from (\ref{eq:invariance}).
Eq. (\ref{eq:oQinvariance}) is shown as
\begin{align}
 M(\hcP^{\osfQ})
 &=
 M(\hcP^{\sfE\sfSigma})
 \notag
 \\
 &=
 M(\hcP^{\sfE})
 \notag
 \\
 &=
 M(\cP)
 \notag
 \\
 &=
 M(\hcP),
\end{align}
where the second equality comes from Lemma~\ref{lem:MPS}
and the last equality comes from the first equality.
\hfill\qed

\subsection{Proof of Lemma~\ref{lem:MP-k+}}
\label{sec:proof-MP-k+}

First, we show (\ref{eq:MP-k}) as
\begin{gather*}
 M(\cP^{-^k})
 =M(\cP^{-^{k-1}})^2
 =[M(\cP^{-^{k-2}})^2]^2
 =
 \cdots
 =
 M(\cP)^{2^k}
\end{gather*}
by applying (\ref{eq:V-}).

Next we show the left hand side inequality of (\ref{eq:MP-+}).
Since $g(\xi)\equiv\max\{\alpha\xi,\beta\}=\max\{\beta,\alpha\xi\}$
is a convex function of $\xi$, we have
\begin{align}
 M(\cP^{-+})
 &=
 \sum_{v_{00},v_{01},v_{10},v_{11}}
 \mu_{v_{00}}\mu_{v_{01}}\mu_{v_{10}}\mu_{v_{11}}
 \max\{
  |\theta_{v_{00}}||\theta_{v_{01}}|,
  |\theta_{v_{10}}||\theta_{v_{11}}|
  \}
 \notag
 \\
 &\geq
 \sum_{v_{00},v_{10}}
 \mu_{v_{00}}\mu_{v_{10}}
 \max\lrb{
  |\theta_{v_{00}}|
  \sum_{v_{01}}\mu_{v_{01}}|\theta_{v_{01}}|,
  |\theta_{v_{10}}|
  \sum_{v_{11}}\mu_{v_{11}}|\theta_{v_{11}}|
 }
 \notag
 \\
 &=
 \sum_{v_{00},v_{10}}
 \mu_{v_{00}}\mu_{v_{10}}
 \max\lrb{
  |\theta_{v_{00}}|
  M(\cP),
  |\theta_{v_{10}}|
  M(\cP)
 }
 \notag
 \\
 &=
 M(\cP)
 \sum_{v_{00},v_{10}}
 \mu_{v_{00}}\mu_{v_{10}}
 \max\lrb{
  |\theta_{v_{00}}|,
  |\theta_{v_{10}}|
 }
 \notag
 \\
 &=
 M(\cP)M(\cP^+),
\end{align}
where the first and third equalities come from the assumption 
(\ref{eq:canonical<}),
and the Jensen inequality is applied twice in the inequality.

Next we show the right hand side inequality of (\ref{eq:MP-+}).
We have
\begin{align}
 M(\cP^{-+})
 &=
 \sum_{v_{00},v_{01},v_{10},v_{11}}\mu_{v_{00}}\mu_{v_{01}}\mu_{v_{10}}\mu_{v_{11}}
 \max\{|\theta_{v_{00}}\theta_{v_{01}}|,|\theta_{v_{10}}\theta_{v_{11}}|\}
 \notag
 \\
 &\leq
 \sum_{v_{00},v_{01},v_{10},v_{11}}\mu_{v_{00}}\mu_{v_{01}}\mu_{v_{10}}\mu_{v_{11}}
 \max\{|\theta_{v_{00}}|,|\theta_{v_{01}}|\}\max\{|\theta_{v_{10}}|,|\theta_{v_{11}}|\}
 \notag
 \\
 &=
 \sum_{v_{00},v_{10}}
 \mu_{v_{00}}
 \mu_{v_{10}}
 \max\{|\theta_{v_{00}}|,|\theta_{v_{10}}|\}
 \sum_{v_{01},v_{11}}
 \mu_{v_{01}}
 \mu_{v_{11}}
 \max\{|\theta_{v_{01}}|,|\theta_{v_{11}}|\}
 \notag
 \\
 &=
 \lrB{\sum_{v_{00},v_{10}}
  \mu_{v_{00}}
  \mu_{v_{10}}
  \max\{|\theta_{v_{00}}|,|\theta_{v_{10}}|\}}
 \lrB{
  \sum_{v_{01},v_{11}}
  \mu_{v_{01}}
  \mu_{v_{11}}
  \max\{|\theta_{v_{01}}|,|\theta_{v_{11}}|\}
 }
 \notag
 \\
 &=
 M(\cP^+)^2
 \notag
 \\
 &=
 M(\cP^{+-}),
\end{align}
where the first equality comes from Lemma~\ref{lem:MP+-}
and the last equality comes from (\ref{eq:V-}).

Next we show the left hand side inequality of (\ref{eq:MP-k+}).
By applying (\ref{eq:MP-+}), we have
\begin{align}
 M(\cP^{-^k+})
 &\geq
 M(\cP^{-^{k-1}})
 M(\cP^{-^{k-1}+})
 \notag
 \\
 &\geq
 M(\cP^{-^{k-1}})
 M(\cP^{-^{k-2}})
 M(\cP^{-^{k-2}+})
 \notag
 \\
 &\quad
 \vdots
 \notag
 \\
 &\geq
 M(\cP^{-^{k-1}})
 \cdots
 M(\cP)
 M(\cP^+).
\end{align}
From (\ref{eq:MP-k}), we have
\begin{align}
 M(\cP^{-^k+})
 &\geq
 \lrB{\prod_{i=0}^{k-1}M(\cP^{-^{i-1}})}
 M(\cP^+)
 \notag
 \\
 &=
 \lrB{\prod_{i=0}^{k-1}M(\cP)^{2^i}}
 M(\cP^+)
 \notag
 \\
 &=
 M(\cP)^{\sum_{i=0}^{k-1}2^i}
 M(\cP^+)
 \notag
 \\
 &=
 M(\cP)^{2^k-1}
 M(\cP^+).
\end{align}

Finally, we show the right hand side inequality of (\ref{eq:MP-k+}).
By applying (\ref{eq:MP-+}), we have
\begin{align}
 M(\cP^{-^k+})
 &\leq
 M(\cP^{-^{k-1}+-})
 \notag
 \\
 &\leq
 M(\cP^{-^{k-2}+-^2})
 \notag
 \\
 &\quad
 \vdots
 \notag
 \\
 &\leq
 M(\cP^{+-^k})
 \notag
 \\
 &=
 M(\cP^+)^{2^k},
\end{align}
where the equality comes from (\ref{eq:MP-k}).
\hfill\qed

\subsection{Proof of Theorem~\ref{thm:degrade}}
\label{sec:proof-degrade}
Let $\cP_i\equiv\{(v,\theta_v)\}_{v=1}^i$.
Similarly to the proof in~\cite{PHTT11}, we have
\begin{align}
 M(\cP_i^+)-M(\cP_{i-1}^+)
 &=
 \min_{v\in\{1,\ldots,i-1\}}\mu_v\mu_{v+1}\lrB{\theta_{v+1}-\theta_v}
 \notag
 \\
 &=
 \lrB{\sqrt[3]{
   \min_{v\in\{1,\ldots,i-1\}}\mu_v\mu_{v+1}\lrB{\theta_{v+1}-\theta_v}
 }}^3
 \notag
 \\
 &=
 \lrB{\min_{v\in\{1,\ldots,i-1\}}
  \sqrt[3]{\mu_v\mu_{v+1}\lrB{\theta_{v+1}-\theta_v}}
 }^3
 \notag
 \\
 &\leq
 \lrB{\frac1{i-1}\sum_{v=1}^{i-1}
  \sqrt[3]{\mu_v\mu_{v+1}\lrB{\theta_{v+1}-\theta_v}}
 }^3
 \notag
 \\
 &\leq
 \frac1{[i-1]^3}
 \lrB{\sum_{v=1}^{i-1}\mu_v}
 \lrB{\sum_{v=1}^{i-1}\mu_{v+1}}
 \lrB{\sum_{v=1}^{i-1}\lrB{\theta_{v+1}-\theta_v}}
 \notag
 \\
 &\leq
 \frac{\theta_i-\theta_1}
 {[i-1]^3},
 \notag
 \\
 &\leq
 \frac{\otheta(\cP)}
 {[i-1]^3}.
 \label{eq:mindiff-d}
\end{align}
where the first inequality comes from the fact that
the minimum is smaller than the average,
and the second inequality comes from the H\"older inequality.

The inequalities (\ref{eq:abs-M-d}) and (\ref{eq:abs-error-d}) are shown
from (\ref{eq:d}) and the fact that
\begin{align}
 M(\cP^+)-M(\cP^{\usfQ+})
 &=
 \sum_{i=Q+1}^{C}
 \lrB{M(\cP_i^+)-M(\cP_{i-1}^+)}
 \notag
 \\
 &\leq
 \sum_{i=Q+1}^{C}
 \frac{\otheta(\cP)}
 {[i-1]^3}
 \notag
 \\
 &\leq
 \otheta(\cP)
 \lrB{\frac1{Q^2}-\frac1{2C^2}}
 \notag
 \\
 &=
 \frac{\otheta(\cP)}
 {Q^2}
 \lrB{1-\frac{Q^2}{2C^2}},
 \label{eq:VS-VQ-d}
\end{align}
where the first inequality comes from (\ref{eq:mindiff-d}).
The left hand side inequality of (\ref{eq:rel-M-d}) is shown as
\begin{align}
 M(\cP^{\usfQ+})
 &\geq
 M(\cP^+)
 -\frac{\otheta(\cP)}{Q^2}\lrB{1-\frac{Q^2}{2C^2}}
 \notag
 \\
 &=
 M(\cP^+)\lrB{1-\frac{\otheta(\cP)}{M(\cP^+)Q^2}\lrB{1-\frac{Q^2}{2C^2}}}
 \notag
 \\
 &\geq
 M(\cP^+)\lrB{1-\frac{\otheta(\cP)}{M(\cP)Q^2}\lrB{1-\frac{Q^2}{2C^2}}},
\end{align}
where the first inequality comes from (\ref{eq:VS-VQ-d})
and the second inequality comes from (\ref{eq:-V+}).
The right hand side inequality of (\ref{eq:rel-error-d}) is shown 
from the fact that
\begin{align}
 1-M(\cP^{\usfQ+})
 &\leq
 \lrB{1-M(\cP^+)}
 \lrB{
  1
  +
  \frac{\otheta(\cP)}{[1-M(\cP^+)]Q^2}
  \lrB{1-\frac{Q^2}{2C^2}}
 }
 \notag
 \\
 &\leq
 \lrB{1-M(\cP^+)}
 \lrB{
  1
  +
  \frac{\otheta(\cP)}{[1-\otheta(\cP)]Q^2}
  \lrB{1-\frac{Q^2}{2C^2}}
 },
 \label{eq:EPQupper-d}
\end{align}
where the first inequality comes from (\ref{eq:VS-VQ-d})
and the last inequality comes from the relation
$M(\cP^+)\leq \otheta(\cP)$ that comes from (\ref{eq:MP+}).
\hfill\qed

\subsection{Proof of Theorem~\ref{thm:degrade1}}
\label{sec:proof-degrade1}
Let $\cP_i\equiv\{(v,\theta_v)\}_{v=1}^i$.
Similarly to (\ref{eq:mindiff-d}), we have
\begin{align}
 M(\cP_i^+)-M(\cP_{i-1}^+)
 &=
 \min_{v\in\{1,\ldots,i-1\}}\mu_v\mu_{v+1}\lrB{\theta_{v+1}-\theta_v}
 \notag
 \\
 &\leq
 \min_{v\in\{1,\ldots,i-1\}}\mu_v\mu_{v+1}\theta_{v+1}\lrB{1-\theta_v}
 \notag
 \\
 &=
 \lrB{\min_{v\in\{1,\ldots,i-1\}}
  \sqrt{
   \mu_{v+1}\theta_{v+1}\mu_v\lrB{1-\theta_v}
  }
 }^2
 \notag
 \\
 &\leq
 \lrB{\frac1{i-1}\sum_{v=1}^{i-1}
  \sqrt{
   \mu_{v+1}\theta_{v+1}\mu_v\lrB{1-\theta_v}
  }
 }^2
 \notag
 \\
 &\leq
 \frac1{[i-1]^2}
 \lrB{\sum_{v=1}^{i-1}\mu_{v+1}\theta_{v+1}}
 \lrB{\sum_{v=1}^{i-1}\mu_v[1-\theta_v]}
 \notag
 \\
 &\leq
 \frac{M(\cP)[1-M(\cP)]}
 {[i-1]^2},
 \label{eq:mindiff-d-1}
\end{align}
where the first inequality comes from the relation $\theta_{v+1}\leq 1$.

The inequalities (\ref{eq:abs-M-d-1}) and (\ref{eq:abs-error-d-1})
are shown from (\ref{eq:d}) and the fact that
\begin{align}
 M(\cP^+)-M(\cP^{\usfQ+})
 &=
 \sum_{i=Q+1}^{C}
 \lrB{M(\cP_i^+)-M(\cP_{i-1}^+)}
 \notag
 \\
 &\leq
 \sum_{i=Q+1}^{C}
 \frac{M(\cP)[1-M(\cP)]}
 {[i-1]^2}
 \notag
 \\
 &\leq
 M(\cP)[1-M(\cP)]
 \lrB{\frac1{Q-1}-\frac1{C-1}}
 \notag
 \\
 &=
 \frac{M(\cP)[1-M(\cP)]}{Q-1}
 \lrB{1-\frac{Q-1}{C-1}},
 \label{eq:VS-VQ-d-1}
\end{align}
where the first inequality comes from (\ref{eq:mindiff-d-1}).
The left hand side inequality of (\ref{eq:rel-M-d-1}) is shown as
\begin{align}
 M(\cP^{\usfQ+})
 &\geq
 M(\cP^+)
 -
 \frac{M(\cP)[1-M(\cP)]}{Q-1}
 \lrB{1-\frac{Q-1}{C-1}}
 \notag
 \\
 &=
 M(\cP^+)\lrB{
  1
  -\frac{M(\cP)[1-M(\cP)]}{M(\cP^+)[Q-1]}
  \lrB{1-\frac{Q-1}{C-1}}
 }
 \notag
 \\
 &\geq
 M(\cP^+)\lrB{
  1
  -\frac{1-M(\cP)}{Q-1}
  \lrB{1-\frac{Q-1}{C-1}}
 },
\end{align}
where the first inequality comes from (\ref{eq:VS-VQ-d-1})
and the second inequality comes from (\ref{eq:-V+}).
The right hand side inequality of (\ref{eq:rel-error-d-1}) is shown as
\begin{align}
 1- M(\cP^{\usfQ+})
 &\leq
 1 - M(\cP^+)
 +
 \frac{M(\cP)[1-M(\cP)]}{Q-1}
 \lrB{1-\frac{Q-1}{C-1}}
 \notag
 \\
 &=
 [1-M(\cP^+)]\lrB{
  1
  +\frac{M(\cP)[1-M(\cP)]}{[1-M(\cP^+)][Q-1]}
  \lrB{1-\frac{Q-1}{C-1}}
 }
\end{align}
where the first inequality comes from (\ref{eq:VS-VQ-d-1})
and the second inequality comes from the relation $M(\cP^+)\leq \otheta(\cP)$
that comes from (\ref{eq:MP+}).
\hfill\qed

\subsection{Proof of Theorem~\ref{thm:upgrade}}
\label{sec:proof-upgrade}
Let $\cP_i\equiv\{(v,\theta_v)\}_{v=1}^i$.
Similarly to the proof in~\cite{PHTT11}, we have
\begin{align}
 M(\cP^+_{i-1})-M(\cP^+_i)
 &=
 \min_{v\in\{2,\ldots,i-1\}}\mu'_v\mu'_{v+1}\lrB{\theta_{v+1}-\theta_{v-1}}
 \notag
 \\
 &=
 \min_{v\in\{2,\ldots,i-1\}}
 \frac{
  \mu_v^2\lrB{\theta_v-\theta_{v-1}}
  \lrB{\theta_{v+1}-\theta_v}
 }{\theta_{v+1}-\theta_{v-1}}
 \notag
 \\
 &\leq
 \min_{v\in\{2,\ldots,i-1\}}
 \mu_v^2\lrB{\theta_v-\theta_{v-1}}
 \notag
 \\
 &=
 \lrB{\sqrt[3]{\min_{v\in\{2,\ldots,i-1\}}\mu_v^2\lrB{\theta_v-\theta_{v-1}}}}^3
 \notag
 \\
 &=
 \lrB{\min_{v\in\{2,\ldots,i-1\}}
  \sqrt[3]{\mu_v^2\lrB{\theta_v-\theta_{v-1}}}
 }^3
 \notag
 \\
 &\leq
 \lrB{\frac1{i-2}\sum_{v=2}^{i-1}
  \sqrt[3]{\mu_v^2\lrB{\theta_v-\theta_{v-1}}}
 }^3
 \notag
 \\
 &\leq
 \frac1{[i-2]^3}
 \lrB{\sum_{v=2}^{i-1}\mu_v}
 \lrB{\sum_{v=2}^{i-1}\mu_v}
 \lrB{\sum_{v=2}^{i-1}\lrB{\theta_v-\theta_{v-1}}}
 \notag
 \\
 &=
 \frac{\theta_{i-1}-\theta_1}
 {[i-2]^3}
 \notag
 \\
 &\leq
 \frac{\otheta(\cP)}
 {[i-2]^3}.
 \label{eq:mindiff-u}
\end{align}
where the first equality comes from (\ref{eq:Mdiff-u}),
the second equality comes from (\ref{eq:mu'-1}) and (\ref{eq:mu'+1}),
the first inequality comes from the fact that
$\theta_v$ is an increasing function of $v$,
the second inequality comes from the fact that
the minimum is smaller than the average,
and the third inequality comes from the H\"older inequality.

The inequalities (\ref{eq:abs-M-u}) and (\ref{eq:abs-error-u}) are shown
from (\ref{eq:d}) and the fact that
\begin{align}
 M(\cP^{\osfQ+})-M(\cP^+)
 &=
 \sum_{i=Q+1}^C
 \lrB{M(\cP_{i-1}^+)-M(\cP_i^+)}
 \notag
 \\
 &\leq
 \sum_{i=Q+1}^C
 \frac{\otheta(\cP)}
 {[i-2]^3}
 \notag
 \\
 &\leq
 \otheta(\cP)
 \lrB{\frac1{[Q-1]^2}-\frac1{2[C-1]^2}}
 \notag
 \\
 &=
 \frac{\otheta(\cP)}
 {[Q-1]^2}
 \lrB{1-\frac{[Q-1]^2}{2[C-1]^2}},
 \label{eq:VS-VQ-u}
\end{align}
where the first inequality comes from (\ref{eq:mindiff-u}).
The right hand side inequality of (\ref{eq:rel-M-u}) are shown as
\begin{align}
 M(\cP^{\osfQ+})
 &\leq
 M(\cP^+)\lrB{
  1
  +\frac{\otheta(\cP)}{M(\cP^+)[Q-1]^2}
  \lrB{1-\frac{[Q-1]^2}{2[C-1]^2}}
 }
 \notag
 \\
 &\leq
 M(\cP^+)\lrB{
  1
  +\frac{\otheta(\cP)}{M(\cP)[Q-1]^2}
  \lrB{1-\frac{[Q-1]^2}{2[C-1]^2}}
 },
\end{align}
where the first inequality comes from (\ref{eq:VS-VQ-u}) and
the second inequality comes from (\ref{eq:-V+}).
The left hand side inequality of (\ref{eq:rel-error-u}) is shown 
from the fact that
\begin{align}
 1-M(\cP^{\osfQ+})
 &\geq
 \lrB{1-M(\cP^+)}\lrB{
  1
  -\frac{\otheta(\cP)}{[1-M(\cP^+)][Q-1]^2}
  \lrB{1-\frac{[Q-1]^2}{2[C-1]^2}}
 }
 \notag
 \\
 &\geq
 \lrB{1-M(\cP^+)}\lrB{
  1
  -\frac{\otheta(\cP)}{[1-\otheta(\cP)][Q-1]^2}
  \lrB{1-\frac{[Q-1]^2}{2[C-1]^2}}
 },
\end{align}
where the first inequality comes from (\ref{eq:VS-VQ-u})
and the last inequality comes from the relation $M(\cP^+)\leq \otheta(\cP)$
that comes from (\ref{eq:MP+}).
\hfill\qed

\subsection{Proof of Theorem~\ref{thm:upgrade1}}
\label{sec:proof-upgrade1}
Let $\cP_i\equiv\{(v,\theta_v)\}_{v=1}^i$.
Similarly to (\ref{eq:mindiff-u}), we have
\begin{align}
 M(\cP^+_{i-1})-M(\cP^+_i)
 &=
 \min_{v\in\{2,\ldots,i-1\}}
 \frac{
  \mu_v^2\lrB{\theta_v-\theta_{v-1}}
  \lrB{\theta_{v+1}-\theta_v}
 }{\theta_{v+1}-\theta_{v-1}}
 \notag
 \\
 &\leq
 \min_{v\in\{2,\ldots,i-1\}}
 \mu_v^2
 [\theta_v-\theta_{v-1}]
 \notag
 \\
 &\leq
 \min_{v\in\{2,\ldots,i-1\}}
 \mu_v^2
 \theta_v
 \notag
 \\
 &=
 \lrB{\sqrt{\min_{v\in\{2,\ldots,i-1\}}\mu_v^2\theta_v}}^2
 \notag
 \\
 &=
 \lrB{\min_{v\in\{2,\ldots,i-1\}}
  \sqrt{\mu_v^2\theta_v}
 }^2
 \notag
 \\
 &\leq
 \lrB{\frac1{i-2}\sum_{v=2}^{i-1}
  \sqrt{\mu_v^2\theta_v}
 }^2
 \notag
 \\
 &\leq
 \frac1{[i-2]^2}
 \lrB{\sum_{v=2}^{i-1}\mu_v}
 \lrB{\sum_{v=2}^{i-1}\mu_v\theta_v}
 \notag
 \\
 &=
 \frac{M(\cP)}
 {[i-2]^2}
 \label{eq:mindiff-u-M}
 \\
 M(\cP^+_{i-1})-M(\cP^+_i)
 &=
 \min_{v\in\{2,\ldots,i-1\}}
 \frac{
  \mu_v^2\lrB{\theta_v-\theta_{v-1}}
  \lrB{\theta_{v+1}-\theta_v}
 }{\theta_{v+1}-\theta_{v-1}}
 \notag
 \\
 &\leq
 \min_{v\in\{2,\ldots,i-1\}}
 \mu_v^2
 [\theta_{v+1}-\theta_v]
 \notag
 \\
 &\leq
 \min_{v\in\{2,\ldots,i-1\}}
 \mu_v^2
 [1-\theta_v]
 \notag
 \\
 &=
 \lrB{\min_{v\in\{2,\ldots,i-1\}}
  \sqrt{\mu_v^2[1-\theta_v]}
 }^2
 \notag
 \\
 &\leq
 \lrB{\frac1{i-2}\sum_{v=2}^{i-1}
  \sqrt{\mu_v^2[1-\theta_v]}
 }^2
 \notag
 \\
 &\leq
 \frac1{[i-2]^2}
 \lrB{\sum_{v=2}^{i-1}\mu_v}
 \lrB{\sum_{v=2}^{i-1}\mu_v[1-\theta_v]}
 \notag
 \\
 &=
 \frac{1-M(\cP)}
 {[i-2]^2}.
 \label{eq:mindiff-u-1-M}
\end{align}

The inequalities (\ref{eq:abs-M-u-1}) and (\ref{eq:abs-error-u-1})
are shown from (\ref{eq:d}) and the fact that
\begin{align}
 M(\cP^{\osfQ+})-M(\cP^+)
 &=
 \sum_{i=Q+1}^C
 \lrB{M(\cP_{i-1}^+)-M(\cP_i^+)}
 \notag
 \\
 &\leq
 \sum_{i=Q+1}^C
 \frac{\min\lrb{M(\cP),1-M(\cP)}}
 {[i-2]^2}
 \notag
 \\
 &\leq
 \min\lrb{M(\cP),1-M(\cP)}
 \lrB{\frac1{Q-2}-\frac1{C-2}}
 \notag
 \\
 &=
 \frac{\min\lrb{M(\cP),1-M(\cP)}}{Q-2}
 \lrB{1-\frac{Q-2}{C-2}},
 \label{eq:VS-VQ-u-1}
\end{align}
where the equality comes from (\ref{eq:Mdiff-u})
and the first inequality comes from
(\ref{eq:mindiff-u-M}) and (\ref{eq:mindiff-u-1-M}).
The right hand side inequality of (\ref{eq:rel-M-u-1}) are shown
from the fact that
\begin{align}
 M(\cP^{\osfQ+})
 &\leq
 M(\cP^+)\lrB{
  1
  +\frac{M(\cP)}{M(\cP^+)[Q-2]}
  \lrB{1-\frac{Q-2}{C-2}}
 }
 \notag
 \\
 &\leq
 M(\cP^+)\lrB{
  1
  +\frac{1}{Q-2}
  \lrB{1-\frac{Q-2}{C-2}}
 },
\end{align}
where the first inequality comes from (\ref{eq:VS-VQ-u-1}) and
the second inequality comes from (\ref{eq:-V+}).
The left hand side inequality of (\ref{eq:rel-error-u-1}) are shown
from the fact that
\begin{align}
 1- M(\cP^{\osfQ+})
 &\geq
 [1-M(\cP^+)]\lrB{
  1
  -\frac{1-M(\cP)}{[1-M(\cP^+)][Q-2]}
  \lrB{1-\frac{Q-2}{C-2}}
 }
 \notag
 \\
 &\geq
 [1-M(\cP^+)]\lrB{
  1
  -\frac{1-M(\cP)}{[1-\otheta(\cP)][Q-2]}
  \lrB{1-\frac{Q-2}{C-2}}
 },
\end{align}
where the first inequality comes from (\ref{eq:VS-VQ-u-1})
and the second inequality comes from the relation $M(\cP^+)\leq \otheta(\cP)$.
\hfill\qed

\section{Concluding Remark}
\label{sec:conclusion}
This paper introduced the methods evaluating the bit error
probabilities.
Furthermore, this paper also proposes the following two open questions:
\begin{enumerate}
 \item
 What measure
 (the bit error probability, the Battacharyya parameter,
  the conditional entropy, the mutual information, and so on)
 is the best for the construction of codes?
 \item
 What measure is the best for the quantization in the early stages of
 polarization?
\end{enumerate}
They are challenges for the future.

\section*{Acknowledgments}
The author thanks Prof.~Y.~Sakai and Prof.~K.~Iwata
for the introduction of the polar codes and helpful discussions.
The author also thanks Dr.~K.~Arai for helpful discussions.

\appendix

\begin{lem}
\label{lem:max}
\begin{equation*}
 \max\{|\theta|,|\theta'|\}
 =
 \frac{|\theta+\theta'|+|\theta'-\theta|}2
\end{equation*}
\end{lem}
\begin{IEEEproof}
 The equality is trivial when $\theta,\theta'\geq 0$ or
 $\theta,\theta'\leq 0$ using the relation
 \begin{equation}
  \max\{\theta,\theta'\}
  =
  \frac{\theta+\theta'+|\theta'-\theta|}2.
  \label{eq:max}
 \end{equation}
 For the rest of the proof, we assume that $\theta\leq 0\leq \theta'$
 without loss of generality.
 In this case, we have
 \begin{align}
  \max\{|\theta|,|\theta'|\}
  &=
  \frac{|\theta|+|\theta'|+||\theta'|-|\theta||}2
  \notag
  \\
  &=
  \frac{-\theta+\theta'+|\theta'-[-\theta]|}2
  \notag
  \\
  &=
  \frac{|-\theta+\theta'|+|\theta'+\theta|}2
  \notag
  \\
  &=
  \frac{|\theta+\theta'|+|\theta'-\theta|}2,
 \end{align}
 where the first equality comes from (\ref{eq:max})
 and the third comes from the fact that $-\theta+\theta'>0$.
\end{IEEEproof}

The following lemma seems to be trivial but we show it for completeness.
\begin{lem}
\label{lem:gth}
Let $\vv\equiv(v_1,\ldots,v_k)$
and $\ttheta_{\vv}=(\theta_{v_1},\ldots,\theta_{v_k})$.
For given function $g$, we have
\begin{equation*}
 \bigcup_{\ttheta:g(\ttheta)=\theta}
 \{\vv:\ttheta_{\vv}=\ttheta\}
 =
 \{\vv:g(\ttheta_{\vv})=\theta\}.
\end{equation*}
for all $\theta$.
\end{lem}
\begin{IEEEproof}
 We prove the lemma by showing
 \begin{align}
  \bigcup_{\ttheta:g(\ttheta)=\theta}
  \{\vv:\ttheta_{\vv}=\ttheta\}
  &\subset
  \{\vv:g(\ttheta_{\vv})=\theta\}
  \label{eq:proof-fth-subset}
  \\
  \bigcup_{\ttheta:g(\ttheta)=\theta}
  \{\vv:\ttheta_v=\ttheta\}
  &\supset
  \{\vv:g(\ttheta_{\vv})=\theta\}.
  \label{eq:proof-fth-supset}
 \end{align}
 
 First, assume that
 \begin{equation*}
  \vv'
  \in
  \bigcup_{\ttheta:g(\ttheta)=\theta}
  \{\vv:\ttheta_{\vv}=\ttheta\}.
 \end{equation*}
 Then there is vector $\ttheta$ such that $g(\ttheta)=\theta$ and $\ttheta_{\vv'}=\ttheta$.
 This implies that
 $g(\ttheta_{\vv'})=\theta$
 and $\vv'\in\{\vv:g(\ttheta_{\vv})=\theta\}$.
 Then we have (\ref{eq:proof-fth-subset}).
 
 Next assume that
 \begin{equation*}
  \vv'\in\{\vv:g(\ttheta_{\vv})=\theta\}.
 \end{equation*}
 Then we have vector $\ttheta\equiv\ttheta_{\vv'}$ such that
 \begin{align}
  g(\ttheta)
  &=
  g(\ttheta_{\vv'})
  \notag
  \\
  &=
  \theta.
 \end{align}
 This implies that $\vv'\in\{\vv:\ttheta_{\vv}=\ttheta\}$
 and (\ref{eq:proof-fth-supset}).
\end{IEEEproof}

\begin{table}
\begin{center}
 \caption{Bit Error Probability of Polar Code with Block Length $2^8=256$}
 \label{table:n=8}
 \begin{minipage}[t]{.3\textwidth}
  \begin{center}
\begin{tabular}[t]{|c||p{21mm}|}
    \hline
    $b^{8}$ & \multicolumn{1}{c|}{Upper bound} \\
    \hline
    \hline
    $00000000$ & $-1.189\times 10^{-28}$ \\
    \hline
    $00000001$ & $-7.710\times 10^{-15}$ \\
    \hline
    $00000010$ & $-7.710\times 10^{-15}$ \\
    \hline
    $00000100$ & $-7.710\times 10^{-15}$ \\
    \hline
    $00001000$ & $-7.710\times 10^{-15}$ \\
    \hline
    $00010000$ & $-7.710\times 10^{-15}$ \\
    \hline
    $00100000$ & $-7.710\times 10^{-15}$ \\
\hline
   \end{tabular}
  \end{center}
 \end{minipage}
 \begin{minipage}[t]{.3\textwidth}
  \begin{center}
   \begin{tabular}[t]{|c||p{21mm}|}
    \hline
    $b^{8}$ & \multicolumn{1}{c|}{Upper bound} \\
    \hline
    \hline
    $00110010$ & $-9.816\times 10^{-3}$ \\
    \hline
    $01010001$ & $-1.041\times 10^{-2}$ \\
    \hline
    $01001100$ & $-1.265\times 10^{-2}$ \\
    \hline
    $10010001$ & $-1.362\times 10^{-2}$ \\
    \hline
    $10001100$ & $-1.602\times 10^{-2}$ \\
    \hline
    $01010010$ & $-1.742\times 10^{-2}$ \\
    \hline
    $00110100$ & $-1.759\times 10^{-2}$ \\
\hline
   \end{tabular}
  \end{center}
 \end{minipage}
\end{center}
\end{table}
\begin{table}
\begin{center}
 \begin{minipage}[t]{.3\textwidth}
  \begin{center}
   \begin{tabular}[t]{|c||p{21mm}|}
    \hline
    $b^{8}$ & \multicolumn{1}{c|}{Upper bound} \\
    \hline
    \hline
    $11010000$ & $\phantom{-}1.657\times 10^{-1}$ \\
    \hline
    $00101111$ & $\phantom{-}1.643\times 10^{-1}$ \\
    \hline
    $10100101$ & $\phantom{-}1.547\times 10^{-1}$ \\
    \hline
    $01101001$ & $\phantom{-}1.521\times 10^{-1}$ \\
    \hline
    $01011100$ & $\phantom{-}1.456\times 10^{-1}$ \\
    \hline
    $10011010$ & $\phantom{-}1.409\times 10^{-1}$ \\
    \hline
    $11000011$ & $\phantom{-}1.218\times 10^{-1}$ \\
\hline
   \end{tabular}
  \end{center}
 \end{minipage}
 \begin{minipage}[t]{.3\textwidth}
  \begin{center}
   \begin{tabular}[t]{|c||p{21mm}|}
    \hline
    $b^{8}$ & \multicolumn{1}{c|}{Upper bound} \\
    \hline
    \hline
    $11110000$ & $\phantom{-}1.077\times 10^{-3}$ \\
    \hline
    $01111001$ & $\phantom{-}9.443\times 10^{-4}$ \\
    \hline
    $11001011$ & $\phantom{-}8.848\times 10^{-4}$ \\
    \hline
    $10110101$ & $\phantom{-}5.891\times 10^{-4}$ \\
    \hline
    $01111010$ & $\phantom{-}5.128\times 10^{-4}$ \\
    \hline
    $11001101$ & $\phantom{-}3.489\times 10^{-4}$ \\
    \hline
    $10110110$ & $\phantom{-}3.133\times 10^{-4}$ \\
\hline
   \end{tabular}
  \end{center}
 \end{minipage}
 \vfill
\end{center}
\end{table}

\begin{table}
\begin{center}
 \caption{Bit Error Probability of Polar Code with Block Length $2^{10}=1024$}
 \label{table:n=10}
 \begin{minipage}[t]{.45\textwidth}
  \begin{center}
   \begin{tabular}[t]{|c||p{21mm}|p{21mm}|}
    \hline
    $b^{10}$ & \multicolumn{1}{c|}{Lower bound} & \multicolumn{1}{c|}{Upper bound} \\
    \hline
    \hline
    $0000000000$ & $-1.599\times 10^{-111}$ & $-1.598\times 10^{-111}$ \\
    \hline
    $0000000001$ & $-2.828\times 10^{-56}$ & $-2.827\times 10^{-56}$ \\
    \hline
    $0000000010$ & $-2.828\times 10^{-56}$ & $-2.827\times 10^{-56}$ \\
    \hline
    $0000000100$ & $-2.828\times 10^{-56}$ & $-2.827\times 10^{-56}$ \\
    \hline
    $0000001000$ & $-2.828\times 10^{-56}$ & $-2.827\times 10^{-56}$ \\
    \hline
    $0000010000$ & $-2.828\times 10^{-56}$ & $-2.827\times 10^{-56}$ \\
    \hline
    $0000100000$ & $-2.828\times 10^{-56}$ & $-2.827\times 10^{-56}$ \\
\hline
   \end{tabular}
  \end{center}
 \end{minipage}
 \begin{minipage}[t]{.45\textwidth}
  \begin{center}
   \begin{tabular}[t]{|c||p{21mm}|p{21mm}|}
    \hline
    $b^{10}$ & \multicolumn{1}{c|}{Lower bound} & \multicolumn{1}{c|}{Upper bound} \\
    \hline
    \hline
    $0000100101$ & $-5.749\times 10^{-14}$ & $-5.748\times 10^{-14}$ \\
    \hline
    $0001000101$ & $-6.141\times 10^{-14}$ & $-6.140\times 10^{-14}$ \\
    \hline
    $0010000101$ & $-6.168\times 10^{-14}$ & $-6.167\times 10^{-14}$ \\
    \hline
    $0100000101$ & $-6.169\times 10^{-14}$ & $-6.168\times 10^{-14}$ \\
    \hline
    $1000000101$ & $-6.169\times 10^{-14}$ & $-6.168\times 10^{-14}$ \\
    \hline
    $0000010110$ & $-6.202\times 10^{-14}$ & $-6.201\times 10^{-14}$ \\
    \hline
    $0000011001$ & $-7.400\times 10^{-14}$ & $-7.399\times 10^{-14}$ \\
\hline
   \end{tabular}
  \end{center}
 \end{minipage}
\end{center}
\end{table}
\begin{table}
\begin{center}
 \begin{minipage}[t]{.45\textwidth}
  \begin{center}
   \begin{tabular}[t]{|c||p{21mm}|p{21mm}|}
    \hline
    $b^{10}$ & \multicolumn{1}{c|}{Lower bound} & \multicolumn{1}{c|}{Upper bound} \\
    \hline
    \hline
    $0001110000$ & $-1.571\times 10^{-9}$ & $-1.570\times 10^{-9}$ \\
    \hline
    $1000100100$ & $-1.836\times 10^{-9}$ & $-1.835\times 10^{-9}$ \\
    \hline
    $0011000010$ & $-2.972\times 10^{-9}$ & $-2.971\times 10^{-9}$ \\
    \hline
    $0010101000$ & $-3.046\times 10^{-9}$ & $-3.045\times 10^{-9}$ \\
    \hline
    $1100000000$ & $-9.032\times 10^{-9}$ & $-9.031\times 10^{-9}$ \\
    \hline
    $0011000100$ & $-1.068\times 10^{-8}$ & $-1.067\times 10^{-8}$ \\
    \hline
    $0100101000$ & $-1.152\times 10^{-8}$ & $-1.151\times 10^{-8}$ \\
\hline
   \end{tabular}
  \end{center}
 \end{minipage}
 \begin{minipage}[t]{.45\textwidth}
  \begin{center}
   \begin{tabular}[t]{|c||p{21mm}|p{21mm}|}
    \hline
    $b^{10}$ & \multicolumn{1}{c|}{Lower bound} & \multicolumn{1}{c|}{Upper bound} \\
    \hline
    \hline
    $0100010101$ & $-7.142\times 10^{-6}$ & $-7.141\times 10^{-6}$ \\
    \hline
    $0001011001$ & $-7.373\times 10^{-6}$ & $-7.372\times 10^{-6}$ \\
    \hline
    $1000010101$ & $-7.446\times 10^{-6}$ & $-7.445\times 10^{-6}$ \\
    \hline
    $0011100000$ & $-7.798\times 10^{-6}$ & $-7.797\times 10^{-6}$ \\
    \hline
    $1010000010$ & $-7.859\times 10^{-6}$ & $-7.858\times 10^{-6}$ \\
    \hline
    $0101010000$ & $-8.940\times 10^{-6}$ & $-8.939\times 10^{-6}$ \\
    \hline
    $0001100101$ & $-1.045\times 10^{-5}$ & $-1.044\times 10^{-5}$ \\
\hline
   \end{tabular}
  \end{center}
 \end{minipage}
\end{center}
\end{table}
\begin{table}
\begin{center}
 \begin{minipage}[t]{.45\textwidth}
  \begin{center}
   \begin{tabular}[t]{|c||p{21mm}|p{21mm}|}
    \hline
    $b^{10}$ & \multicolumn{1}{c|}{Lower bound} & \multicolumn{1}{c|}{Upper bound} \\
    \hline
    \hline
    $0011001010$ & $-6.635\times 10^{-4}$ & $-6.634\times 10^{-4}$ \\
    \hline
    $0100101100$ & $-7.359\times 10^{-4}$ & $-7.358\times 10^{-4}$ \\
    \hline
    $1001000101$ & $-7.412\times 10^{-4}$ & $-7.411\times 10^{-4}$ \\
    \hline
    $0000011111$ & $-7.891\times 10^{-4}$ & $-7.890\times 10^{-4}$ \\
    \hline
    $1100000100$ & $-7.983\times 10^{-4}$ & $-7.982\times 10^{-4}$ \\
    \hline
    $0101000110$ & $-8.068\times 10^{-4}$ & $-8.067\times 10^{-4}$ \\
    \hline
    $0010110100$ & $-8.798\times 10^{-4}$ & $-8.797\times 10^{-4}$ \\
\hline
   \end{tabular}
  \end{center}
 \end{minipage}
 \begin{minipage}[t]{.45\textwidth}
  \begin{center}
   \begin{tabular}[t]{|c||p{21mm}|p{21mm}|}
    \hline
    $b^{10}$ & \multicolumn{1}{c|}{Lower bound} & \multicolumn{1}{c|}{Upper bound} \\
    \hline
    \hline
    $0001101101$ & $-9.555\times 10^{-3}$ & $-9.554\times 10^{-3}$ \\
    \hline
    $1000011011$ & $-1.010\times 10^{-2}$ & $-1.009\times 10^{-2}$ \\
    \hline
    $0010011101$ & $-1.071\times 10^{-2}$ & $-1.070\times 10^{-2}$ \\
    \hline
    $0001110011$ & $-1.156\times 10^{-2}$ & $-1.155\times 10^{-2}$ \\
    \hline
    $0110001010$ & $-1.229\times 10^{-2}$ & $-1.228\times 10^{-2}$ \\
    \hline
    $0001101110$ & $-1.231\times 10^{-2}$ & $-1.230\times 10^{-2}$ \\
    \hline
    $0011100100$ & $-1.245\times 10^{-2}$ & $-1.244\times 10^{-2}$ \\
\hline
   \end{tabular}
  \end{center}
 \end{minipage}
\end{center}
\end{table}
\begin{table}
\begin{center}
 \begin{minipage}[t]{.45\textwidth}
  \begin{center}
   \begin{tabular}[t]{|c||p{21mm}|p{21mm}|}
    \hline
    $b^{10}$ & \multicolumn{1}{c|}{Lower bound} & \multicolumn{1}{c|}{Upper bound} \\
    \hline
    \hline
    $0110010100$ & $-5.523\times 10^{-2}$ & $-5.522\times 10^{-2}$ \\
    \hline
    $1000110011$ & $-5.845\times 10^{-2}$ & $-5.844\times 10^{-2}$ \\
    \hline
    $0101001011$ & $-5.940\times 10^{-2}$ & $-5.939\times 10^{-2}$ \\
    \hline
    $0000111111$ & $-5.947\times 10^{-2}$ & $-5.946\times 10^{-2}$ \\
    \hline
    $0011001110$ & $-5.962\times 10^{-2}$ & $-5.961\times 10^{-2}$ \\
    \hline
    $1001011000$ & $-6.010\times 10^{-2}$ & $-6.009\times 10^{-2}$ \\
    \hline
    $1100000110$ & $-6.063\times 10^{-2}$ & $-6.062\times 10^{-2}$ \\
\hline
   \end{tabular}
  \end{center}
 \end{minipage}
 \begin{minipage}[t]{.45\textwidth}
  \begin{center}
   \begin{tabular}[t]{|c||p{21mm}|p{21mm}|}
    \hline
    $b^{10}$ & \multicolumn{1}{c|}{Lower bound} & \multicolumn{1}{c|}{Upper bound} \\
    \hline
    \hline
    $0100111100$ & $-1.783\times 10^{-1}$ & $-1.782\times 10^{-1}$ \\
    \hline
    $0011101001$ & $-1.797\times 10^{-1}$ & $-1.796\times 10^{-1}$ \\
    \hline
    $1110000000$ & $-1.840\times 10^{-1}$ & $-1.839\times 10^{-1}$ \\
    \hline
    $0011011100$ & $-1.857\times 10^{-1}$ & $-1.856\times 10^{-1}$ \\
    \hline
    $0001111011$ & $-1.874\times 10^{-1}$ & $-1.873\times 10^{-1}$ \\
    \hline
    $1000111010$ & $-1.893\times 10^{-1}$ & $-1.892\times 10^{-1}$ \\
    \hline
    $1100010010$ & $-1.919\times 10^{-1}$ & $-1.918\times 10^{-1}$ \\
\hline
   \end{tabular}
  \end{center}
 \end{minipage}
\end{center}
\end{table}
\begin{table}
\begin{center}
 \begin{minipage}[t]{.45\textwidth}
  \begin{center}
   \begin{tabular}[t]{|c||p{21mm}|p{21mm}|}
    \hline
    $b^{10}$ & \multicolumn{1}{c|}{Lower bound} & \multicolumn{1}{c|}{Upper bound} \\
    \hline
    \hline
    $0110100101$ & $\phantom{-}1.596\times 10^{-1}$ & $\phantom{-}1.597\times 10^{-1}$ \\
    \hline
    $1010110000$ & $\phantom{-}1.576\times 10^{-1}$ & $\phantom{-}1.577\times 10^{-1}$ \\
    \hline
    $1100001011$ & $\phantom{-}1.569\times 10^{-1}$ & $\phantom{-}1.570\times 10^{-1}$ \\
    \hline
    $1100100100$ & $\phantom{-}1.568\times 10^{-1}$ & $\phantom{-}1.569\times 10^{-1}$ \\
    \hline
    $1010010101$ & $\phantom{-}1.560\times 10^{-1}$ & $\phantom{-}1.561\times 10^{-1}$ \\
    \hline
    $1001100110$ & $\phantom{-}1.540\times 10^{-1}$ & $\phantom{-}1.541\times 10^{-1}$ \\
    \hline
    $0010111110$ & $\phantom{-}1.497\times 10^{-1}$ & $\phantom{-}1.498\times 10^{-1}$ \\
\hline
   \end{tabular}
  \end{center}
 \end{minipage}
 \begin{minipage}[t]{.45\textwidth}
  \begin{center}
   \begin{tabular}[t]{|c||p{21mm}|p{21mm}|}
    \hline
    $b^{10}$ & \multicolumn{1}{c|}{Lower bound} & \multicolumn{1}{c|}{Upper bound} \\
    \hline
    \hline
    $1100010101$ & $\phantom{-}5.508\times 10^{-2}$ & $\phantom{-}5.509\times 10^{-2}$ \\
    \hline
    $0011101101$ & $\phantom{-}5.434\times 10^{-2}$ & $\phantom{-}5.435\times 10^{-2}$ \\
    \hline
    $1011000011$ & $\phantom{-}5.406\times 10^{-2}$ & $\phantom{-}5.407\times 10^{-2}$ \\
    \hline
    $1011100000$ & $\phantom{-}5.296\times 10^{-2}$ & $\phantom{-}5.297\times 10^{-2}$ \\
    \hline
    $0110101100$ & $\phantom{-}5.047\times 10^{-2}$ & $\phantom{-}5.048\times 10^{-2}$ \\
    \hline
    $0101111000$ & $\phantom{-}4.863\times 10^{-2}$ & $\phantom{-}4.864\times 10^{-2}$ \\
    \hline
    $0111001001$ & $\phantom{-}4.848\times 10^{-2}$ & $\phantom{-}4.849\times 10^{-2}$ \\
\hline
   \end{tabular}
  \end{center}
 \end{minipage}
\end{center}
\end{table}
\begin{table}
\begin{center}
 \begin{minipage}[t]{.45\textwidth}
  \begin{center}
   \begin{tabular}[t]{|c||p{21mm}|p{21mm}|}
    \hline
    $b^{10}$ & \multicolumn{1}{c|}{Lower bound} & \multicolumn{1}{c|}{Upper bound} \\
    \hline
    \hline
    $1001101011$ & $\phantom{-}9.556\times 10^{-3}$ & $\phantom{-}9.557\times 10^{-3}$ \\
    \hline
    $0111010100$ & $\phantom{-}9.159\times 10^{-3}$ & $\phantom{-}9.160\times 10^{-3}$ \\
    \hline
    $1010110100$ & $\phantom{-}8.158\times 10^{-3}$ & $\phantom{-}8.159\times 10^{-3}$ \\
    \hline
    $1110100000$ & $\phantom{-}8.045\times 10^{-3}$ & $\phantom{-}8.046\times 10^{-3}$ \\
    \hline
    $0011111010$ & $\phantom{-}7.860\times 10^{-3}$ & $\phantom{-}7.861\times 10^{-3}$ \\
    \hline
    $1011010001$ & $\phantom{-}7.256\times 10^{-3}$ & $\phantom{-}7.257\times 10^{-3}$ \\
    \hline
    $1101000101$ & $\phantom{-}6.846\times 10^{-3}$ & $\phantom{-}6.847\times 10^{-3}$ \\
\hline
   \end{tabular}
  \end{center}
 \end{minipage}
 \begin{minipage}[t]{.45\textwidth}
  \begin{center}
   \begin{tabular}[t]{|c||p{21mm}|p{21mm}|}
    \hline
    $b^{10}$ & \multicolumn{1}{c|}{Lower bound} & \multicolumn{1}{c|}{Upper bound} \\
    \hline
    \hline
    $1100111000$ & $\phantom{-}7.321\times 10^{-4}$ & $\phantom{-}7.322\times 10^{-4}$ \\
    \hline
    $1001110110$ & $\phantom{-}7.147\times 10^{-4}$ & $\phantom{-}7.148\times 10^{-4}$ \\
    \hline
    $1110001001$ & $\phantom{-}6.668\times 10^{-4}$ & $\phantom{-}6.669\times 10^{-4}$ \\
    \hline
    $1101010010$ & $\phantom{-}6.418\times 10^{-4}$ & $\phantom{-}6.419\times 10^{-4}$ \\
    \hline
    $1001111001$ & $\phantom{-}4.968\times 10^{-4}$ & $\phantom{-}4.969\times 10^{-4}$ \\
    \hline
    $0111001101$ & $\phantom{-}4.671\times 10^{-4}$ & $\phantom{-}4.672\times 10^{-4}$ \\
    \hline
    $1011101000$ & $\phantom{-}4.576\times 10^{-4}$ & $\phantom{-}4.577\times 10^{-4}$ \\
\hline
   \end{tabular}
  \end{center}
 \end{minipage}
\end{center}
\end{table}
\begin{table}
\begin{center}
 \begin{minipage}[t]{.45\textwidth}
  \begin{center}
   \begin{tabular}[t]{|c||p{21mm}|p{21mm}|}
    \hline
    $b^{10}$ & \multicolumn{1}{c|}{Lower bound} & \multicolumn{1}{c|}{Upper bound} \\
    \hline
    \hline
    $0011111110$ & $\phantom{-}1.460\times 10^{-5}$ & $\phantom{-}1.461\times 10^{-5}$ \\
    \hline
    $1010111010$ & $\phantom{-}1.424\times 10^{-5}$ & $\phantom{-}1.425\times 10^{-5}$ \\
    \hline
    $1111000010$ & $\phantom{-}1.306\times 10^{-5}$ & $\phantom{-}1.307\times 10^{-5}$ \\
    \hline
    $1110101000$ & $\phantom{-}1.125\times 10^{-5}$ & $\phantom{-}1.126\times 10^{-5}$ \\
    \hline
    $0111100110$ & $\phantom{-}1.024\times 10^{-5}$ & $\phantom{-}1.025\times 10^{-5}$ \\
    \hline
    $1100110011$ & $\phantom{-}9.559\times 10^{-6}$ & $\phantom{-}9.561\times 10^{-6}$ \\
    \hline
    $0111011100$ & $\phantom{-}9.456\times 10^{-6}$ & $\phantom{-}9.457\times 10^{-6}$ \\
\hline
   \end{tabular}
  \end{center}
 \end{minipage}
 \begin{minipage}[t]{.45\textwidth}
  \begin{center}
   \begin{tabular}[t]{|c||p{21mm}|p{21mm}|}
    \hline
    $b^{10}$ & \multicolumn{1}{c|}{Lower bound} & \multicolumn{1}{c|}{Upper bound} \\
    \hline
    \hline
    $1110010011$ & $\phantom{-}2.723\times 10^{-8}$ & $\phantom{-}2.724\times 10^{-8}$ \\
    \hline
    $1101011010$ & $\phantom{-}2.676\times 10^{-8}$ & $\phantom{-}2.677\times 10^{-8}$ \\
    \hline
    $1101100101$ & $\phantom{-}2.269\times 10^{-8}$ & $\phantom{-}2.270\times 10^{-8}$ \\
    \hline
    $1011110100$ & $\phantom{-}2.266\times 10^{-8}$ & $\phantom{-}2.267\times 10^{-8}$ \\
    \hline
    $1001111110$ & $\phantom{-}1.501\times 10^{-8}$ & $\phantom{-}1.502\times 10^{-8}$ \\
    \hline
    $1101011100$ & $\phantom{-}1.340\times 10^{-8}$ & $\phantom{-}1.341\times 10^{-8}$ \\
    \hline
    $0110111101$ & $\phantom{-}1.230\times 10^{-8}$ & $\phantom{-}1.231\times 10^{-8}$ \\
\hline
   \end{tabular}
  \end{center}
 \end{minipage}
\end{center}
\end{table}
\begin{table}
\begin{center}
 \begin{minipage}[t]{.45\textwidth}
  \begin{center}
   \begin{tabular}[t]{|c||p{21mm}|p{21mm}|}
    \hline
    $b^{10}$ & \multicolumn{1}{c|}{Lower bound} & \multicolumn{1}{c|}{Upper bound} \\
    \hline
    \hline
    $1111011000$ & $\phantom{-}7.385\times 10^{-13}$ & $\phantom{-}7.386\times 10^{-13}$ \\
    \hline
    $1011011101$ & $\phantom{-}7.041\times 10^{-13}$ & $\phantom{-}7.042\times 10^{-13}$ \\
    \hline
    $1011100111$ & $\phantom{-}6.420\times 10^{-13}$ & $\phantom{-}6.421\times 10^{-13}$ \\
    \hline
    $0111110101$ & $\phantom{-}5.090\times 10^{-13}$ & $\phantom{-}5.091\times 10^{-13}$ \\
    \hline
    $0101111111$ & $\phantom{-}4.916\times 10^{-13}$ & $\phantom{-}4.917\times 10^{-13}$ \\
    \hline
    $1100111011$ & $\phantom{-}4.683\times 10^{-13}$ & $\phantom{-}4.684\times 10^{-13}$ \\
    \hline
    $1111100100$ & $\phantom{-}3.775\times 10^{-13}$ & $\phantom{-}3.776\times 10^{-13}$ \\
\hline
   \end{tabular}
  \end{center}
 \end{minipage}
 \begin{minipage}[t]{.45\textwidth}
  \begin{center}
   \begin{tabular}[t]{|c||p{21mm}|p{21mm}|}
    \hline
    $b^{10}$ & \multicolumn{1}{c|}{Lower bound} & \multicolumn{1}{c|}{Upper bound} \\
    \hline
    \hline
    $1111010101$ & $\phantom{-}6.046\times 10^{-24}$ & $\phantom{-}6.047\times 10^{-24}$ \\
    \hline
    $1111010110$ & $\phantom{-}3.023\times 10^{-24}$ & $\phantom{-}3.024\times 10^{-24}$ \\
    \hline
    $1111100011$ & $\phantom{-}1.605\times 10^{-24}$ & $\phantom{-}1.606\times 10^{-24}$ \\
    \hline
    $1111011001$ & $\phantom{-}1.513\times 10^{-24}$ & $\phantom{-}1.514\times 10^{-24}$ \\
    \hline
    $1111011010$ & $\phantom{-}7.565\times 10^{-25}$ & $\phantom{-}7.566\times 10^{-25}$ \\
    \hline
    $1111100101$ & $\phantom{-}4.018\times 10^{-25}$ & $\phantom{-}4.019\times 10^{-25}$ \\
    \hline
    $1011011111$ & $\phantom{-}3.796\times 10^{-25}$ & $\phantom{-}3.797\times 10^{-25}$ \\
\hline
   \end{tabular}
  \end{center}
 \end{minipage}
 \vfill
\end{center}
\end{table}

\begin{table}
 \caption{Maximum Relative Error and Number of Unclarified Relations}
 \label{table:vs}
 \begin{center}
  \begin{tabular}[t]{|r||l|l|l|l|r|r|}
   \multicolumn{7}{c}{Block length $2^8=256$}
   \\ \hline
   & \multicolumn{2}{c|}{Lower bound (Upgraded)}
   & \multicolumn{2}{c|}{Upper bound (Degraded)}
   & \multicolumn{2}{c|}{Unclarified Relations}
   \\ \cline{2-7}
   \multicolumn{1}{|c||}{$Q$}
   & \multicolumn{1}{c|}{\makebox[14mm]{Proposed}}
   & \multicolumn{1}{c|}{\makebox[14mm]{Conventional}}
   & \multicolumn{1}{c|}{\makebox[14mm]{Proposed}}
   & \multicolumn{1}{c|}{\makebox[14mm]{Conventional}}
   & \multicolumn{1}{c|}{\makebox[14mm]{Proposed}}
   & \multicolumn{1}{c|}{\makebox[14mm]{Conventional}}
   \\ \hline\hline
   $2$
   & $1.00\times 10^0$
   & $1.00\times 10^0$
   & $7.68\times 10^{30}$
   & $1.63\times 10^{35}$
   & $200$ & $202$
   \\ \hline
   $4$
   & $9.95\times 10^{-1}$
   & $9.95\times 10^{-1}$
   & $3.65\times 10^8$   
   & $4.05\times 10^{28}$
   & $110$ & $129$
   \\ \hline
   $8$
   & $4.58\times 10^{-1}$
   & $8.30\times 10^{-1}$
   & $6.82\times 10^0$
   & $8.22\times 10^{20}$
   & $10$ & $44$
   \\ \hline
   $16$
   & $8.29\times 10^{-2}$
   & $3.55\times 10^{-1}$
   & $9.04\times 10^{-2}$
   & $3.10\times 10^{12}$
   & $0$ & $7$
   \\ \hline
   $32$
   & $9.08\times 10^{-3}$
   & $1.27\times 10^{-1}$
   & $5.86\times 10^{-3}$
   & $2.54\times 10^{4}$
   & $0$ & $2$
   \\ \hline
   $64$
   & $4.94\times 10^{-3}$
   & $2.66\times 10^{-2}$
   & $1.52\times 10^{-3}$
   & $1.40\times 10^{2}$
   & $0$ & $0$
   \\ \hline
   $128$
   & $1.36\times 10^{-4}$
   & $5.60\times 10^{-3}$
   & $5.46\times 10^{-5}$
   & $5.52\times 10^{0}$
   & $0$ & $0$
   \\ \hline
   $256$
   & $2.15\times 10^{-5}$
   & $9.39\times 10^{-4}$
   & $8.26\times 10^{-6}$
   & $3.23\times 10^{-1}$
   & $0$ & $0$
   \\ \hline
   $512$
   & $1.59\times 10^{-5}$
   & $1.15\times 10^{-4}$
   & $1.47\times 10^{-6}$
   & $9.17\times 10^{-4}$
   & $0$ & $0$
   \\ \hline
  \end{tabular}
  \\[5mm]
  \begin{tabular}[t]{|r||l|l|l|l|r|r|}
   \multicolumn{7}{c}{Block length $2^{10}=1024$}
   \\\hline
   & \multicolumn{2}{c|}{Lower bound (Upgraded)}
   & \multicolumn{2}{c|}{Upper bound (Degraded)}
   & \multicolumn{2}{c|}{Unclarified Relations}
   \\ \cline{2-7}
   \multicolumn{1}{|c||}{$Q$}
   & \multicolumn{1}{c|}{\makebox[14mm]{Proposed}}
   & \multicolumn{1}{c|}{\makebox[14mm]{Conventional}}
   & \multicolumn{1}{c|}{\makebox[14mm]{Proposed}}
   & \multicolumn{1}{c|}{\makebox[14mm]{Conventional}}
   & \multicolumn{1}{c|}{\makebox[14mm]{Proposed}}
   & \multicolumn{1}{c|}{\makebox[14mm]{Conventional}}
   \\ \hline\hline
   $2$
   & $1.00\times 10^0$
   & $1.00\times 10^0$
   & $8.63\times 10^{151}$
   & $1.15\times 10^{163}$
   & $883$ & $883$
   \\ \hline
   $4$
   & $1.00\times 10^0$
   & $1.00\times 10^0$
   & $1.04\times 10^{62}$
   & $4.41\times 10^{145}$
   & $699$ & $720$
   \\ \hline
   $8$
   & $9.90\times 10^1$
   & $1.00\times 10^0$
   & $3.24\times 10^{19}$
   & $1.00\times 10^{127}$
   & $431$ & $512$
   \\ \hline
   $16$
   & $7.70\times 10^{-1}$
   & $9.84\times 10^{-1}$
   & $6.52\times 10^{-2}$
   & $6.40\times 10^{107}$
   & $124$ & $280$
   \\ \hline
   $32$
   & $3.08\times 10^{-1}$
   & $8.87\times 10^{-1}$
   & $4.27\times 10^{-1}$
   & $2.50\times 10^{88}$
   & $22$ & $144$
   \\ \hline
   $64$
   & $5.29\times 10^{-2}$
   & $6.56\times 10^{-1}$
   & $3.12\times 10^{-2}$
   & $2.34\times 10^{61}$
   & $7$ & $101$
   \\ \hline
   $128$
   & $8.35\times 10^{-3}$
   & $4.80\times 10^{-1}$
   & $4.89\times 10^{-3}$
   & $2.81\times 10^{36}$
   & $0$ & $72$
   \\ \hline
   $256$
   &$1.99\times 10^{-3}$
   &$2.91\times 10^{-1}$
   &$1.10\times 10^{-3}$
   &$6.24\times 10^{35}$
   & $0$ & $48$
   \\ \hline
   $512$
   &$4.12\times 10^{-4}$
   &$1.54\times 10^{-1}$
   &$4.12\times 10^{-4}$
   &$3.08\times 10^{33}$
   & $0$ & $34$
   \\ \hline
  \end{tabular}
\end{center}
\end{table}

\begin{table}
 \caption{CPU Time in Seconds}
 \label{table:vs-cpu}
 \begin{center}
  \begin{tabular}[t]{|r||l|l|l|l|}
   \multicolumn{5}{c}{Block length $2^8=256$}
   \\ \hline
   & \multicolumn{2}{c|}{Lower bound (Upgraded)}
   & \multicolumn{2}{c|}{Upper bound (Degraded)}
   \\ \cline{2-5}
   \multicolumn{1}{|c||}{$Q$}
   & \multicolumn{1}{c|}{\makebox[14mm]{Proposed}}
   & \multicolumn{1}{c|}{\makebox[14mm]{Conventional}}
   & \multicolumn{1}{c|}{\makebox[14mm]{Proposed}}
   & \multicolumn{1}{c|}{\makebox[14mm]{Conventional}}
   \\ \hline\hline
   $2$
   & $7.51\times 10^0$
   & $1.36\times 10^1$
   & $7.78\times 10^0$
   & $1.21\times 10^1$
   \\ \hline
   $4$
   & $1.18\times 10^1$
   & $4.40\times 10^1$
   & $1.80\times 10^1$   
   & $3.52\times 10^1$
   \\ \hline
   $8$
   & $2.19\times 10^1$
   & $9.93\times 10^1$
   & $1.66\times 10^1$
   & $8.48\times 10^1$
   \\ \hline
   $16$
   & $3.18\times 10^1$
   & $1.46\times 10^2$
   & $2.77\times 10^1$
   & $1.25\times 10^2$
   \\ \hline
   $32$
   & $4.65\times 10^1$
   & $2.10\times 10^2$
   & $4.19\times 10^1$
   & $1.72\times 10^2$
   \\ \hline
   $64$
   & $7.91\times 10^1$
   & $3.79\times 10^2$
   & $7.12\times 10^1$
   & $2.92\times 10^2$
   \\ \hline
   $128$
   & $1.36\times 10^2$
   & $7.13\times 10^2$
   & $1.28\times 10^2$
   & $5.50\times 10^2$
   \\ \hline
   $256$
   & $2.89\times 10^2$
   & $1.85\times 10^3$
   & $2.47\times 10^2$
   & $1.21\times 10^3$
   \\ \hline
   $512$
   & $7.13\times 10^2$
   & $6.55\times 10^3$
   & $6.08\times 10^2$
   & $3.33\times 10^3$
   \\ \hline
  \end{tabular}
  \\
  The CPU time of the proposed method with $Q=\infty$ is $1.87\times 10^4$.
  \\[5mm]
  \begin{tabular}[t]{|r||l|l|l|l|}
   \multicolumn{5}{c}{Block length $2^{10}=1024$}
   \\\hline
   & \multicolumn{2}{c|}{Lower bound (Upgraded)}
   & \multicolumn{2}{c|}{Upper bound (Degraded)}
   \\ \cline{2-5}
   \multicolumn{1}{|c||}{$Q$}
   & \multicolumn{1}{c|}{\makebox[14mm]{Proposed}}
   & \multicolumn{1}{c|}{\makebox[14mm]{Conventional}}
   & \multicolumn{1}{c|}{\makebox[14mm]{Proposed}}
   & \multicolumn{1}{c|}{\makebox[14mm]{Conventional}}
   \\ \hline\hline
   $2$
   & $3.29\times 10^1$
   & $1.23\times 10^2$
   & $2.96\times 10^1$
   & $7.08\times 10^1$
   \\ \hline
   $4$
   & $1.50\times 10^2$
   & $6.56\times 10^2$
   & $9.68\times 10^1$   
   & $4.83\times 10^2$
   \\ \hline
   $8$
   & $3.41\times 10^2$
   & $1.34\times 10^3$
   & $2.38\times 10^2$
   & $9.37\times 10^2$
   \\ \hline
   $16$
   & $5.65\times 10^2$
   & $2.08\times 10^3$
   & $4.09\times 10^2$
   & $1.43\times 10^3$
   \\ \hline
   $32$
   & $9.71\times 10^2$
   & $3.99\times 10^3$
   & $6.47\times 10^2$
   & $2.09\times 10^3$
   \\ \hline
   $64$
   & $2.13\times 10^3$
   & $1.46\times 10^4$
   & $1.10\times 10^3$
   & $3.57\times 10^3$
   \\ \hline
   $128$
   & $8.52\times 10^3$
   & $1.02\times 10^5$
   & $2.09\times 10^3$
   & $7.98\times 10^3$
   \\ \hline
   $256$
   & $5.13\times 10^4$
   & $8.42\times 10^5$
   & $5.18\times 10^3$
   & $2.29\times 10^4$
   \\ \hline
   $512$
   & $3.21\times 10^5$
   & $7.08\times 10^6$
   & $1.60\times 10^4$
   & $7.72\times 10^4$
   \\ \hline
  \end{tabular}
\end{center}
\end{table}

\end{document}